\documentclass{emulateapj}
\usepackage{amssymb, amsmath, epsfig}

\newcommand{\Mpc}{{\rm Mpc}}

\newcommand{\cmeter}{{\rm cm}}
\newcommand{\Msun}{{M_{\odot}}}
\newcommand{\eV}{{\rm eV}}
\newcommand{\Ry}{{\rm Ry}}
\newcommand{\second}{{\rm s}}
\newcommand{\Kelvin}{{\rm K}}

\newcommand{\Myr}{{\rm Myr}}
\def\apj{ApJ}
\def\apjl{ApJL}
\def\apjs{ApJS}
\def\aj{AJ}
\def\nat{Nature}
\def\mnras{MNRAS}

\def\pasj{PASJ}

\def\aap{{A\&A}}

\begin{document}

\title{HeII Reionization and its Effect on the IGM}

%\author[M. McQuinn et al.]{Matthew
%McQuinn$^1$\thanks{mmcquinn@cfa.harvard.edu}, Adam Lidz$^1$, Matias
%Zaldarriaga$^{1,2}$, Lars Hernquist$^{1}$,\newauthor
%Suvendra Dutta$^{1}$,
%Philip F. Hopkins$^1$\\ $^{1}$ Harvard-Smithsonian Center for
%Astrophysics, 60 Garden St., Cambridge, MA 02138\\ $^2$ Jefferson
%Laboratory of Physics, Harvard University, Cambridge, MA 02138\\ }

 \author{Matthew McQuinn\altaffilmark{1}, Adam Lidz\altaffilmark{1},
Matias Zaldarriaga\altaffilmark{1,2}, Lars Hernquist\altaffilmark{1},
Philip F. Hopkins\altaffilmark{1}, Suvendra Dutta\altaffilmark{1},
Claude-Andr{\'e} Faucher-Gigu{\`e}re\altaffilmark{1}}

\altaffiltext{1} {Harvard-Smithsonian Center for Astrophysics, 60
Garden St., Cambridge, MA 02138; mmcquinn@cfa.harvard.edu}

\altaffiltext{2} {Jefferson Laboratory of Physics, Harvard University,
Cambridge, MA 02138}

%\pubyear{2006} \volume{000} \pagerange{1}

%\maketitle\label{firstpage}

\begin{abstract}

Observations of the intergalactic medium (IGM) suggest that quasars
reionize HeII in the IGM at $z \approx 3$.  We have run a set of $190$
and $430$ comoving Mpc simulations of HeII being reionized by quasars
to develop an understanding of the nature of HeII reionization and its
potential impact on observables.  We find that HeII reionization heats
regions in the IGM by as much as $25,000 \, \Kelvin$ above the
temperature that is expected otherwise, with the volume-averaged
temperature increasing by $\sim 12,000 \, \Kelvin$ and with large
temperature fluctuations on $\sim 50$ Mpc scales.  Much of this
heating occurs far from quasars by photons with long mean free paths.  We find a
temperature-density equation of state of $\gamma -1 \approx 0.3$
during HeII reionization, but with a wide dispersion in this relation
having $\sigma_{T} \sim 10^4$ K.  HeII reionization by the observed
population of quasars cannot produce an inverted relation ($\gamma - 1
< 0$).  Our simulations are consistent with the observed evolution in
the mean transmission of the HeII Ly$\alpha$ forest.  We argue that
the heat input from HeII reionization is unable to cause the observed
depression at $z \approx 3.2$ in the HI Ly$\alpha$ forest opacity as
has been suggested.  We investigate how uncertainties in the
properties of QSOs and of HeII Lyman-limit systems influence our
predictions.

\end{abstract}

%\begin{keywords}
%cosmology: theory  --
%intergalactic medium  --
%galaxies: high redshift
%\end{keywords}


\section{Introduction}
\label{firstpage}

In the standard picture for the reionization history of the
Universe, radiation from Population II stars ionized the intergalactic
HI at $z>6$ as well as the HeI, converting the vast majority of the
intergalactic helium to HeII.  However, these stars cannot ionize
HeII, and at $z \approx 3$ quasars, with their harder UV spectrum,
doubly ionize the intergalactic helium.  To test this model, many
observations are targeting high redshifts to probe hydrogen reionization
(e.g., \citealt{fan02, taniguchi05, bouwens07, kashikawa06, stark07,
totani06}).  In this picture, HeII reionization occurs at redshifts
for which there is substantially more data on the state of the
intergalactic medium (IGM).

% A definitive identification of the
%redshift of HeII reionization would provide important constraints on
%reionization.  For example, if HeII reionization is found to occur at
%$z < 5$, it is most probable that the sources that reionize the
%hydrogen are sufficiently soft to not be able to doubly ionize the
%helium.

In fact, a number of observations suggest that HeII reionization
happened at $z \sim 3$.  Two measurements of the mean transmission in
the HI Ly$\alpha$ forest have noted an upward bump at $z \approx 3.2$
\citep{bernardi03, faucher07}, which \citet{theuns02} interpreted as
arising from a temperature increase of the IGM during HeII
reionization (but see \citet{faucher07} for alternative explanations).
An increase in the average temperature of the IGM would also decrease
the small-scale fluctuations in the HI Ly$\alpha$ forest.
\citet{ricotti00} and \citet{schaye00} measured the temperature from
the widths of the narrowest lines in the HI Ly$\alpha$ forest and
claimed to have detected a sudden increase in the temperature of
$\Delta T \sim 10^4$ K between $z = 3.5$ and $3$.  Photo-heating
during HeII reionization is the only known process that could be
responsible for such an effect (e.g., \citealt{miralda94, abel99}).

However, a subsequent study by \citet{mcdonald01b} using a similar
method and \citet{zaldarriaga01} using the HI Ly$\alpha$ forest power
spectrum did not confirm this sudden increase in temperature, but
rather found a constant temperature at mean density of $T_0 \approx
17,000$ K for $2 < z < 4$.  
%Although, the measurements of
%\citet{mcdonald01b} and \citet{zaldarriaga01} were taken in only three
%bins centered at $z = 2.4, 3.0$ and $3.9$, which could obscure a
%sudden temperature change.  
Temperatures of $17,000\; K$ are difficult
to explain without HeII reionization occurring at $z \sim 3$
\citep{hui03}, and it is unclear whether as sudden an increase in
temperature as \citet{ricotti00} and \citet{schaye00} find is even
expected theoretically.

If a substantial fraction of the helium is in HeII ($\gtrsim 1\%$),
this would produce a Gunn-Peterson absorption trough in the spectra of
high-redshift quasars at wavelengths blueward of HeII Ly$\alpha$.
Observations of HeII Ly$\alpha$ forest absorption at $2.8 < z < 3.3$
find $10$s of comoving Mpc regions with no detected transmission
\citep{jakobsen94, davidsen96, hogan97, reimers97, heap00}.  These
troughs may signify the presence of diffuse intergalactic HeII.
However, current data, which consist of only a few quasar sight-lines,
do not rule out the intergalactic HeII being primarily ionized and in
photo-ionization equilibrium with a weak background \citep{giroux97,
fardal98,heap00}. The Cosmic Origins Spectrograph, which NASA plans to
install on the Hubble Space Telescope in 2009, will
increase the quantity and quality of HeII Ly$\alpha$ forest
sight-lines.

As the intergalactic HeII becomes progressively more ionized, the
extragalactic UV background will harden around the ionization energy
of HeII at $54.4 \; \eV$.  This hardening will affect the ionization
state of intergalactic metals.  \citet{songaila98} observed a sharp
evolution at $z \approx 3$ in the column density ratios in SIV
(Ionization Potential $= 45.1$ eV) to CIV ($64.5$ eV) absorbers.
\citet{boksenberg03} found evidence for a more gradual hardening of
the background between $2 < z <4$ from the column density ratios of NV
($98 \, {\rm eV}$) to CIV.  Finally, by simultaneously fitting to
multiple metal lines that originate from the same absorption systems,
\citet{agafonova05} and \citet{agafonova07} inferred a background
spectrum that is hardening at $z \approx 3$ near $4$ Ry.

%Furthermore,
%observations by \citet{cowie98} of the metalicity of the IGM between
%$2.8 < z < ?$ find that the CIV/HI column density ratio of an
%inter-galactic cloud is independent of $N_{\rm HI}$ for $3 \times
%10^{13} < N_{\rm HI} <10^{17}$ cm$^{-2}$.  While this result is
%surprising and implies a metallicity that is independent of $N_{\rm
%HI}$ if there is no background above $4$ Ry.  If HeII reionization
%occurred at earlier redshifts, then this would imply that the
%metalicity is higher in lower column density systems that are
%optically thin at $4$ Ry.

%It is compelling that several different observations find evidence for
%a transition in some properties of the IGM at the epoch
%theoretically predicted for HeII reionization by quasars
%\citep{madau99, miralda00, sokasian02}.  It is important to
%understand whether past ``detections'' of HeII reionization are
%consistent with theoretical models.

Measurements from the $z \sim 3$ HI Ly$\alpha$ forest ignore the
effects of a patchy HeII reionization process.  For example, estimates
of the photo-ionizing background and the IGM temperature from the
forest assume a power-law
temperature-density relation. Cosmological parameter studies from the
HI Ly$\alpha$ forest power spectrum make a similar assumption.
Different regions can have vastly different HeII reionization
histories, resulting in a more complicated distribution of
temperatures and pressure-smoothing scales than is commonly adopted.
Realistic simulations of HeII reionization will help quantify the
level at which this process biases these measurements.\footnote{Of note,
\citet{lai06} used simple models for HeII reionization to show that it
has a surprisingly small effect on the Ly$\alpha$ forest power
spectrum on large scales, modifying it at a $ \lesssim 5\%$ level for
wavevectors $k \lesssim 5 \, {\rm comoving} \; \Mpc^{-1}$.}

The aim of this paper is to run realistic simulations of HeII
reionization to understand the morphology of this process as well as
its effect on observables. We concentrate primarily on its impact on
quantities that are sensitive to the IGM temperature, but we also
study the effect of HeII reionization on the transmission in the HeII
Ly$\alpha$ forest.

\citet{sokasian02} and \citet{paschos07} have performed the most
realistic simulations of patchy HeII reionization to date.  There are
several differences between our work and these earlier
investigations.  Both of these studies employed volumes $\leq 100^3$
comoving $\Mpc^3$.  Here, we examine HeII reionization in $186^3$ and
$429^3$ comoving Mpc$^3$ volumes, providing a more representative
cosmic sample.  However, both \citet{sokasian02} and \citet{paschos07}
simulated HeII reionization as a post-processing step on top of
cosmological simulations that included gas dynamics.  Our study instead
uses N-body simulations, which result in a less realistic model for
the gas distribution, but afford a larger dynamic range.  Furthermore,
\citet{sokasian02} assumed sharp ionizing fronts and ignored the
detailed temperature evolution. \citet{paschos07} did not calculate
the gas temperature self-consistently.  Our calculations capture the
width of the ionization fronts and the temperature in a consistent manner.
Finally, in contrast to previous studies, our work presents a large
set of radiative transfer simulations in order to survey the parameter
space.

%There have been several analytic studies of HeII reionization and its
%impact on observations.  An analytic model for HeII reionization has
%been developed recently bin \citet{furlanetto07a} and
%\citet{furlanetto07b}, extending the hydrogen reionization framework
%formulated in \citet{furlanetto04a} that has been remarkably
%successful at reproducing the results of simulations \citep{zahn07}.
%In addition, \citet{gleser05} presented a semi-analytic study of this
%process.  Where possible, we compare our results with those obtained
%in these works.

In Section \ref{code}, we describe the details of our code.  The
models for the quasar sources are described in Section \ref{sources}.
Section \ref{sims} presents the simulations.  Finally, Section
\ref{observations} addresses the implications HeII reionization has on
observations of the HI and HeII Ly$\alpha$ forests.

Throughout, we use a $\Lambda$CDM cosmology with $n_s = 1$, $\sigma_8
= 0.8$, $\Omega_m = 0.27$, $\Omega_{\Lambda} = 0.73$, $\Omega_b =
0.046$, and $h = 0.7$, consistent with the most recent WMAP results
\citep{komatsu08}.  All distances are in comoving units unless
specified otherwise.  An overbar over a variable signifies a volume
average, and $x_Y$ is the fraction of helium/hydrogen that is in
ionization state $Y$.
 
\section{Algorithm}
\label{code}

Our work employs a new ray-tracing code that has been adapted
significantly from the one originally presented in \citet{sokasian01}
and refined in \citet{mcquinn07}.  This code performs
cosmological radiative transfer as a post-processing step on top of
N-body or Smooth Particle Hydrodynamics density fields.  In this study
we use N-body simulations.  
%Radiative transfer in post-processing
%misses the back-reaction that the radiation has on the gas
%distribution.
%\footnote{HeII reionization roughly doubles the IGM temperature, which causes the gas smoothing scale to increase by $%\sim 2^{1/2}$.  \citet{gnedin98} showed that this pressure relaxation time
%for a system takes almost a Hubble time at the mean density.
%Therefore, the factor of $\sim 2^{1/2}$ is likely an over-estimate for regions that were recently ionized.} 
The differences between our code and those described in
\citet{sokasian01} and \citet{mcquinn07} include that it tracks
photons in multiple frequency bins, that it calculates the temperature
of the gas, that it does not assume sharp ionizing fronts, and that it
is parallelized over shared memory.  This section describes the details
of our code, and Appendix \ref{ap:tests} discusses various tests of
it.

At the beginning of each timestep, the code grids the particles from
an N-body snapshot, and inputs a list of source positions and
luminosities.  Then, the code adjusts the neutral fraction to account
for the total number of recombinations that will occur over the
ensuing timestep.  Next, the code casts rays from every source,
randomizing the order of the sources in this loop as well as the
direction in which the rays are cast.  Every ray carries a set number
of photons.  Initially, $N_{\rm ray} = 12\times 4^L$ rays are cast
with $L = 5$ or $6$ for an isotropically emitting source, with these
rays uniformly pixelating the unit sphere using the HealPIX algorithm
as described in \citet{abel02}.\footnote{While the value for $N_{\rm
ray}$ may seem large, most of the computation time is spent tracing
rays that are far from the source so that the speed of the algorithm
depends weakly on the number of rays that are initially cast.}  Each
source must wait for all of the other sources to send a ray before it
sends a second ray.  Rays are split adaptively as they traverse the
box.  At a minimum, one ray from a source intersects every cell face
for cells within the light travel time of a ray. (Although cells that
are tens of Mpc from a quasar typically receive multiple rays from the
source in a timestep, owing to the large numbers of rays that are
initially sent.)

A ray travels either until its photons are expended\footnote{The
termination criterion is that the number of photons in the ray be
$\epsilon$ times the number of atoms in a fully neutral cell with
baryonic overdensity equal to zero, where we set $\epsilon \sim
10^{-4} - 10^{-5}$ with the exact value depending on the run in
question.  The photons from terminated rays are placed in the
background.  We have run a simulation with with $\epsilon = 10^{-6}$
to test convergence (simulation L1d in Table 1) and find no notable
differences with simulation L1 for which $\epsilon = 8 \times
10^{-4}$.  In fact, the temperature and ionization fields are almost
visually indistinguishable between these two simulations.} or until it
has traveled a distance $c \, \Delta t$, where $\Delta t$ is the
simulation timestep.  Rays that have traveled $c \, \Delta t$ are
stored onto disk until the following timestep, at which time the
stored rays are redshifted in frequency, randomly ordered, and recast.
Rays that have traveled a total of $1.5$ box lengths are terminated
and their photons are added to the background (Section
\ref{background}.  Most cosmological radiative transfer codes allow
rays to travel further within a timestep than $c \, \Delta t$.  To
study HeII reionization, it is crucial to capture light-travel
effects.

  For the calculations presented here, rays carry photons with
energies between the ionization potential of HeII ($E_{\rm HeII}
= 54.4~\eV$) and the energy that has a mean free path (m.f.p.)  equal
to $3.5$ times the size of the $186$ Mpc simulation box ($1.8$ times
for the $429$ Mpc box), which works out to $E_{\gamma} \sim 500$ eV in
the $186$ Mpc box at $z = 3$.  The number $3.5$, while somewhat arbitrary,
assures that the flux is uniform on the box scale for photon energies
that are not tracked by the rays.  Photons up to $5$ keV are put into
an ionizing background (\S \ref{background}).  Many of the background
photons are never absorbed because photons with $E_{\gamma} \approx
0.8 \times [(1+z)/4]^{1/2} \; \bar{x}_{\rm HeII}^{1/3} \; {\rm keV}$
have a m.f.p. equal to the Hubble scale.
% The energy bins are specified such that either all bins start off with an
%equal number of photons, an equal total energy, or some compromise
%between these two extremes.  For this study, we use a compromise that
%is between these two cases: a bin with higher mean photon energy has
%fewer photons than a bin at lower photon energies, but it has a higher
%total energy.

Each ray carries some number of photons with energy at $E_i$ for $1
\leq i \leq n_{\nu}$.  This study typically uses $n_{\nu} = 5$.  The
value of $E_{\gamma}$ of the photons within an energy bin is specified
to conserve photon number and total energy given the source spectrum.
The centers of the energy bins that are tracked by the rays are
roughly $(60, 70, 100, 140, 270)$ eV for our fiducial spectrum in the
$186$ Mpc box.  In Appendix \ref{ap:tests},
we show that $5$ is the
minimum number of frequency bins that provides adequate convergence.
The final volume-averaged temperature of a simulation with a much
larger number of energy bins is within $300 \; {\rm K}$ of the same
run with $n_{\nu} = 5$ energy bins, and the ionization and temperature
fields are similar.

%We use as few frequency bins in a ray as is possible to converge to
%the proper solution.  In Appendix \ref{tests} we show that $n_{\nu} =
%5$ provides adequate convergence.  When a ray reaches a cell, it
%breaks up into $m$ sub-rays each with $N_{\gamma}(\nu_i)/m$ photons
%per frequency bin, and then randomly cycles through each of the
%frequency bins in each sub-ray to solve for the ionization state of
%the cell. In the limit that the ray does not appreciable change the
%ionization state of the cell, only one cycle ($m = 1$) is necessary.
%Whereas, if the ray changes the ionization state of the cell
%appreciably, many cycles are needed.  We set $m = 1 + N_{\gamma,
%tot}/\bar{N}_{\rm He}$, where the second term on the RHS is the number
%of photons in the ray divided by the number of helium atoms in a cell
%at mean density.  We find that this method converges quickly to the
%correct solution.

Ideally, radiative transfer codes would employ a timestep that is much
smaller than the photo-ionization equilibrium timescale.  For
cosmological ray tracing codes, such small timesteps can be
computationally intractable (although, see \citealt{trac06}).  Our
algorithm is designed to operate with a timestep that can be much
larger than the equilibration timescale; although, in this study it is
typically used in the regime where the two are comparable.
The timestep used in our simulations is $\Delta
t \approx 10$ Myr.  Compare this to the photo-ionization equilibration
timescale $(\Gamma_{\rm HeII} + \alpha \, n_e)^{-1}$, where
$\Gamma_{\rm HeII}$ is the HeII photo-ionization rate and $\alpha$ is
the recombination rate coefficient.  This timescale, which in most cases is
dominated by the $\Gamma_{\rm HeII}$ term, is observed to be $3-30$ Myr at $2.5
< z < 3$ (i.e. $-15 <\log_{10} \Gamma_{\rm HeII} < -14$ in c.g.s.),
but it can be much shorter near quasars.  An accurate solution can be
obtained with our time steps as demonstrated in Appendix
\ref{ap:correction}.

Converging to the solution also requires capturing the correct order
that rays intersect a cell.  The number of absorptions a ray
experiences depends on whether it arrives before or after previous
rays that have intersected a cell.  If the timestep is small enough,
such that the ionizing front does not move significantly over a
timestep, the order rays hit a cell does not matter.  Our code is not
always operating in this limit.  It attempts to reach a converged
solution by having many rays from a source intersect a cell that is within
$\sim 30$ Mpc from the QSO and by randomizing the order in which the
rays are cast.
 
%This criteria is
%relevant to make sure that the balance of photons, ionizations, and
%recombinations is adequately captured.

%However, a
%code needs to track ${x}_{\rm HeII}$ within the overdense cells to a
%part in $\sim 10^3$ to be able to correctly filter the radiation
%field. A different method is employed for overdense regions that is
%described in \S \ref{LLS} that attempts to capture how these systems
%filter the ionizing radiation field.

%For HeII Ly$\alpha$ absorption statistics, it is also important to
%track $\bar{x}_{\rm HeII}$ in underdense regions to better than one part in
%$10^3$.  Fortunately, this is still possible using a coarse radiative
%transfer timestep.  To do so, we correct $\bar{x}_{\rm HeII}$ using
%the true solution for $\bar{x}_{\rm HeII}$, which takes as input the
%value for $\Gamma_{\rm HeII}$ that our simulation supplies over a
%timestep (Appendix \ref{ap:correction}).

\subsection{Background}
\label{background}
HeII-ionizing photons with $\lambda_{\rm mfp} > 3.5 \, L_{\rm box}$,
where $\lambda_{\rm mfp}$ is calculated assuming a homogeneous IGM
with $x_{\rm HeII} = 1$, are put into a pool of background photons,
which we specify with $10$ frequency bins.\footnote{This  number is for the $186$ Mpc box.  We use the criterion
$\lambda_{\rm mfp} > 1.8\;L_{\rm box}$ for the $429$ Mpc box.}  Five
keV is the maximum photon energy that is included in the background.
Averaged over a timestep, each cell receives a background flux of
\begin{equation}
  f_{\nu} = c \; \frac{N_{\rm bk, \nu}}{L_{\rm box}^3 \; \tau_\nu} \; \left[1 - \exp(-\tau_\nu)\right],
\end{equation}
where $N_{\rm bk, \nu}$ is the total number of background photons in
frequency bin $\nu$, $L_{\rm box}$ is the size of the box, and
$\tau_\nu$ is the average optical depth of a photon traveling a
distance $c \; \Delta t$ in the simulation volume.\footnote{A separate background is also included to determine the ionization
state of hydrogen.  While we could include this ionizing background
self-consistently by extrapolating the $4$ Ry luminosity of quasars in
the box to $1$ Ry, there is mounting evidence that there is a
substantial stellar contribution to the ionizing background at these
redshifts \citep{madau99, steidel01, boksenberg03, sokasian03,
faucher08b}.  Therefore, we instead adopt an empirical approach,
utilizing observations of the HI photo-ionization rate ($\Gamma_{\rm
HI}$) from the HI Ly$\alpha$ forest.\\
\indent \citet{bolton05}, \citet{becker06}, and especially \citet{faucher08b} measured the hydrogen photo-ionization rate ($\Gamma_{\rm HI}$) to be flat as a
function of redshift using the flux decrement method on $2 < z < 4$ HI
Ly$\alpha$ forest data and to within a factor of $\sim 2$ given
by
\begin{equation}
\Gamma_{\rm HI}(z) = 10^{-12} \;{\rm s^{-1}}.
\label{eqn:bkgd}
\end{equation}
We use this function for $\Gamma_{\rm HI}(z)$ throughout this paper
and for all simulated redshifts.  The background level primarily
affects our calculations regarding the HI Ly$\alpha$ forest opacity
and does not influence our conclusions.\\
\indent The approximation of a global background is excellent for tracking the
photo-ionization of HI at the redshifts relevant for HeII reionization
because the m.f.p. for hydrogen ionizing photons is measured to be
$l_{\rm mfp} = 300 \; [(1+z)/4]^{-3.2} \; \Mpc$ \citep{meiksin04},
with present uncertainty at a factor of two level \citep{faucher08b}.
A rough estimate is that there is $1$ galaxy per Mpc$^3$ and one
quasar per $30^3$ Mpc$^3$, such that there are thousands of
sources that contribute to the background within $l_{\rm mfp}$.  This
implies that spatial fluctuations in the background are small at $z
\lesssim 5$ \citep{croft04, mcdonald05}.  For similar reasons, this
approximation is also valid for the hard HeII-ionizing photons that
are treated as a global background by our code.}

%For photons that are treated as part of the uniform background, their
%ionizing proximity region around QSOs is not modeled correctly.  The
%size the proximity region around a quasar -- defined here as the
%region at which the ionizing flux from the quasar is larger than the
%ionizing flux of the background -- is $r_p \approx (4 \pi \; n_{\rm
%QSO} \; l_{\rm mfp})^{-0.5}$ (assuming quasars are the dominant source
%for the ionizing background).  The value of $r_p$ for the HI proximity
%region is $\approx 3$ Mpc, assuming $n_{\rm QSO} = [30\; {\Mpc}]^{-3}$
%at $z = 3$ (and $r_p$ is similar for HeII ionizing photons that are
%treated as a uniform background).\footnote{This crudely estimated
%value of $r_p= 3$ is smaller than the measured proximity region size for many
%bright quasars.  Because the proximity region size scales as the
%square root of luminosity, the rare bright quasars for which this has
%been observed have large proximity regions, but these regions
%encapsulate a small fraction of the total volume of the Universe.}  For $r_p= 3$ Mpc, %a small fraction ($\approx 0.3\%$) of the box
%volume is within the proximity region of a QSO.\footnote{In contrast,
%for photons near the HeII Lyman-limit, the m.f.p. is comparable to the
%distance between QSOs such that a significant fraction of space is in
%a QSO proximity region.}

\subsection{Temperature}
\label{temperature}

Our code tracks the temperature evolution of the gas, which is
governed by the differential equation (e.g., \citealt{hui97})
\begin{equation}
\frac{dT}{dt} = -2 \; H \; T + \frac{2 \;T}{3\;\Delta_b} \; \frac{d\delta_b}{dt} - \frac{T}{\sum \tilde{X}_i}\;\frac{d \sum \tilde{X}_i}{dt} + \frac{2}{3 k_B n_{\rm tot}}\;\frac{dQ}{dt},
\label{eqn:temp}
\end{equation}
where $\tilde{X}_i$ is defined such that the number density in species
$i$ is $(1+\delta) \; \tilde{X}_i \; \rho_b/m_p$, $dQ/dt$ is the
heating rate, and $n_{\rm tot}$ is the total number of gas
particles.\footnote{Several published studies that solve for $T$ use
an incorrect form for the third term on the R.H.S. of equation
(\ref{eqn:temp}) -- a factor of $-2/3$ different from what appears in
this equation. This mistake stems essentially from writing the first
law of thermodynamics as $dq_p = d\epsilon_p + P \, d(1/n)$, where
$q_p$ is the external heating per particle, $\epsilon_p$ is the energy
density per particle, and $P$ is the pressure.  The correct form for
this equation is $dq_p = d\epsilon_p + P \, d(1/(\mu m_p n))$, where
$\mu$ is the mean molecular weight and $m_p$ is the proton mass.
Unlike in the former equation, the second term in the correct version
does not depend on the ionization state of the gas.  This mistake
results in the temperature of the gas increasing during ionization
processes by too large an amount.}  The first and second
terms on the right-hand side of equation (\ref{eqn:temp}) account for
the effect of adiabatic heating/cooling, the third term describes the
change in the number of gas particles owing to ionizations and
recombinations, and the final term accounts for radiative heating and
cooling.  The most important process that contributes to $dQ/dt$ is
photo-heating, but atomic cooling, recombination cooling, dielectric
recombination cooling, collisional cooling, and Compton cooling off of
CMB photons are also included in our calculation.\footnote{We assume a
power-law of $1.6$ in energy flux to calculate the HI photo-heating.
We neglect HeI photo-heating in this calculation.  The heat input from
HeI photo-heating is $\approx 25\%$ that of the HI photo-heating.
Uncertainties in the spectrum of the photo-ionizing radiation result
in larger uncertainties in the photo-heating rate than the HeI
contribution to it.}  We use an implicit solver to obtain a solution
to equation (\ref{eqn:temp}).

To evaluate the Lagrangian time derivatives in equation
(\ref{eqn:temp}) within a cell, the locations of particles at the
beginning of the timestep are used to calculate the initial values for
the ionization, density, and temperature.  Namely, each particle
initially has the ionization, density, and temperature of the cell
that it was in at the end of the previous timestep. Summing up these
previous values for all the particles within the cell (with the
appropriate weighting to conserve energy) gives the initial values of
these scalars.
%\footnote{Since
%the timesteps in our simulation are coarse, we assume that the
%evolution of the photo-heating contribution to $dQ/dt$, of $\delta_b$,
%and of the $x_{\rm HeII}$ are respectively constant, linear, and
%linear during a timestep.  The assumptions of constancy and linearity
%do not affect the conclusions.}

%The solution to equation (\ref{eqn:temp}) is generated on
%a coarse grid.  If we average equation (\ref{eqn:temp}),
%weighting by $\Delta_b$ over a cell volume, this equation becomes
%\begin{eqnarray}
%\frac{d\langle T \rangle_M}{dt} &=& -2 \; H \; \langle T \rangle_M + \langle \frac{2}{3} \;\frac{T}{\Delta_b} \;\frac{d\delta_b}{dt} \rangle_M \nonumber \\
%& &- \langle \frac{T}{\sum \tilde{X}_i}\;\frac{d \sum \tilde{X}_i}{dt}\rangle_M + \frac{2}{3 k_B}\;\frac{dq}{dt},
%\label{eqn:temp2}
%\end{eqnarray}
%where $\langle ... \rangle_M$ denotes the mass-averaged quantity and
%$q = \langle Q/n_{\rm tot} \rangle_M$.  All contributions to $Q$ aside
%from atomic cooling processes are linear in $\Delta_b$.  Fortunately,
%atomic cooling is a subdominant contribution to the RHS of equation
%(\ref{eqn:temp2}) for densities relevant to the IGM.  Therefore, only
%the second term on the RHS of equation (\ref{eqn:temp2}) -- an
%adiabatic cooling/heating term -- depends significantly on subgrid
%fluctuations.

The density field is calculated from the gridded N-body particles. The dynamics of N-body particles are different from that
of gas particles.  Therefore, it is questionable whether this
algorithm will be able to predict the correct temperature.  However,
\citet{hui97} demonstrated that the temperature evolution produced
from smoothing dark matter simulations at the Jeans scale provides
good agreement with the temperature evolution seen in hydrodynamic
simulations.  In addition, \citet{hui97} showed that even using linear perturbation
theory to evolve the density field provides
reasonable agreement with the evolution of the $T$-$\Delta_b$ relation
in hydrodynamical simulations, suggesting that exactly capturing the
Jeans scale is not crucial.  A comparison of the temperature evolution
our code predicts with the \citet{hui97} analytic formula is
presented in Appendix \ref{ap:tests}.

The simulations in our study capture the Jeans scale to varying
degrees. [Although, more appropriate may be the
filtering scale, which is typically a factor $\sim 2$ smaller
\citep{gnedin98}.]  The Jeans mass is given by
\begin{equation}
M_{\rm J} = 9.6 \times 10^9 \; \Delta_b^{-1/2} \; \left(\frac{T}{10^4 \; {\rm K}} \right)^{3/2} \; \left(\frac{1+z}{4} \right)^{-3/2} \, \Msun.
\label{eqn:jeans}
\end{equation}
Compare this mass to the mass within a grid cell in our simulations:
\begin{equation}
M_{\rm cell} = 1.4 \times 10^{10} \; \Delta_b \;\left(\frac{L_{\rm box}}{186 \; \Mpc} \; \frac{256}{N_{\rm cells}} \right)^3 \; \Msun.
\label{eqn:Mcell}
\end{equation}
These masses are comparable for $\Delta_b = 1$ for our fiducial
resolution and box size ($L_{\rm box} = 186 \; \Mpc$ and $N_{\rm
cells}=256$).  For the $512^3$ runs in the $186 \; \Mpc$ box, $M_{\rm
cell} \ll M_{\rm J}(z = 3)$ for $\Delta_b = 1$, and, in the $429 \;
\Mpc$ box, $512^3$ resolution achieves $M_{\rm cell} \approx M_{\rm
J}(z = 3)$ for $\Delta_b = 1$.  Most important, our conclusions do not
change if we use the $512^3$ mesh rather than $256^3$.

\subsection{Secondary Photo-ionizations and Heating}

We assume that the excess energy above $E_{\rm HeII}$ from a photon that photo-ionizes a HeII ion goes
into heating the gas.  In reality, the secondary electron
produced by the photo-ionization may go on to collisionally ionize another
atom, excite a bound electron, or suffer coulomb collisions that heat
the gas \citep{shull85}.

However, \citet{shull85} find that $99\%$ of the excess energy of a
$3$ keV X-ray photon that photo-ionizes a HeII ion goes into heating
the gas for a situation in which $\bar{x}_{\rm HI} = 0.05$,
$\bar{x}_{\rm HeI} = 0.05$, and $\bar{x}_{\rm HeII} = 0.95$.  This
percentile will increase for the smaller $\bar{x}_{\rm HI}$,
$\bar{x}_{\rm HeI}$, and photon energies that are most relevant during
HeII reionization.  Therefore, the assumption that all the excess
energy goes into heating the gas is excellent.  Furthermore,
\citet{shull85} find that interactions with HeII are unimportant in
their calculations.  Most of the excess energy is converted into heat
because collisions with electrons dominate the energy loss of the
secondary particles once hydrogen is reionized.

Finally, Compton cooling of the energetic electrons off of the CMB,
which was not included in the calculations of \citet{shull85}, is a
subdominant energy loss mechanism at $z\sim 3$ for the relevant gas
densities and photon energies \citep{madau04}.

\subsection{Heat Input Estimates}
\label{heat_input}

 Let us develop an understanding for how much heating is expected from
 HeII reionization. An HeII ionization front heats up the gas behind
 it by (e.g., \citealt{abel99})
\begin{eqnarray}
    k_b \, \Delta T &\approx& \frac{2 \, Y_{\rm He}}{3 \, (8 - 5 \; Y_{\rm He})} ~ ( \int_{E_{\rm HeII}}^\infty \frac{d E}{E} \times \nonumber\\
 &&  \left(E - E_{\rm HeII}\right) \; \sigma_{\rm HeII}(E) \; J_i(E) \; e^{-\tau_E}) \nonumber \\
    & & / \left( \int_{E_{\rm HeII}}^\infty \frac{dE}{E} \; \sigma_{\rm HeII}(E) \; J_{i}(E) \; e^{-\tau_E} \right),
\label{eqn:DeltaT}
\end{eqnarray}
where $J_i$ is the unabsorbed intensity emitted by the source,
$J_i(E_{\gamma}) \exp(-\tau_E)$ is the incident spectrum, $\tau_E$ is
the optical depth from the source to the gas parcel, $Y_{\rm He}$ is
the primordial helium mass fraction, and $\sigma_{\rm HeII}$ is the
HeII photo-ionization cross section.

For $\tau_E \ll 1$ and using that $\sigma_{\rm HeII} \sim (E_{\gamma} - E_{\rm HeII})^{-3}$,
equation (\ref{eqn:DeltaT}) becomes
\begin{equation}
\Delta T \approx \frac{2}{3} \, \frac{E_{\rm HeII}}{27 \,k_b \, (2 +\alpha_{\rm UV})} = 4500 \, \left(\frac{3.5}{2 + \alpha_{\rm UV}} \right)\; {\rm K} ,
 \label{eqn:Tsmallr}
\end{equation}
where $\alpha_{\rm UV}$ is the spectral index of $J_i(E_\gamma)$,
$E_{\rm HeII}/(2 +\alpha_{\rm UV})$ is the average excess energy of a
photon above $E_{\rm HeII}$, and $27$ is the number of particles over
which this energy is distributed assuming $Y_{\rm He} = 0.25$.  
%Modulo cooling, the value for $\Delta T$ in equation
%(\ref{eqn:Tsmallr}) determines the $r =0$ value for the heating for
%the curves in Figure \ref{fig:num_freq}, which is discussed in
%Appendix \ref{ap:tests}.  These curves describe the heating and
%ionization around a single QSO embedded in a homogeneous medium.
If $\tau_E$ is appreciable, equation (\ref{eqn:Tsmallr}) still
has some relevance, except replace $\alpha_{\rm UV}$ with the
effective spectral index of the incident spectrum, which will be
harder, resulting in larger $\Delta T$. 

% This effect is also seen in Figure
%\ref{fig:num_freq}.  The temperature at the ionization front edge is
%higher as time proceeds because the radiation is
%hardening as it travels away from the source.

The value $4500 \; {\rm K}$ in equation (\ref{eqn:Tsmallr}) is quite small, and it might be difficult
to reconcile such a small $\Delta T$ with measurements of $\Delta T$ in the Lyman-$\alpha$ forest (assuming HeII reionization is the cause of the temperature increase). It turns out
that this number is an underestimate for the total heating during HeII
reionization. If we assume all photons up to $E_{\rm max}$ are
absorbed then the heat injection is
\begin{equation}
 \Delta T \approx 31,000\, \left(\frac{0.5}{\alpha_{\rm UV} - 1} \right) \, \left(1-  \frac{\alpha_{\rm UV} \, E_{\rm HeII}^{\alpha_{\rm UV} - 1}}{E_{\rm max}^{\alpha_{\rm UV} - 1}}\right)  \; {\rm K},
 \label{eqn:Toptthick}
\end{equation}
%\begin{eqnarray}
%& \Delta T &\approx \frac{2}{3} \, \frac{E_{\rm HeII}}{27 \,k_b (\alpha_{\rm UV} - 1)} \, \left(1-  \frac{\alpha_{\rm UV} \; E_{\rm HeII}^{\alpha_{\rm UV} - 1}}{E_{\rm max}^{\alpha_{\rm UV} - 1}}\right) \nonumber \\
%&=& 31,000\, \left(\frac{0.5}{\alpha_{\rm UV} - 1} \right) \, \left(1-  \frac{\alpha_{\rm UV} \, E_{\rm HeII}^{\alpha_{\rm UV} - 1}}{E_{\rm max}^{\alpha_{\rm UV} - 1}}\right)  \; {\rm K},
% \label{eqn:Toptthick}
%\end{eqnarray}
where we have kept only the leading order in ${E_{\rm HeII}}/{E_{\rm
max}}$.  This equation can be derived by setting $\sigma_{\rm HeII} =
1$ and $\tau_E = 0$ in equation (\ref{eqn:DeltaT}) and performing the
integrals.  The heating implied by equation (\ref{eqn:Toptthick}) is
much larger than by equation (\ref{eqn:Tsmallr}). Several hundred eV
photons have optical depth unity on roughly the $186$ Mpc box scale
assuming a homogeneous IGM with $x_{\rm HeII} = 1$, so these photons
will be absorbed somewhere in the IGM if they are produced during HeII
reionization (a photon travels $850$ Mpc between $z =4$ and
$3$, roughly the interval over which HeII reionization occurs in our
simulations).  

If we set $E_{\rm max} = 350$ eV [which has
$\tau \sim 1$ over $200$ Mpc for $\bar{x}_{\rm HeII} = 1$] yields
$\Delta T = 13,000 \; {\rm K}$ using equation (\ref{eqn:Toptthick})
and $\alpha_{\rm UV} = 1.5$ (and $15,000\; {\rm K}$ if we had not
expanded in ${E_{\rm HeII}}/{E_{\rm max}}$). This number is comparable
to the average temperature increase seen in our simulations discussed
in Section \ref{sims}.

%Finally, the different temperature inceases
%given in the optically thin limit (eqn. \ref{eqn:Tsmallr}) and the
%optically thick limit (eqn. \ref{eqn:Toptthick}) suggest that there
%will be large temperature fluctuations in the IGM -- regions near
%quasars where the quasar light is unobscured will be heated less than
%regions where

\subsection{Recombination Rates}
\label{ss:rec_rates}

The recombination timescale for HeIII is $t_{\rm rec,
HeIII} = [\alpha_{\rm A}(T) \; {n}_e(z)]^{-1}$, where
$\alpha_{\rm A}$ is the relevant Case A recombination coefficient and
$\bar{n}_e$ is the mean electron density. If $t_{\rm rec, HeIII}$ is
expressed in terms of the age of the Universe, $t_{\rm uni} \approx
2/3 \; H(z)^{-1}$, it yields
\begin{equation}
\frac{t_{\rm rec, HeIII}}{t_{\rm uni}} \approx 0.6 ~ \left(\frac{T}{10^4
      \; {\rm K}}\right)^{0.7} \; \left(\frac{1+z}{4}\right)^{-3/2} \;
      \Delta_b^{-1},
\label{eqn:rec_time}
\end{equation}
where $\Delta_b$ is one plus $\delta_b$ -- the overdensity in gas --
and we have assumed that the intergalactic hydrogen is fully ionized.
The HeII recombination timescale is $5.4$ times shorter than this
timescale for hydrogen at $T= 2\times 10^4 \; {\rm K}$.  This fact is
important because additional recombinations require more ionizations
to reionize the IGM, which requires more energy injection.  Although,
the recombination timescale is comparable to the gas cooling timescale, so
that the additional heating does not lead to significantly larger temperatures
in regions that recombine and are reionized \citep{bolton-HeII}.

%Also, note that if we use the Case B rate, then equation
%(\ref{eqn:rec_time}) is the same but with $0.6 \rightarrow 0.9$. 

  Since the number of HeIII recombinations during HeII reionization is
large, recombination radiation could contribute significantly to
$\Gamma_{\rm HeII}$.\footnote{Ground state recombination radiation
results in a line profile of the form $\exp[-(E_{\gamma} - E_{\rm
HeII})/kT]$ for $E_{\gamma} > E_{\rm HeII}$, such that most photons
have excess energies of $\sim 1$ eV above $E_{\rm HeII}$ for
characteristic temperatures of $T \sim 10^4$ K.  A photon will travel
$80 \; [(1 +z)/4]^{-1/2} \; ([E_{\gamma} - E_{\rm HeII}]/1 \, {\rm
eV})$ Mpc prior to redshifting below $E_{\rm HeII}$.  This distance is
longer than the m.f.p. for such photons to be absorbed, and,
therefore, they will typically be absorbed either in dense clouds or
by diffuse intergalactic HeII.}  Unfortunately, HeII-ionizing photons
produced by recombinations to the ground state of HeII are not
followed by our radiative transfer code.  \citet{fardal98} showed that
this radiation could contribute to the photo-ionization rate at the
$\sim 20\%$ level.  Tracking this radiation with our code would be
prohibitively expensive because it would require treating all cells as
sources of ionizing radiation.\footnote{Two recent ray tracing codes
outlined in \citet{pawlik08} and \citet{altay08} have developed
techniques to alleviate this problem.}  Instead, our code either
assumes that this recombination radiation contributes locally to
ionizations, which it does by using the Case B recombination
coefficient for HeII, or that these photons do not contribute to
ionizations of the diffuse IGM, which it does by using the Case A
recombination coefficient.  The code also uses the corresponding
recombination cooling rates for these two cases.  Treating HeII
recombinations with the Case A coefficient results in a hotter IGM;
more ionizing photons from QSOs are required to ionize the IGM, and
the QSO photons are harder than the photons from recombinations to the
ground state.  Neither Case A nor Case B is correct in detail, and the
most appropriate choice for the recombination coefficient depends on
the ionization state of the gas.

%After HeII reionization, the HeII-ionizing recombination radiation has
%a shorter mean free path compared to a typical ionizing photon
%produced by a quasar.  This could result in recombination radiation
%contributing negligibly to the global radiation background, making it
%more appropriate to use the Case A recombination coefficient in the
%low density IGM.  

%However, this is true in the limit that the
%m.f.p. of ground state recombination photons is much shorter than that
%of the ionizing photons from QSOs so that they contribute negligibly
%to the HeII-ionizing radiation background, which is never really the case.

During HeII reionization, a large fraction of the ionizing photons
from recombinations will ionize the diffuse IGM since the HeII bubble
size is comparable to the m.f.p. to be absorbed in dense systems.  In
this case, Case B is the better choice.  Since this paper is most
interested in studying the HeII reionization process, most of our
simulations use the Case B recombination coefficient.  However, we
have run a case that uses the Case A coefficient (simulation D4 in
Table 1).  This simulation results in a slightly higher
volume-averaged temperature at the end of HeII reionization, with the
difference being $\Delta \bar{T} \approx 1000$ K when compared to a
similar simulation that uses the Case B rate.  The character of the
ionization and temperature fields are similar between these two
simulations.

\subsection{Mean Free Path of Ionizing Photons}
\label{LLS}

At our grid scale, subgrid fluctuations may play an important role in
absorbing and hardening the typical spectrum -- ``filtering'' -- of
the HeII-ionizing radiation.  In Appendix \ref{ap:LLS}, we investigate
how dense clumps filter the radiation, and we quantify how well this
effect is captured in our simulations.  Systems that have HI column
densities $N_{\rm HI} \sim 10^{15} \, \cmeter^{-2}$ are responsible
for filtering HeII ionizing radiation in HeIII regions during HeII
reionization.  These systems are much less overdense and more diffuse
than those that limit the m.f.p. of HI Lyman-limit photons, having
overdensities of $\delta_b \sim 10$ at $z \sim 3$ \citep{schaye01},
and, therefore, they can be captured with lower resolution simulations
than their counterparts for HI Lyman-limit photons, which have
$\delta_b \sim 100$.  In Appendix \ref{ap:NHI}, we show that our
gridded density field has some success reproducing the column density
distribution of these systems when compared to high-resolution
hydrodynamic simulations.  However, even though the column density
distribution is reproduced, the bias and density of these systems is
altered compared to high resolution hydrodynamic simulations. In
addition, our radiative transfer code may systematically over-ionize
these systems, as described in Appendix \ref{ap:NHI}.  Therefore, we
supplement our calculations with two prescriptions (Methods A and B) to
study the effect that dense systems have on filtering the ionizing
radiation.  These methods are described in detail in Appendix
\ref{ap:LLS} and briefly here.

Method A assumes that the high-column density HeII systems that are
resolved in our simulations are in photo-ionization equilibrium.  We
argue that this approach is reasonable in Appendix \ref{ap:LLS}.
Method B uses a \citet{haardt96}-like model that takes the
distribution of $N_{\rm HI}$ measured from Lyman-$\alpha$ forest
spectra and, given a model for the density of these systems and the
local value of the photo-ionization rates in the simulation, infers
the HeII column density distribution and,
therefore, the opacity in these systems.  Further, this model assumes
that these absorbers are associated with halos in our simulation that
have $m > 7\times 10^9 \, \Msun$ and places the absorbers at the
locations of these halos in the simulation box.  The reader might be
wary of such methods for supplementing the resolution, but we find
that these prescriptions have only a minor impact on the final
results (Section \ref{ss:filtering}).

If much of the intergalactic helium is in HeII, the m.f.p. is often
limited by diffuse gas rather than by dense systems.  In the limit
that diffuse regions near the mean density dominate the opacity, the
m.f.p for a photon with energy $E_{\gamma}$ is given by
\begin{equation}
\lambda_{\rm mfp} \approx 5 \; \bar{x}_{\rm HeII}^{-1} \, \left(\frac{E_{\gamma}}{100 \; {\rm eV}}\right)^{3} \; \left(\frac{1+z}{4}\right)^{-2} ~~ \Mpc.
\label{eqn:mfp}
\end{equation}
The value of $\lambda_{\rm mfp}$ scales strongly with
$E_{\gamma}$ in this limit.
%\footnote{Note that the value of
%$\lambda_{\rm mfp}$ scales less strongly with $E_{\gamma}$ after HeII
%reionization when the diffuse gas has been ionized.
%\citet{furlanetto07a}, using the formula derived in \citet{zuo93}, found
%that if the number of HeII absorbers per unit redshift scales as the
%HeII column density to the $-3/2$ power (roughly appropriate for
%ionization equilibrium), then $\lambda_{mfp} \propto%\
%E_{\gamma}^{-3/2}$ rather than $E_{\gamma}^{-3}$ as one finds in the
%homogeneous case.  The approximate scaling can also be inferred from
%Figure \ref{fig:abs_prob} in Appendix \ref{ap:LLS}.}  
A photon with $E_{\gamma} = 55$ eV has $\lambda_{\rm mfp} = 0.8 \;
\Mpc$ at $z = 3$, whereas one with $E_{\gamma}= 200$ eV has
$\lambda_{\rm mfp} = 40$ Mpc.  The average energy of a photon for our
fiducial UV power-law index of $\alpha_{\rm UV} = 1.6$ is $150$ eV.
However, half of the HeII-ionizing photons have $54.4 < E_{\gamma} <
84$ eV.

\section{Ionizing Source Model}
\label{sources}

Quasars are the leading candidates for reionizing HeII.  Cooling
radiation from massive halos is the next most likely contender
\citep{miniati04}.  However, the large fluctuations in the HeII to HI
column density ratios at $z\sim 3$ strongly disfavor cooling radiation
as being the dominant source of HeII-ionizing photons.  Too many
sources of cooling radiation are present within one mean free path to
source these large fluctuations \citep{bolton06}.  Finally, there are
other more exotic sources that could potentially ionize HeII at $z
\lesssim 6$.  Possibilities include an unknown source of $>4$ Ry
radiation that is generating the HeII $1640 \; \AA$ recombination line
observed in the composite spectrum of $z \approx 3$ Lyman break
galaxies \citep{shapley03,
furlanetto07b}\footnote{\citet{furlanetto07b} find that only if HeII
ionizing photons can escape with $f_{\rm esc, HeII} \gtrsim 0.5$ from
galaxies can they contribute a significant fraction of the ionizing
background. These values are much larger than the theoretical
expectation for $f_{\rm esc, HeII}$ from galaxies (e.g.,
\citealt{gnedin07}).}, or the bi-products of decaying/annihilating
light dark matter particles.  These more exotic scenarios are probably also in
conflict with observations for the same reason that cooling radiation
is disfavored.  Our study concentrates on HeII reionization by
quasars.

Detailed observations of quasars in the past couple decades have provided
many clues into the nature of these extremely luminous objects.  We
now know that quasars are powered by accretion onto super-massive
black holes.  Presently, the $z \lesssim 3$ luminosity function of QSOs is well
constrained over a few decades in luminosity in both the optical and
the X-ray.  The clustering of these objects reveals that optically
selected quasars reside in halos with $m
\approx 10^{12}\; \Msun$, independent of their luminosity
\citep{porciani04, croom05}.  Although, clustering is
currently measured over only about a decade in luminosity.  For the
analysis in this study, the biggest uncertainties in modeling QSOs are
their lifetimes, $\tau_{\rm QSO}$, the fraction of HeII ionizing
photons that escape the environs of a QSO into the IGM as a function
of direction, and their spectrum between $4$ Ry and $1$ keV.

\begin{figure}
\epsfig{file=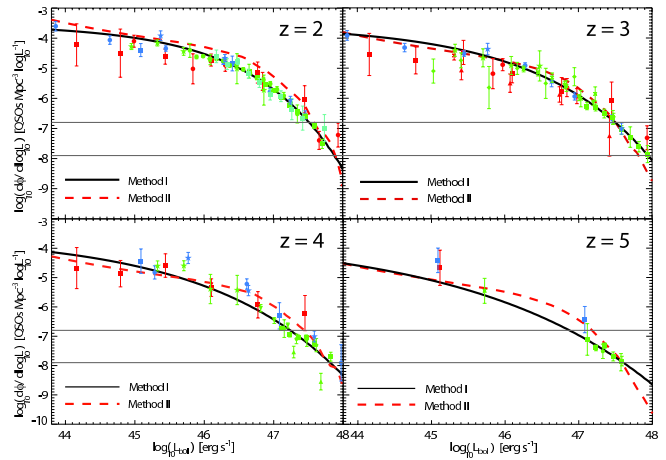, width=8.6cm}
\caption{Comparison of models and measurements of the bolometric
QSO luminosity function at $4$ redshifts.  The data points are from a
compilation of observations across different wavebands
\citep{hopkins07a}, using a spectral template to infer the bolometric
luminosity function. The solid curves are the \citet{hopkins07a}
best-fit luminosity function (which the quasars in Method I are drawn
from) and the dashed curves are the luminosity functions generated
with Method II.  The solid horizontal lines represent the number densities at which $1$ QSO is in the $186$ and $429$ Mpc boxes.
\label{fig:quasar_lf}}
\end{figure}

Figure \ref{fig:quasar_lf} plots the bolometric luminosity function of
quasars for four redshifts.  The black, solid curves are the \citet{hopkins07a} best-fit
model.\footnote{The \citet{hopkins07a} luminosity function is derived
by simultaneously fitting to the infrared, optical, and X-ray
luminosity functions.  These fits model the observed quasar spectra
with a bolometric luminosity-dependent intrinsic spectral template
plus a model for the column-density distributions in $N_{\rm HI}$ and
dust.  This procedure results in a good fit to the observed luminosity
function in all bands \citep{hopkins07a}.}
%While the luminosity function at $4-40$ Ry is more relevant for this study than the bolometric luminosity function, the luminosity function %at these $E_{\gamma}$ only differs from the bolometric luminosity function by a redshift-independent factor in our models.  The data %points in Figure Figure \ref{fig:quasar_lf} represent the most recent observations.  
The cyan points with error-bars are infrared observations, the green
are optical, the blue are soft X-ray, and the red are hard X-ray.  The
references for these observations are given in \citet{hopkins07a}, and
the symbols follow the same conventions as described in their Table 1.
Observations in different bands are included in this figure by
assuming an intrinsic spectral template and an average column density
distribution for the obscuring matter that is motivated by X-ray
measurements of $N_{\rm HI}$.  At $z > 3$, the luminosity function
falls in amplitude.  At $z >5$ there is not much data to
constrain the models.  The abundance of QSOs between $3 \lesssim z
\lesssim 4.5$ shapes the duration and morphology of reionization in
our simulations.  Insufficient HeII-ionizing photons are produced at
higher redshifts to substantially ionize the HeII with the best-fit
luminosity function model.  In this model, $\sim 2$ HeII ionizing
photons per helium atom are produced by $z =3$, about enough to reionize
the HeII.

The faint-end slope of the the luminosity function $d \phi/d \log L$
is constrained to be rather flat at $z \approx 3$, with power-law
index $-0.5$.  This flatness means that most of the HeII-ionizing
photons are produced by $L \sim L_*$ quasars.  The horizontal lines in
the top panel in Figure \ref{fig:quasar_lf} demark the number
densities that yield $1$ quasar in the $186$ and $429$ Mpc simulation
boxes.  Our simulations are large enough to contain many $L_*$ QSOs.

Quasars are placed in our simulations with one of two methods:

\subsection{QSO Method I}
 The lifetime of all quasars is fixed to be $\tau_{\rm QSO}$, and
quasars are placed in halos of mass $m \propto L^{4/3}$.  This scaling
assumes QSOs shine at their Eddington Luminosity ($L_{\rm edd} \propto
M_{\rm BH}$) with some duty cycle and that $M_{\rm bh} \sim \sigma^4
\sim m^{4/3}$.\footnote{The specific mapping we use is $m = 7.9 \times
10^{12} \; [(5 \, \nu L_\nu)/10^{46} \, {\rm erg \,s}^{-1}]^{3/4} \;
[(1 +z)/4]^{1.5} \; \Msun$, where $\nu L_\nu$ is evaluated at $1$ Ry.
Our results depend weakly on the mapping between $L$ and $m$ because
the morphology of HeII reionization turns out to be determined
primarily by Poisson fluctuations in the abundance of quasars rather
than by their clustering.} The number of QSOs in the box at a given
luminosity is drawn from a Poisson distribution with mean set by the
observed luminosity function. We use the \citet{hopkins07a} luminosity
function evaluated at $912\; \AA$ and extrapolate across the relevant
band with a power-law $\alpha_{\rm UV}$.

\subsection{QSO Method II}

  Recently, simulations of merging galaxies have been used to predict
quasar light curves \citep{hopkins05, hopkins06}, and these
predictions have been remarkably successful at reconciling
observations of the quasar luminosity function, quasar clustering, and
the unresolved X-ray background (e.g., \citealt{hopkins06}).  The
bolometric light curve of the quasars in merger simulations can be roughly parametrized as
\begin{eqnarray}
L_{\rm bol}(t) =  0.8 \;L_{\rm edd}(m_{\rm bh}) \times \; \biggl\{ \begin{array}{ll}
   \exp[t/\tau_S] & ~~~t < 0 \\
   \left(1 + t/\tau_S \right)^{-b} & ~~~t > 0, \end{array}
\label{eqn:QSOlum}
\end{eqnarray}
where $m_{\rm bh}$ is the super-massive black hole mass, $b = 1.7 +
0.7\times \log_{10}(L_{\rm edd}/ 10^{12} \; L_{\sun})$, $L_{\rm edd}$
is the Eddington luminosity, and $\tau_S = 40$ Myr.  To map from
$L_{\rm bol}(t)$ to the intrinsic luminosity at the HI Lyman-limit, we
take the \citet{hopkins07a} $L_{\rm bol}$-dependent spectral template.

Next, equation (\ref{eqn:QSOlum}) plus a relationship between halo
mass and $L_{\rm edd}(m_{\rm bh})$ allows one to deconvolve from the
observed QSO luminosity function the triggering rate per unit volume
of quasars as a function of time.  The triggering rate is defined as
the rate quasars shine at their peak luminosity.  We use the Magorrian
relationship to relate a galaxy's stellar mass to its super-massive
black hole mass \citep{hopkins07b, hopkins07c}, and the halo
occupation model of \citet{tinker05, conroy06,vale06} (for details,
see \citet{hopkins08a}) to map from halo mass to
stellar mass.\footnote{As a final ingredient, since there is scatter
in the relationship between $\bar{m}_{\rm bh}$ and $M_{\rm stellar}$,
we assume that the actual $m_{\rm bh}$ is log-normal function with
dispersion $0.3$ dex and logarithmic mean given by the Magorrian
relationship.  This step is necessary in order to reproduce the bright
end of the luminosity function.} Figure \ref{fig:quasar_lf} show the
luminosity function that this model generates (red, dashed curves).
While it differs somewhat from the best-fit luminosity function
(black, solid curves), it agrees well with the observations.  To
implement this model in our simulations, we switch on quasars in the
simulation box using the triggering rate per unit halo mass described above,
accounting for Poisson scatter.

%The QSO triggering rate allows one to calculate $\bar{N}_{\rm QSO}(m,
%\Delta m, z)$ -- the average number of halos with mass between $m$ and
%$m + \Delta m$ that are triggered within the simulation box.  We then
%draw the number of triggerings in our box $N_{\rm QSO}(m, \Delta m, z)$ from
%a Poisson distribution with mean $\bar{N}_{\rm QSO}(m, \Delta m, z)$,
%and we randomly select $N_{\rm QSO}(m,\Delta m,z)$ halos to be triggered
%within the simulation box that are in this mass range. 

\subsection{Discussion of QSO Models}

\begin{figure}
\rotatebox{-90}{\epsfig{file=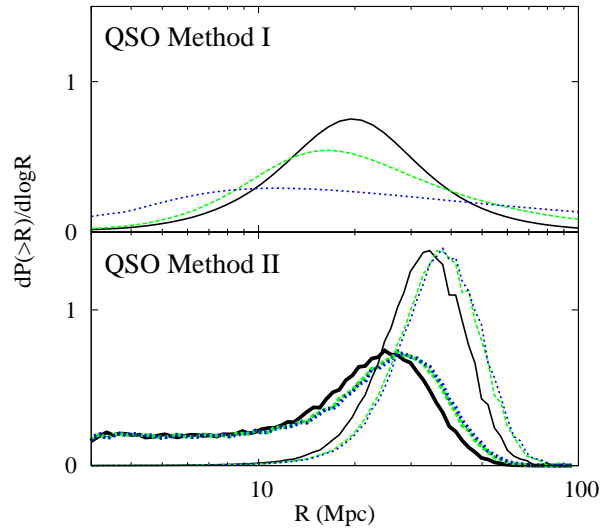, height=8.3cm}}
\caption{HeIII bubble size distributions in our two QSO models
assuming $\alpha_{\rm UV} = 1.6$, isotropic emission, $1$ QSO per
bubble, and no recombinations.  The black solid curves correspond to
$z= 2$, the green dashed to $z = 4$, and the blue dotted to $z= 6$.
Top panel: Thin curves are the volume-weighted probability
distribution of bubble radii in Method I.  These curves assume
$\tau_{\rm QSO} = 40$ Myr. Bottom panel: Thick curves are the
volume-weighted probability distribution in Method II, and the thin
curves are the number-weighted probability distribution.
\label{fig:quasars_Rb}}
\end{figure}

Figure \ref{fig:quasars_Rb} shows the HeIII bubble size probability
distribution function (PDF) in Method I and Method II.  These bubble
sizes correspond to the size of a bubble if all the QSOs sit in
separate spherical HeIII regions and there are no recombinations.  All
curves are calculated assuming $\alpha_{\rm UV} = 1.6$.
For the thin curves, the PDF is volume-weighted, and, for the thick
curves in the bottom panel, it is number-weighted.  Method I has an
infinite number of quasars with $L < L_*$ and so the number-weighted
PDF is not well-defined.

The HeIII bubbles from the quasars in Method II have a characteristic
size of $35$ Mpc in the volume-weighted PDF, and the probability
distribution of $R_b$ does not vary significantly with redshift.  With
Method I, the characteristic radius is a bit smaller, $20$ Mpc for
$\tau_{\rm QSO} = 40$ Myr, and the dispersion in $R_b$ is larger.
Method I has a non-negligible probability for extremely large bubble
sizes to exist, bubbles with $R_b \gtrsim 70$ Mpc.  These large
bubbles are averted in Method II because brighter quasars have
the shorter lifetimes in this method.  The properties of HeII
reionization do not change substantially if we employ Method I or
Method II in our simulations.

Given the large observational uncertainty in the mean and distribution
of $\alpha_{\rm UV}$ (see Appendix \ref{ap:qso}), we leave these
quantities as free parameters in both QSO Models.  For each quasar in
the simulation box, $\alpha_{\rm UV}$ is drawn from a Gaussian
distribution with mean $\bar{\alpha}_{\rm UV}$ and
s.d. $\sigma_{\alpha}$.  Our fiducial model assumes $\bar{\alpha}_{\rm
UV} = 1.6$ and $\sigma_{\alpha} = 0.2$, which is motivated by the
results of \citet{telfer02} and \citet{steffen06} [Appendix
\ref{ap:qso}].  We find that the amount of heating from HeII
reionization \emph{is} sensitive to the spectrum of QSOs.
%Note that our power-law parametrization for
%the UV spectrum can only crudely account for intrinsic HeII absorption
%in the host system (which could be present), which will obscure
%preferentially low-energy photons.

We also include a parameter $\zeta$, a suppression factor for
the number of HeII-ionizing photons that are emitted by QSOs.
This parameter gives us a knob to tune in order to have HeII
reionization occur at a desired redshift, and, given current
uncertainties in the QSO luminosity function, can be adjusted at a
factor of $2$ level.  
%Finally, we present one simulation in which the
%QSO emission is beamed, where the beaming model is described in Appendix
%\ref{ap:qso}.

\section{Simulations}
\label{sims}

\begin{table*}
\begin{center}
\caption{HeII Reionization simulations.\label{table1}}

\end{center}
\begin{center}
\begin{tabular}{l c c c c c c l}
\hline Sim.$^a$ & $L_{\rm box}$ (Mpc) & LLS$^b$ & QSO Model & Emission$^{c}$ & $\zeta$ &
Grid & Comments\\
\hline
D1 & $186$ & & II &  iso & $0.75$ & $256$ &  \\
D2 & $186$ & & II & iso & $0.75$ & $256$ & no density fluctuations ($\delta_b = 0$)\\
D3 & $186$ & & II & iso & $0.75$ & $256$ &  twice as many frequency bins as D1\\
D4 & $186$ & & II & iso & $0.75$ & $256$ & case A recombinations for HeII \\
\hline
L1 & $186$ & A & II & iso & $0.75$ & $256$ &  \\
L1b & $186$ & A & II & iso & $0.75$ & $256$ & $\gamma = 0$ at $z = 6$ \\
L1c & $186$ & A & II & iso & $1.5$ & $256$ & slightly earlier reionization than L1 \\

L1d & $186$ & A & II & iso & $0.75$ & $256$ & extremely conservative photon termination \\
L2 & $186$ & A & II & iso & $1$ & $512$ &  high resolution run \\
L3 & $186$ & B & II & iso & $0.75$ & $256$ &  includes $N_{\rm HI} > 10^{14.5} \, \cmeter^{-2}$\\
\hline
X1 & $186$ & A & II & iso & $0.7$ & $256$ &  omits photons in background\\
\hline
S1 & $186$ & A & II & beamed & $0.75$ & $256$ &  beam uses \citet{gilli07} model\\
S2 & $186$ & A & I & iso & $2$ & $256$ & $\tau_{\rm QSO} = 10 ~\Myr$\\
S3 & $186$ & A & II & iso & $0.75$ & $256$ & assumes $\bar{\alpha}_{\rm UV} = 1.2$\\
S4 & $186$ & A & II & iso & $0.75$ & $256$ & assumes $\bar{\alpha}_{\rm UV} = 0.6$\\
S4b & $186$ & A & II & iso & $1.25$ & $256$ & assumes $\bar{\alpha}_{\rm UV} = 0.6$, no background\\
\hline
B1 & $429$ &  & II & iso & $1$ & $256$ & \\
B2 & $429$ & A & II & iso & $1$ & $512$ & \\ \hline
\end{tabular}
\end{center}
$^a$ \, Unless specified otherwise, the initial conditions for these
simulations are $\gamma - 1 = 0.3$ and $T_0 = 10^4$ K at $z = 6$, and
$\bar{\alpha}_{\rm UV} = 1.6$.  All simulations allow for a random
dispersion in $\alpha_{\rm UV}$ with $\sigma_{\alpha} = 0.2$.  At fixed $\zeta$, the number of photons per timestep is the same in all the simulations that use  QSO Model II, independent of $\bar{\alpha}_{\rm UV}$.\\ $^b$ \,
{The ``LLS'' column specifies the method for capturing the dense absorbers,
as discussed in Appendix \ref{ap:LLS}.\\} $^c$ \, {iso = isotropic
emission\\ }

\end{table*}

\begin{figure}
\rotatebox{-90}{\epsfig{file=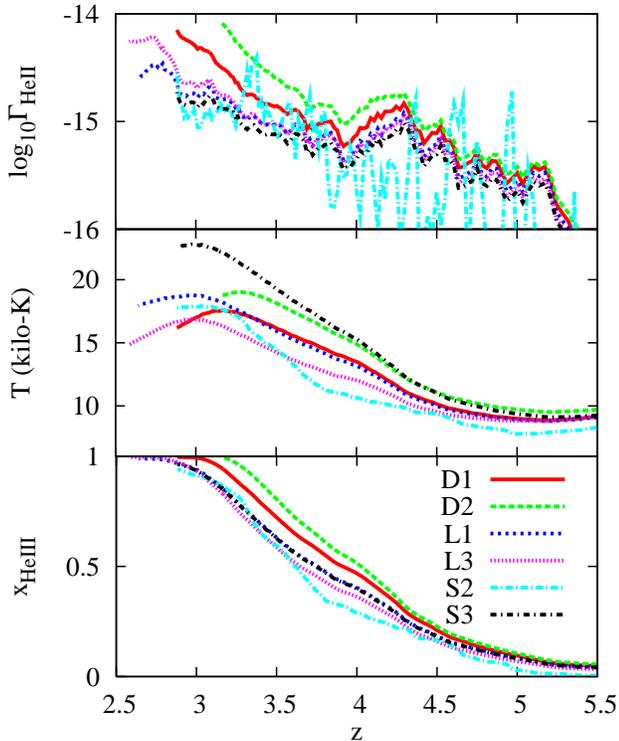, height=8.6cm}}
\caption{Volume-averaged global history of
$\Gamma_{\rm HeII}$ in $s^{-1}$, $T$ in kilo-Kelvin, and
$\bar{x}_{\rm HeIII}$, in six of our simulations. 
\label{fig:global_hist}}
\end{figure}

We have run a series of simulations to investigate the morphology of
HeII reionization.  All of our simulations track quasars to $L_{\rm bol} >
10^{43}$ or $10^{44}$ erg s$^{-1}$, approximately $3$ decades
below $L_*$.  Table \ref{table1} lists the simulations discussed in
this paper, and Figure \ref{fig:global_hist} plots the volume-averaged
history of $\Gamma_{\rm HeII}$, $T$, and ${x}_{\rm HeIII}$ in six of
these simulations.  The simulations in Figure \ref{fig:global_hist}
use different prescriptions to include the QSOs and to capture the
high column-density absorption systems.  Yet, with a minor amount of
tuning, the end of HeII reionization occurs near $z \sim 3$ in all of
the simulations.\footnote{Note
that we tune the parameter $\zeta$, which normalizes the total
ionizing emissivity, between $0.75$ and $2$ in the simulations
presented here in order for HeII reionization to end at $z \approx 3$.
Given observed uncertainty in the QSO luminosity function and spectral
behavior near the HeII Lyman-limit, such tuning is acceptable.  This
aids comparison between simulations, and it also allows the
simulations to address better the set of observations that indicate
that HeII reionization is nearing completion at $z \approx 3$.}  The known
population of quasars is able to ionize the HeII in the Universe by
$2.5 < z < 3.5$, approximately the redshift range where several
observations suggest that HeII reionization is ending.  In
the end, $2.2$ photons per helium atom are required to reionize the
HeII to $\bar{x}_{\rm HeII} = 0.95$ in simulation D2 (which employed no
filtering method) and $3.4$ photons per helium atom are required in
simulations L1 and L3 (which used filtering methods A and B).

 The middle and top panels in Figure \ref{fig:global_hist} show the
history of $\bar{T}$ and $\bar{\Gamma}_{\rm HeII}$.  In these
simulations the temperature of the IGM peaks near the end of HeII
reionization, and $\bar{\Gamma}_{\rm HeII}$ increases during this
process. The jaggedness of $\Gamma_{\rm HeII}(z)$ owes to sample
variance.\footnote{The jaggedness is particularly acute in simulation
S2 because it uses the shortest quasar lifetimes, it has the fewest
quasars at high redshifts, and its brightest quasars have the same
lifetime as its dimmest ones.}

%Simulations B1 and B2, whose curves are not shown in Figure
%\ref{fig:global_hist}, are in the $430$ Mpc simulation box rather than
%the $186$ Mpc. The evolution of $\bar{\Gamma}_{\rm HeII}$ is smoother
%in these simulations owing to the larger sample of the Universe.  We
%find that the evolution of $\bar{T}$ and $\bar{x}_{\rm HeIII}$ is
%similar in the bigger volume, suggesting that the $186$ Mpc box is
%sufficiently large to make predictions for these volume-averaged
%quantities.

The remainder of this section discusses our simulations in detail.
But first, it is important to highlight a drawback of our radiative
transfer method.  Time steps are set to $\Delta t = 10$ Myr, such that
a ray will travel a distance $c \, \Delta t$ over a timestep before
being held in memory.  This introduces a characteristic scale of $c \,
\Delta t = 3$ proper Mpc that can appear in the simulations.  In
particular, the effect of a finite timestep is sometimes apparent in
the $\Gamma_{\rm HeII}$ field, through rings that show sharp gradients
in $\Gamma_{\rm HeII}$.  These small artifacts do not affect our
conclusions.

% Second, there is
%a trade-off between the computation time and the number of rays that
%are cast.  The more rays, the less noise in the calculations,
%especially for quantities like $\Gamma_{\rm HeII}$ that depend on the
%number of photons that reach a cell during a particular
%timestep. Figure \ref{fig:homogeneous}, which is run with our fiducial
%quasar model in a homogeneous density field, gives some sense for the
%amount of noise that is present in our simulations.  This effect can
%be discerned visually by looking for deviations from sphericity in
%locations where a single QSO ionizes its HeIII region.  None of the
%conclusions presented in this study are altered by increasing the
%number of rays.

\subsection{Homogeneous Universe}
\label{ss:delta_homo}

\begin{figure*}
\begin{center}
\epsfig{file=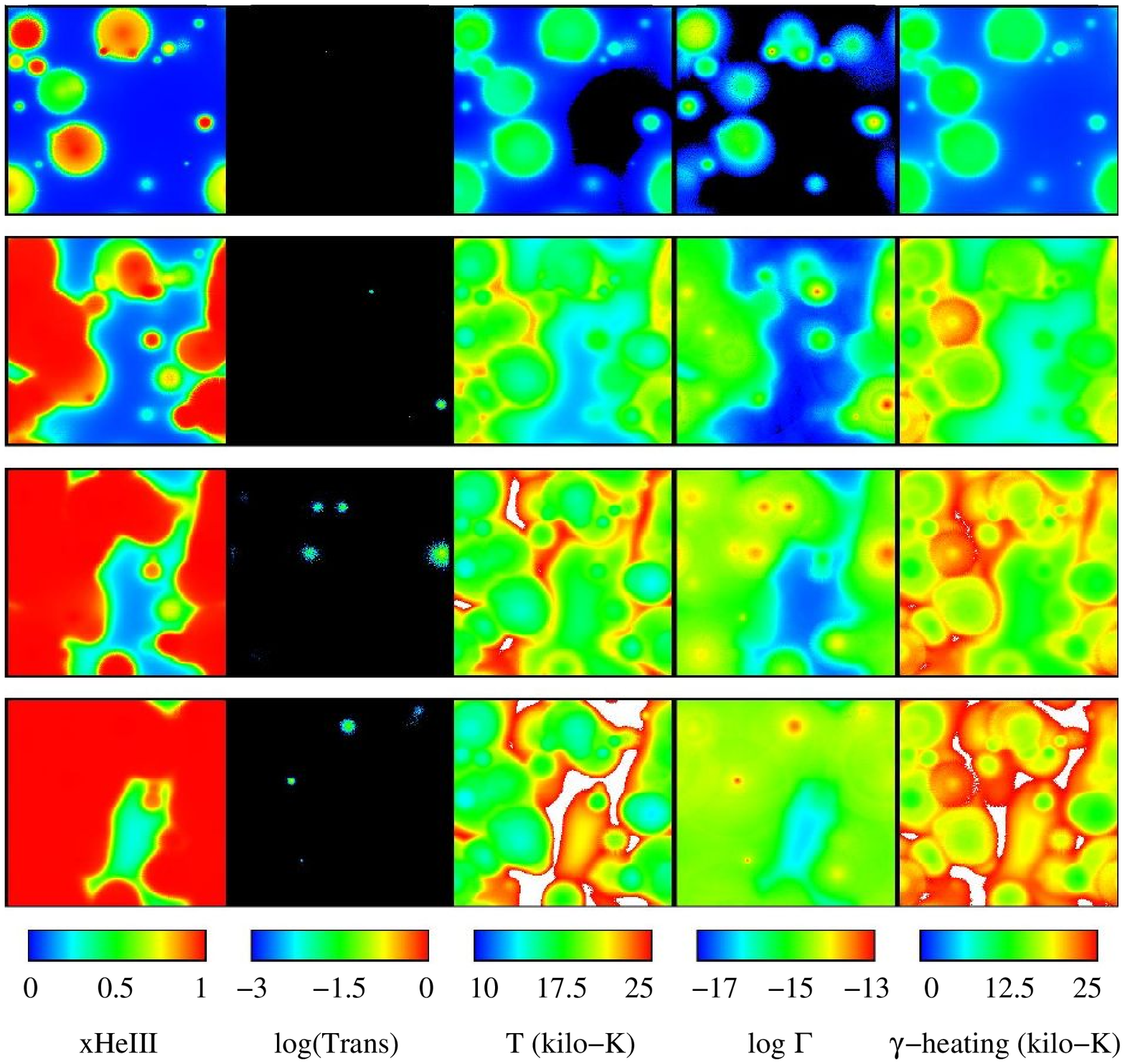, width=14cm}
\end{center}
\caption{ Slices through simulation D2, which assumes a homogeneous
 IGM ($\Delta_b = 1$).  Each slice has breadth $186$ Mpc and width
 $3.5$ Mpc (except for the second column, which has a width of $1$
 grid cell or $0.7$ Mpc). From top to bottom, the panels in each row
 feature snapshots at $z = 4.3, 3.5, 3.2$ and $3$ respectively,
 corresponding to $\bar{x}_{\rm HeIII}= 0.1$, $0.5$, $0.8$ and $0.99$.
 The first column shows the spatial distribution of $x_{\rm HeIII}$.
 The second column depicts $\log_{10}$ of the transmission in the HeII
 Ly$\alpha$ forest. The middle column shows the temperature of the
 IGM.  The fourth column plots $\log_{10}$ of $\Gamma_{\rm HeII}$, and
 the right-most column shows the cumulative amount of heat deposited
 by HeII photo-ionizations.  White regions represent higher values
 than shown in the scale and black represent lower values.
\label{fig:homogeneous}}
\end{figure*}

 Simulation D2 is run in the $186$ Mpc box using the positions of the
halos in the N-body code and Method II for populating these halos with
quasars.  In order to isolate some interesting physical effects, this
simulation is run in a homogeneous density field
($\Delta_b(\boldsymbol{x}, z) = 1$) rather than using the density
field from the N-body simulation as in the other cases.  The left-most
panels in Figure \ref{fig:homogeneous} depict the HeII ionization
field at several different times.  Initially,
the HeIII bubbles around each QSO are isolated spheres with
characteristic circular radii of $\approx 20 \; \Mpc$.\footnote{This
value for $R_b$ may seem somewhat smaller than those projected in
Figure \ref{fig:quasars_Rb} for several reasons: (1) A random slice
through a bubble will on average have a factor of $2$ smaller radius
than the 3-D radius; (2) active quasar bubbles are smaller than the
final bubble size -- the quantity plotted in Figure
\ref{fig:quasars_Rb} (the quasars in Method II have long lifetimes
that are comparable to the Salpeter time); (3) a fraction of the
ionizing photons are deposited ahead of the front and some reionize
recombining atoms; (4) $\zeta < 1$ in D2.}

The edges of the HeIII regions are not sharp, with the region over
which $0.1 < x_{\rm HeIII} < 0.9$ spanning several Mpc and with a
lower ionization tail extending much further.  When the Universe is
roughly a few tenths ionized, the spherical HeIII regions begin to merge and form larger non-spherical HeIII
regions.  In addition, the fractionally ionized regions that extend
far ahead of the ionizing front overlap, creating a global
ionization floor.  By the time $\bar{x}_{\rm HeIII} > 0.8$ in
simulation D1, $x_{\rm HeIII} > 0.2$ everywhere.

Since the recombination timescale at mean density is shorter than the
Hubble time and comparable to the HeII reionization timescale, relic
HeIII regions recombine significantly over the course of HeII
reionization.  Relic HeIII regions in which the HeIII has
substantially recombined are present in the top-right panel in Figure
\ref{fig:homogeneous}.  However, these regions persist only
while the HeII ionizing background has $\Gamma_{\rm HeII} \lesssim
\alpha_{\rm B} \, \bar{n}_e \approx 2 \times 10^{-17} \; {\rm s}^{-1}$
at $z \approx 4$, where $\alpha_{\rm B}$ is the Case B recombination
coefficient.  Otherwise, ionizations will be sufficient to balance recombinations.  Relic HeIII regions that have substantially
recombined exist in our simulations primarily during the early phases
of HeII reionization ($\bar{x}_{\rm HeIII} \lesssim 0.5$).  The fourth
column in Figure \ref{fig:homogeneous} displays $\Gamma_{\rm HeII}$.
The condition $\Gamma_{\rm HeII} > 2\times 10^{-17} \; {\rm s}^{-1}$
is met in most regions in the IGM at times when $\bar{x}_{\rm HeII} \gtrsim
0.5$.

The panels in the second column of Figure \ref{fig:homogeneous}
show $\log_{10}$ of the transmission in the HeII Ly$\alpha$ forest.
For a homogeneous Universe, there is little transmission in the HeII
Ly$\alpha$ forest, even at times near the end of HeII reionization.
The inclusion of density inhomogeneities significantly enhances the 
transmission level (Section \ref{delta_inhomo}).

The IGM temperature field has a different morphology from the
$x_{\rm HeIII}$ field (compare the third column in Fig.
\ref{fig:homogeneous} with the first).  Early on in HeII reionization,
the morphologies of both the $x_{\rm HeIII}$ and $T$ fields are driven
by isolated HeIII regions.  The temperature of an isolated HeIII
region is $\Delta T \approx 7,000$ K above the average temperature
in HeII regions, with hotter regions at the HeIII front edge and
slightly colder regions inside the front.  Once ionized, the
HeIII region begins to cool adiabatically (owing to cosmic expansion).
Regions that were ionized earlier are cooler than regions that have
been ionized more recently. The volume-averaged temperature increase
of $10,000\, {\rm K}$ during HeII reionization is consistent with
estimates that assume that the IGM absorbs all photons with energies
less than a few hundred eV during HeII reionization (Section 2.3).
%A second effect that
%accentuates these holes is that the outskirts of the HeII regions are
%heated up by harder photons than inside the front owing to the
%filtering of the ionizing radiation field as it travels.  This effect
%is most clear in the left-most column, which displays the temperature
%increase from HeII photo-heating.  The temperature difference between
%the relic HeIII regions and their outskirts is partially mitigated by
%including recombinations, which allow extra heat to be injected into
%the relic HeIII as recombined atoms are again re-ionized.
 
The right-most column in Figure \ref{fig:homogeneous} displays the
cumulative amount of HeII photo-ionization heating measured in Kelvin.
In the temperature panels it is apparent that the coolest regions are
the first regions to be ionized, which we justified by the fact that
these regions had more time to cool.  However, in these photo-heating
panels, we see that the regions that are ionized first also have the
least heat injection from HeII photo-heating.  Regions that are
ionized at later times have absorbed more of the hard radiation
background before being fully ionized.

%Finally, the total photo-heating also accounts
%for the photo-heating after the gas is ionized.  Over a recombination
%timescale, there will be at least several thousand Kelvin injected
%into an ionized gas parcel.

\subsection{Density Fluctuations}
\label{delta_inhomo}

\begin{figure*}
\begin{center}
\epsfig{file=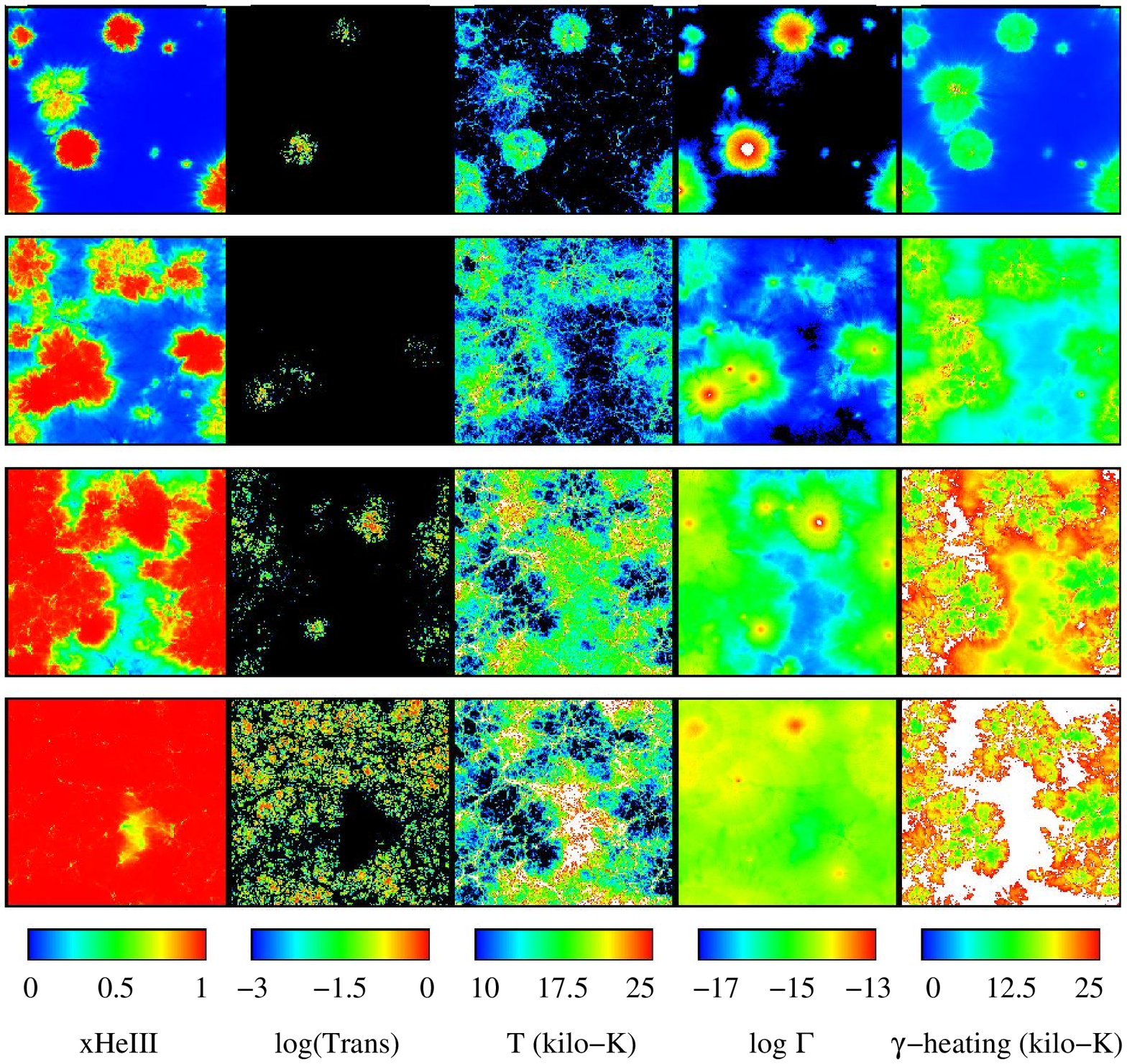, width=14cm}
\end{center}
\caption{Same as Fig. \ref{fig:homogeneous} but using simulation D1
(which includes density fluctuations).  The panels in each row are
chosen to match the $\bar{x}_{\rm HeIII}$ fractions quoted in the
caption of Fig. \ref{fig:homogeneous}.
\label{fig:densityfluc}}
\end{figure*}

Figure \ref{fig:densityfluc} is the same as Figure
\ref{fig:homogeneous}, but features simulation D1 which includes
density inhomogeneities using the $256^3$ gridded N-body field.  The
small-scale structure of HeII reionization is changed significantly by
including density inhomogeneities (compare the left-most panels in
Figures \ref{fig:homogeneous} and \ref{fig:densityfluc}).  However,
the large-scale morphology of HeII reionization is similar between the
homogeneous and inhomogeneous cases.

The structure of the temperature fluctuations is affected the most by
density inhomogeneities compared to the other quantities shown in
Figure \ref{fig:densityfluc}.  Adiabatic heating and cooling from
structure formation imprints additional small-scale features into the
temperature field in simulation, and it makes more
regions cooler in D1 than in D2 because most of the volume in the IGM is
underdense and expanding faster than if it were in the Hubble flow.

The second column in Figure \ref{fig:densityfluc} shows the total HeII
Ly$\alpha$ forest transmission.  The transmission level is
significantly higher when density inhomogeneities are included.
Underdense regions are responsible for most of the transmission in the
$z \approx 3$ HeII Ly$\alpha$ forest.  Section \ref{forest} shows that
simulation D2 is consistent with measurements of the HeII Ly$\alpha$
forest mean transmission.

\begin{figure}
\epsfig{file=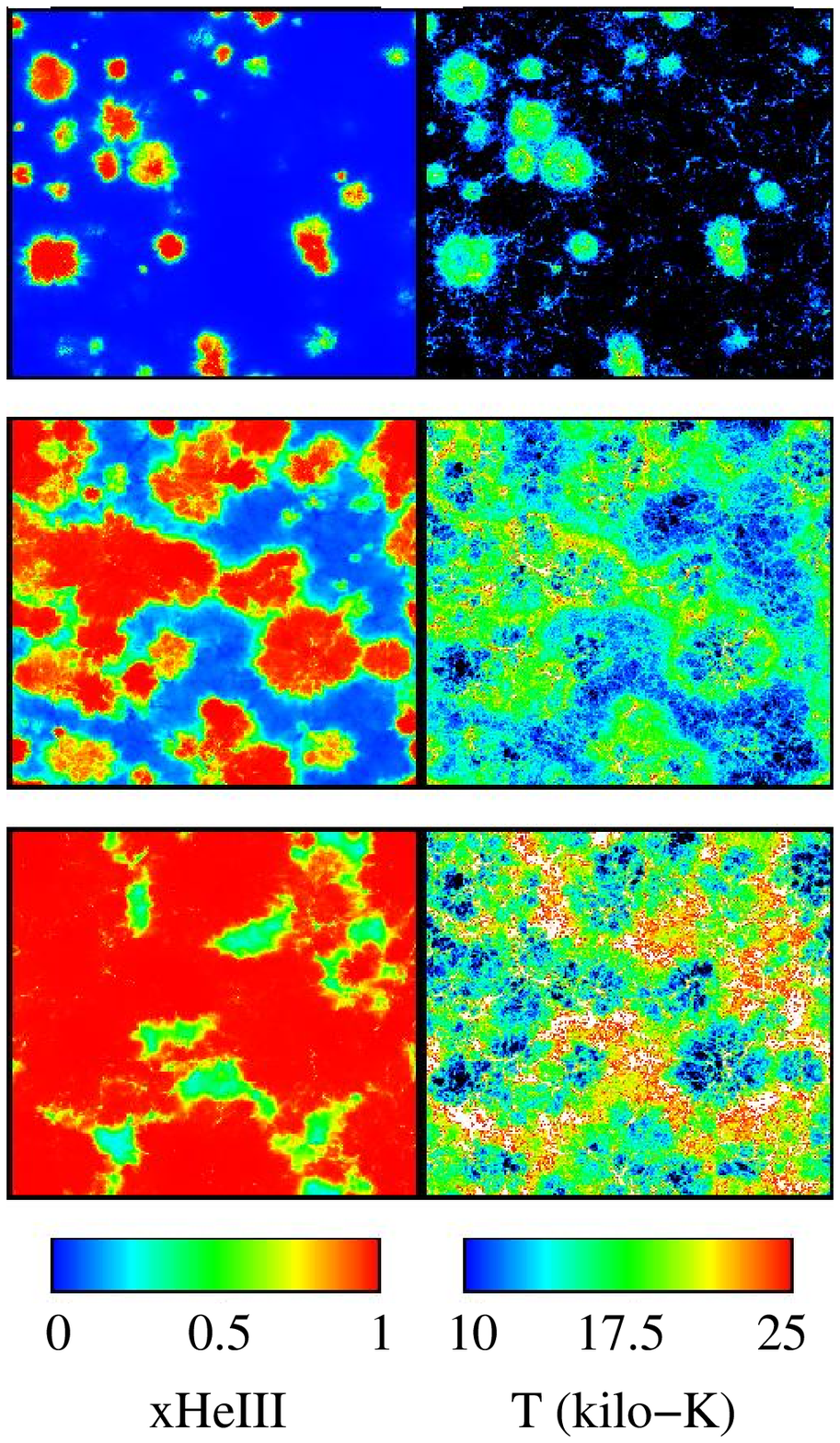, width=8.5cm}
\caption{Slices through simulation B1, which employs the same run
specifications as simulation D1 (Fig. \ref{fig:densityfluc}), but
uses the $429$ Mpc box.  Each panel shows a $429 \times 429 \times
8.4$ Mpc slice through this simulation volume. From top to bottom,
$\bar{x}_{\rm HeIII} =0.1, 0.5$, and $0.9$.
\label{fig:densityfluc2}}
\end{figure}

Figure \ref{fig:densityfluc2} presents simulation B1, which has the
same run specifications as simulation D1, but in the larger $429$ Mpc
box.  The larger box provides a better sample of the structures during
HeII reionization.  On these larger scales, QSO clustering is evident
in the maps.

\subsection{The Effect of Dense Clumps}
\label{ss:filtering}

\begin{figure*}
\begin{center}
\epsfig{file=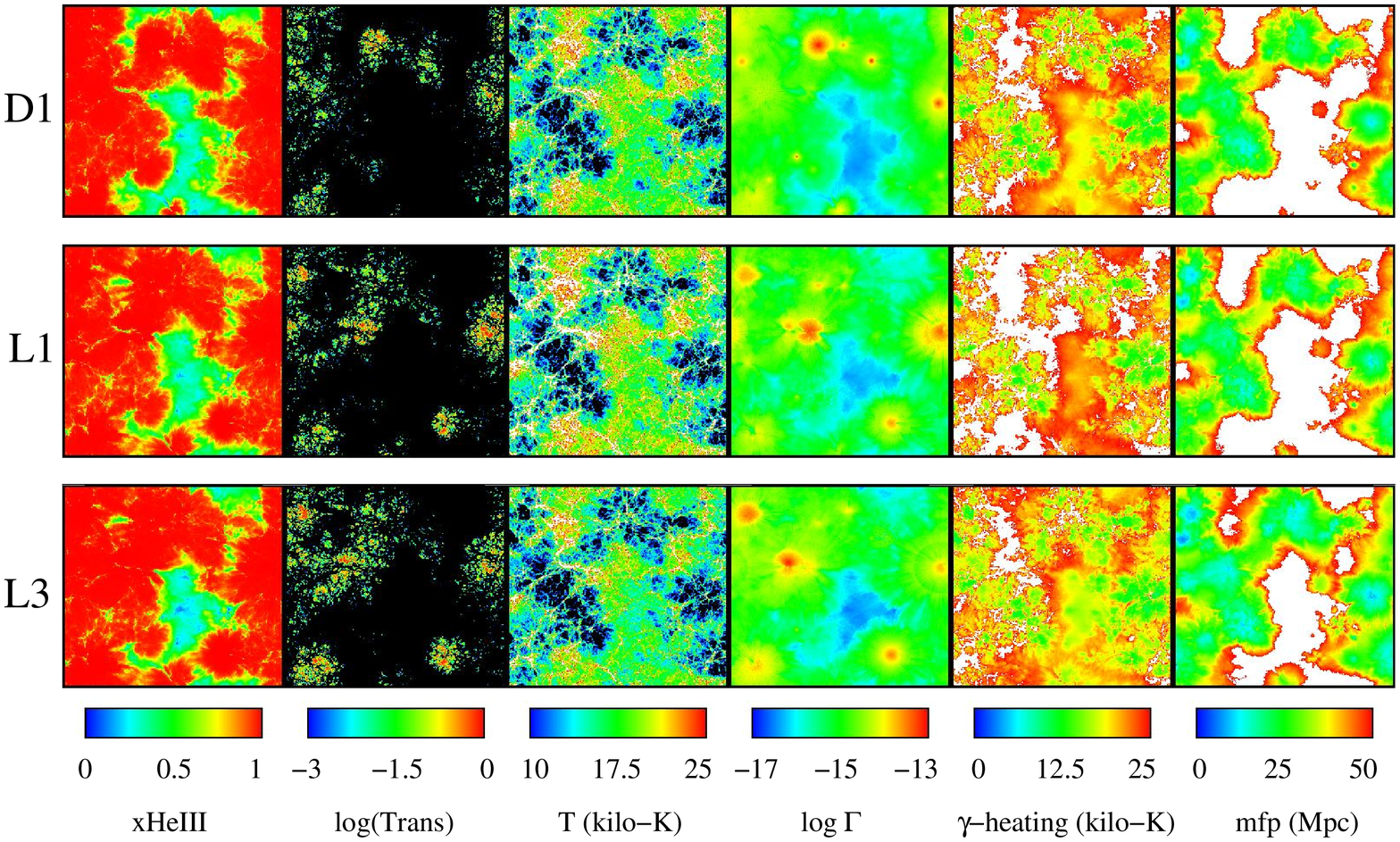, width=16cm}
\end{center}
\caption{Three rows use snapshots at $\bar{x}_{\rm HeIII} = 0.85$.
The top panels are from a simulation that does not employ an
additional filtering method, the middle panels are from a simulation
that uses filtering method A, and the bottom panels are from a
simulation that uses filtering Method B.
\label{fig:filtering}}
\end{figure*}

Figure \ref{fig:filtering} compares three simulations that employ
different methods for including the effect of dense clumps on filtering
the ionizing radiation.  Each row features a snapshot from its
respective simulation with $\bar{x}_{\rm HeIII} = 0.85$.  The top panels
are from simulation D1 (which does not employ an additional filtering
method), the middle panels are from simulation L1 (filtering Method
A), and the bottom panels are from simulation L3 (Method B).  All
three simulations are run on the $256^3$ grid using quasar Model II.
Simulation D1 reaches $\bar{x}_{\rm HeIII} = 0.85$ at $z = 3.3$, and
simulations L1 and L3 approach this mark at slightly later times, at
$z \approx 3.15$, owing to the additional filtering.  For the most
part, the $x_{\rm HeIII}$ and $T$ fields show similar patterns in the
three simulations, which suggests that our results depend weakly on
the filtering method.  The volume-averaged temperatures in the three
simulations are within $2000$ K of each other at fixed $\bar{x}_{\rm
HeII}$.  Furthermore, the distribution of temperatures at mean density
are similar between these simulations, but with the distribution in L1
extending to a few thousand Kelvin higher temperatures than those in
the other simulations. 
%We have also looked at power spectrum
%statistics of the the ionization and temperature fields: there are no
%significant differences among the three simulations at fixed
%$\bar{x}_{\rm HeIII}$.

%Simulation L2 uses the $512^3$ grid and filtering Method A.  This
%simulation, which is not shown in Figure \ref{fig:filtering}, has the
%most filtering because of its higher resolution grid and, therefore,
%results in the most heating.  The volume averaged temperature at
%$\bar{x}_{\rm HeIII} = 0.85$ is $700$ K more than in L1 -- the hottest
%of the three simulations in Figure \ref{fig:filtering} -- and the
%temperature distribution for cells at mean density is broader,
%extending several thousand Kelvin higher in temperature at the end of
%HeII reionization.  As discussed in Appendix \ref{ap:LLS}, filtering
%Method A should overpredict the amount of filtering and $512^3$
%resolution on dark matter produces slightly more high column density
%systems than gas simulations (Appendix \ref{ap:gdist}), and so the
%amount of heating in L2 should be considered an upper limit.

The right-most column in Figure \ref{fig:filtering} illustrates how
the filtering methods affect the m.f.p. of ionizing photons, where the
m.f.p. is tabulated by averaging the distance that absorbed photon travel (where this average is weighted by the
energy it deposits).  The average m.f.p. decreases
slightly from simulation D1 to L1 as expected, and again from L1 to
L3.

It is surprising that filtering from dense systems does not have a more significant
impact, particularly on the temperature fields, which are sensitive to
the hardness of the ionizing radiation.  If the bubble sizes were much
smaller than the photon m.f.p., then it would not be surprising that
filtering by dense regions is unimportant -- radiation would be
filtered minimally by dense clumps prior to reaching diffuse neutral
gas.  However, the bubble sizes are comparable to the m.f.p. (Appendix
A.), and we know that about $1/3$ of the photons are absorbed in dense
systems in L1 and L3 since HeII reionization in these simulations
requires $3.4$ photons per HeII ion as opposed to $2.2$ in simulation
D1.  If we assume that the filtering removes the bottom third of the
HeII-ionizing photons in energy and that the IGM is ionized by the
remaining, harder radiation, the temperature increase of the IGM would
be a factor of $2.4$ larger for $\bar{\alpha}_{\rm UV} = 1.6$ than if
the filtering had not occurred (at least if $\Delta T$ is computed in
the optically thin limit; eqn. \ref{eqn:DeltaT}).

Why is filtering in our simulations different from this toy scenario?  The answer is that not just the lowest
energy photons are filtered.  The effective optical depth of
dense systems scales as $\sim E_{\gamma}^{-1.5}$ if the $N_{\rm HeII}$
distribution scales as $-1.5$ \citep{zuo93}. 
%(However, the scaling is
%steeper than $E_{\gamma}^{-1.5}$ for $E_{\gamma} \gtrsim 150$ eV
%because the $N_{\rm HeII}$ distribution becomes steeper than $-1.5$
%once $N_{\rm HeII} \gtrsim 10^{19} \; \cmeter^{-2}$, as noted in
%Appendix \ref{ap:LLS}.)  
This absorption contrasts with the absorption from diffuse neutral
gas, for which the effective optical depth will scale closer to $\sim
E_{\gamma}^{-3}$.  Therefore, the spectral shape of the ionizing
radiation is not as altered by filtering by dense systems as by
diffuse IGM gas, and it is certainly not as hardened by dense systems
as in the above toy model.  Furthermore, equation (\ref{eqn:Tsmallr})
shows that the amount of heating depends weakly on the spectral index
of the incident spectrum.

Much of the filtering that hardens the radiation field in the HeII
  photo-heating panel in Figure \ref{fig:filtering} occurs in the
  diffuse gas rather than in dense clumps.  This is why the
  temperatures in simulation D2, for which there are no density
  fluctuations, are similar to the other simulations.  Even in the
  HeIII bubbles in Figure \ref{fig:filtering}, there are significant
  amounts of neutral gas which do not reside in very overdense
  locations.  Furthermore, photons that penetrate into neutral regions
  will typically encounter a higher total HeII column density in
  diffuse gas than in dense systems.  These photons are responsible
  for heating the hottest regions (which are the last regions to be
  ionized).

\begin{figure}
\begin{center}
\rotatebox{-90}{\epsfig{file=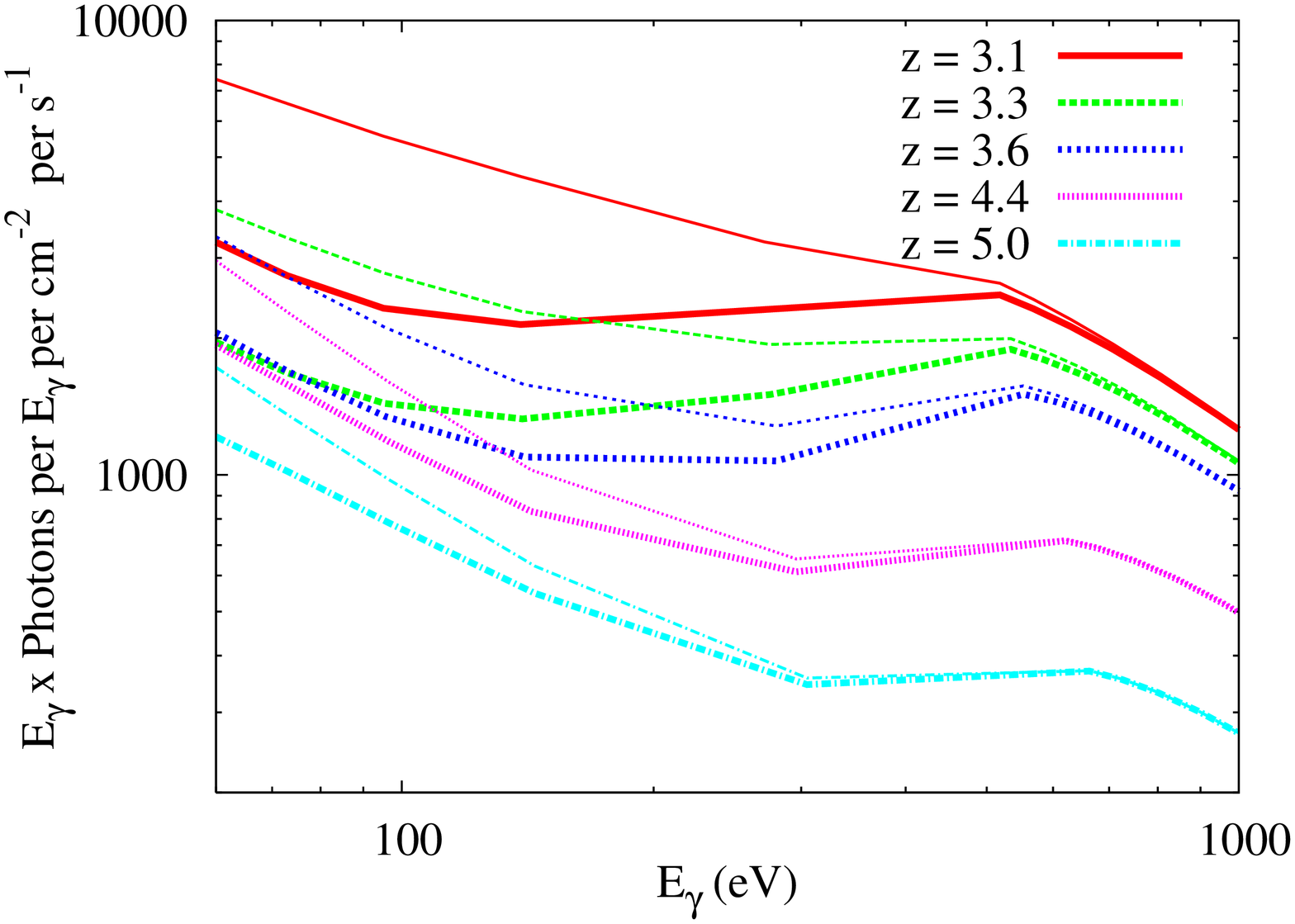, height=8cm}}
\end{center}
\caption{Globally averaged radiation
spectrum at various times during our
simulations, computed by tabulating the incident spectrum in many randomly
selected cells and then averaging.  The thin
curves are for simulation D1 and the thick curves are for L1.  L1
differs from D1 in that it uses filtering method A.
\label{fig:spectrum}}
\end{figure}

Figure \ref{fig:spectrum} compares the globally averaged spectrum in
simulations D1 (no filtering) and L1 (filtering Method A).  This
global average is computed by tabulating the incident spectrum in
randomly selected cells in the simulation volume and then averaging.  (The global spectrum in L3, which is not
shown, is similar to this spectrum in L1.)  Reionization ends
($\bar{x}_{\rm HeIII} = 0.95$) at $z = 3.3$ in D1 and at $z = 3.1$ in
L1.  The average spectrum has a similar shape, particularly at the
beginning of HeII reionization when the impact of filtering is
minimal.  At later times, the difference is a factor of $\lesssim 2$, and
the slope is not significantly changed.  Near the end of HeII
reionization, the amplitude of the spectrum
evolves by order unity in both simulation D1 and L1 on timescales of
$\Delta z \approx 0.2$.  This fast evolution suggests that diffuse gas
is playing a role in limiting the mean free path in these simulations
even at the end of HeII reionization.

%In the optically thin
%photo-heating limit (equation \ref{??})), all of the spectra in Figure
%\ref{fig:spectrum} yield about $13$ eV per ionization, with a
%dispersion that is less than $1$ eV. 

%Rather than compare the
%spectra at the same redshifts, a more relevant comparison may be at
%the same values of $\bar{x}_{\rm HeIII}$.  The $z=3.3$ curve from
%simulation L1 in Figure \ref{fig:spectrum} and the $z = 3.6$ curve
%from D1 are both at $\bar{x}_{\rm HeIII} \approx 0.7$.  The difference
%in the spectra between these two curves is smaller than, for example,
%between the two $z=3.3$ curves.  

%Finally, we have computed the average
%energy of an ionization for an optically thin gas parcel using the
%spectra in Figure \ref{fig:spectrum} (see equation \ref{??}).  The
%average energy injection from an ionization is $12.9$ eV for the
%global spectrum for simulation D1 and $13.2$ eV for the global
%spectrum in simulation D2 at both $z = 3.1$ and $z = 3.3$ -- an
%insignificant difference between the two simulations.

%A second reason why filtering may not have a significant effect on the
%temperatures is that reionization takes longer in the simulations with
%filtering by $\Delta z \approx 0.2$ because these simulations require
%more photons to reach percolation.  This allows gas in the Hubble flow to
%cool by an additional $10\%$ owing to expansion, and it provides
%time for additional softer photons to be produced by the QSOs.  These photons contribute to the ionization of the diffuse gas, in contrast to the toy case.

After HeII reionization, the filtering method affects the spectrum more because the m.f.p. is limited by
dense systems and not by diffuse gas.  The value of $\bar{\Gamma}_{\rm
HeII}$, which in a large part determines the transmission in the HeII
Ly$\alpha$ forest, is proportional to $\lambda_{\rm mfp}$.   The
photo-ionization rate at the end of the simulation is $2-4\times
10^{-15}$ s$^{-1}$ in L1 and L2 whereas it is $7\times 10^{-15}$
s$^{-1}$ and increasing steadily when simulation D1 was terminated at
$z = 2.9$.

\subsection{Soft X-ray Photons}
\label{ss:xrayphotons}
Simulation X1 is the same as L1 except that photons with m.f.p. larger
than $3.5$ times box size, background
photons with roughly $E_{\gamma} \gtrsim 500$ eV, are not included in
the calculation.  How much heating and ionizations do these
high-energy photons contribute?  When the Universe is $90\%$ ionized
in HeII, model X1 is $1400$ K cooler than model L1.  Furthermore, the
fraction of ionizations that arise from photons with $E_{\gamma}
\gtrsim 500$ eV is small.  The heating from the background is largest
in the regions that are ionized last, the regions that are exposed to
the background for the longest period of time.  The background heating
adds an additional several thousand Kelvin in these regions.

The relative unimportance of heating from photons with $E_{\gamma}
\gtrsim 500$ eV is reassuring, because we do not model the obscured
quasar contribution to the ionizing background.  Models of the
obscured contribution to the diffuse X-ray background find that
obscured quasars (defined as having QSOs with obscuring columns of
$N_{\rm HI} > 10^{21} \; \cmeter^{-2}$) begin to dominate the
background above approximately $2$ keV (e.g., \citealt{gilli07}).

%Is ignoring the contribution of obscured QSOs to the ionizing
%background justified?  Photons with $E_{\gamma} \gtrsim 0.8$ keV have
%an optical depth of less than unity to be absorbed in the diffuse IGM
%prior to HeII reionization.  In the optically thin limit, the heating
%from these photons is proportional to $\approx E_{\gamma} \,
%I(E_{\gamma}) \, \sigma_{\rm HeII}(E_{\gamma})$, where $I(E_{\gamma})$
%is the specific intensity.  As long as $E_{\gamma}^{-2} \;
%I(E_{\gamma})$ is a decreasing function (as is the case even for the
%$z =0 $ soft X-ray background, where $E_{\gamma} I(E_{\gamma}) \propto
%E_{\gamma}^{0.33}$ for $E_{\rm gamma} > 1$ keV and is flatter at lower
%energies), $1$ keV energy photons are a more significant contribution
%to the heating than $2$ keV or $5$ keV photons.  Therefore, the
%contribution to the HeII photo-heating from obscured sources is likely
%subdominant.  {\bf weave into simulation X1 better}

\subsection{QSO Model}
\label{QSO_model}

\begin{figure*}
\begin{center}
\epsfig{file=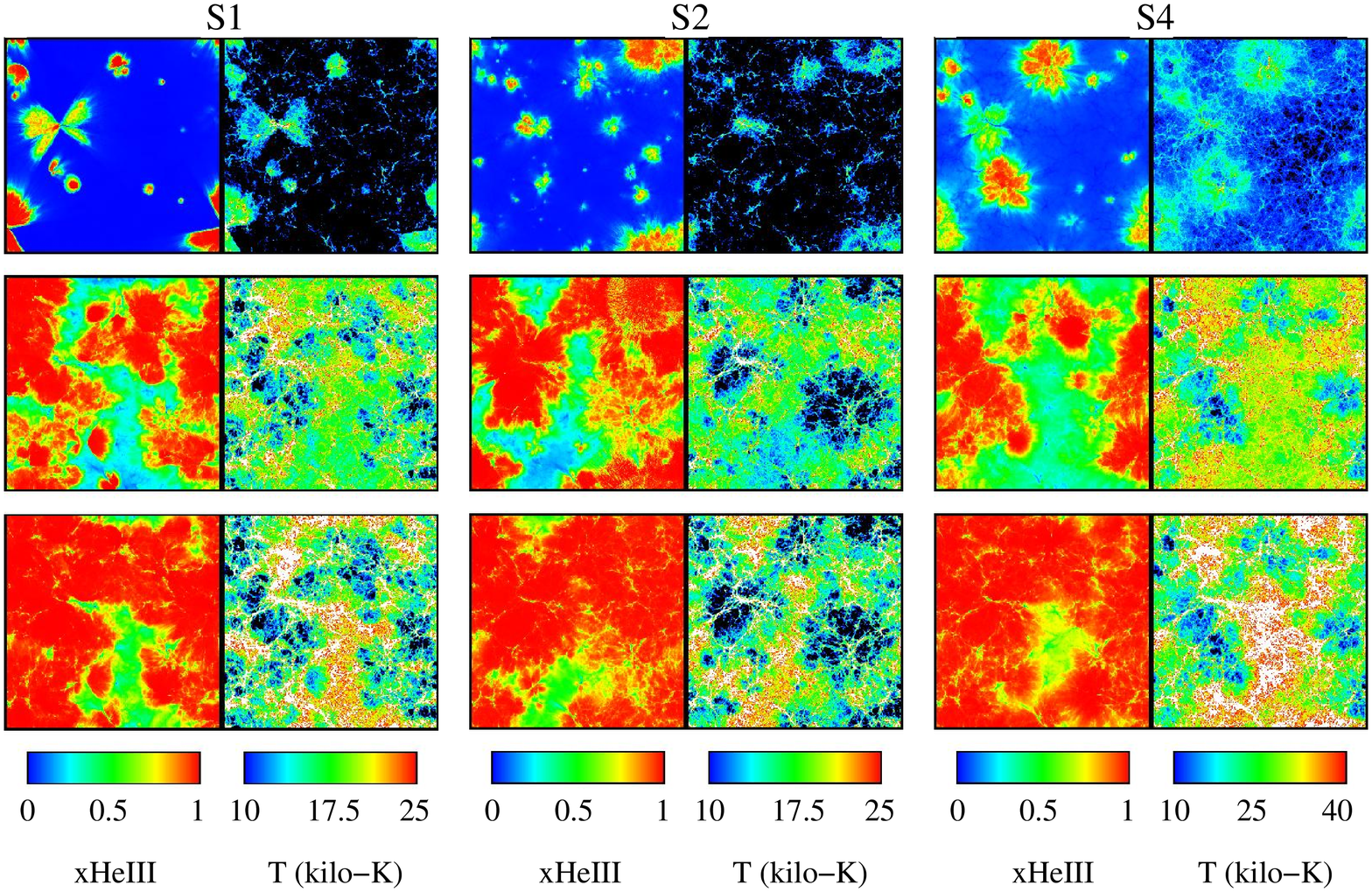, width=18cm}
\end{center}
\caption{Morphology of HeII reionization for three possible source
models.  In S1 the emission from the quasars is beamed as described in
the text, in simulation S2 the quasars have a light bulb behavior with
$\tau_{\rm QSO} = 10$ Myr, and in simulation S4 the quasars have a
hard spectrum with $\bar{\alpha}_{\rm UV} = 0.6$.  From top to bottom,
the panels represent snapshots with $\bar{x}_{\rm HeIII} \approx 0.1,
0.75$, and $0.9$. \emph{Note that temperature panels for simulation S4
have an adjusted range that extends to $40,000$ K compared to $25,000$
K for the other simulations.}
\label{fig:models}}
\end{figure*}

The major uncertainties
associated with modeling QSOs are their lifetimes, spectra, and
whether their emission is beamed.  Figure \ref{fig:models} considers
$3$ QSO models, to illustrate how these uncertainties influence our
results.  From top to bottom, the panels represent snapshots with
$\bar{x}_{\rm HeIII} \approx 0.1, 0.75$, and $0.9$.  Beamed emission
(see simulation S1 in Fig. \ref{fig:models} which uses the beaming model described in Appendix E)
enhances the structure of the ionization field earlier on in HeII
reionization.  Similarly a light bulb model with $\tau_{\rm QSO} = 10$
Myr (simulation S2 in Fig. \ref{fig:models}) results in more relic
HeIII regions compared to our fiducial case and in slightly smaller
bubbles.  This fact owes to the shorter quasar lifetime compared to
the fiducial case.

The IGM temperatures in S4 and S4b, which have a harder UV spectral
index of $0.6$ compared to the fiducial value of $1.6$ (chosen to
better agree with the results of \citealt{scott04}), are significantly
higher than the other simulations analyzed in this paper. This result
agrees with that of \citet{tittley07}, who also find a strong
dependence of the temperature on the spectral index of the sources.
In S4 (right panels in Fig. \ref{fig:models}), the hottest regions are
typically the least ionized.  In the other simulations featured in Figure
\ref{fig:models}, the hottest regions are typically the most ionized
until $\bar{x}_{\rm HeIII} \gtrsim 0.8$.  

Why is the morphology of the temperature so different in S4 whereas
the $\bar{x}_{\rm HeIII}$ field is similar to the other simulations
(except for a higher ionization floor)?  For $\bar{\alpha}_{\rm UV} =
0.6$, the average energy photon that is absorbed is much higher than
in the fiducial model, injecting more heat.  However, near quasars it
is still the softest photons that are absorbed, resulting in similar
structures in the ${x}_{\rm HeIII}$ field compared to the fiducial
model. The hardest photons (which are more plentiful in S4) are
absorbed in the neutral regions outside of the HeIII bubbles, leading
to the inverted behavior between ${x}_{\rm HeIII}$ and $T$.

%It is possible that the temperatures achieved in simulation S4 are
%ruled out by current data.  However, it is not obvious that this is
%the case given that there are relatively cold regions present in the
%IGM with $T < 15,000 \, {\rm K}$.  Even though $\bar{T}\approx 35,000
%\, {\rm K}$ in this simulation at the end of HeII reionization,
%measurements of the temperature from the narrowest lines in the
%Ly$\alpha$ forest (and even with the small-scale Ly$\alpha$ power
%spectrum) are most sensitive to the coldest regions.

%To make the comparison more quantitative, we have calculated the power
%spectrum from our simulations.  The average power spectrum is
%discussed in detail in Section \ref{power_spectrum}.  Here, we discuss
%relative differences when comparing simulations S1, S2, and S4.  The
%power spectrum of $x_{\rm HeII}$ at all times agrees to $\lesssim
%50\%$ at fixed ionized fraction between these three simulations.
%Simulation S1 has more power than L1 by $\approx 30\%$ on scales
%smaller than the bubbles owing to the beam structure enhancing
%fluctuations, but on large scales the power spectrum is similar
%between these two simulations.  The $x_{\rm HeII}$ power spectra of
%simulation S2 agrees with those of L1 well when compared at fixed
%$x_{\rm HeII}$, with the biggest discrepancy occurring at the end of
%HeII reionization, where the fluctuations are smaller by $\approx
%20\%$.  Finally, the $x_{\rm HeII}$ power is similar between
%simulations S4 and L1, but the temperature power differs
%significantly.

\subsection{Power Spectrum}
\label{power_spectrum}

\begin{figure}
\rotatebox{-90}{\epsfig{file=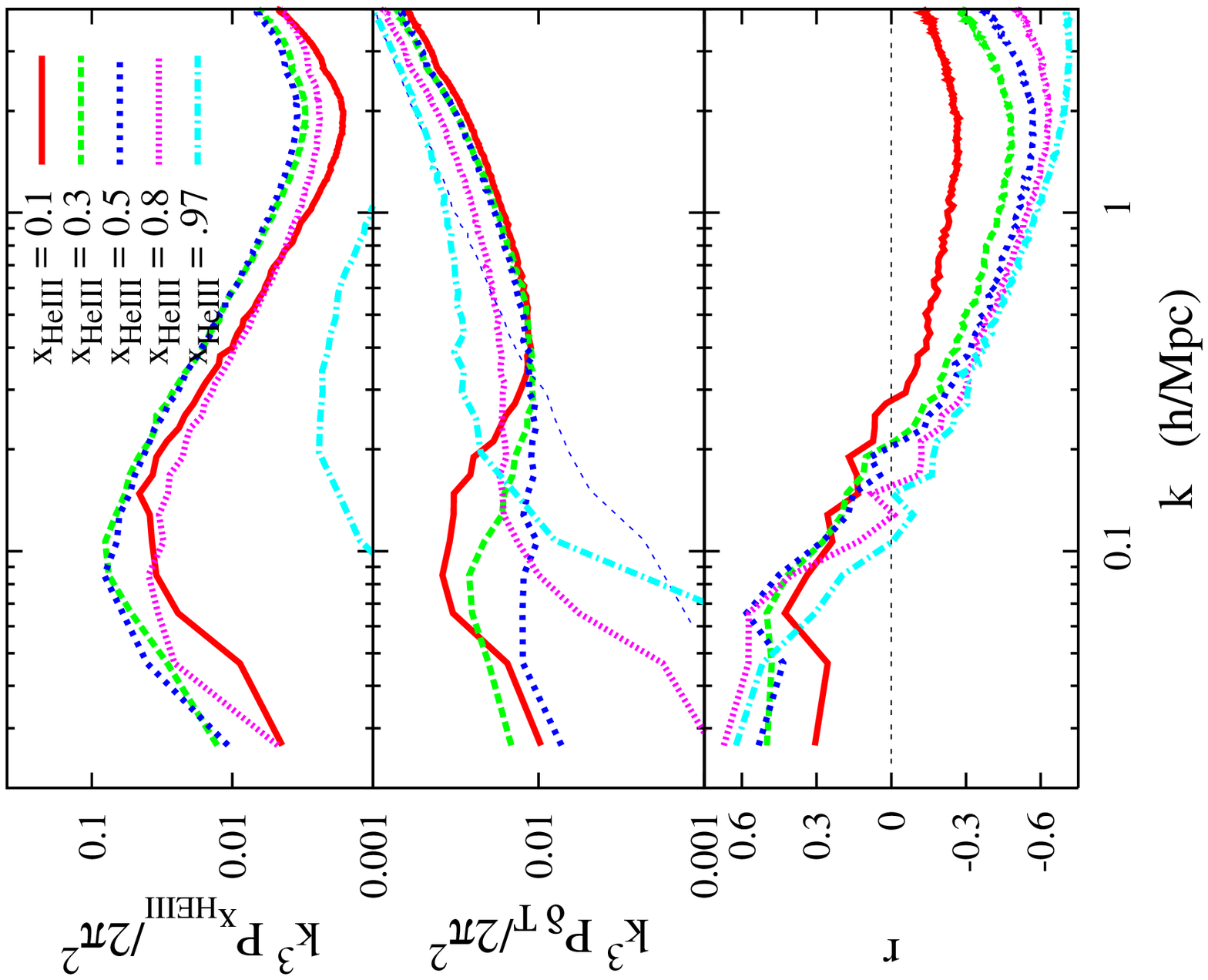, height=9cm}}
\caption{Power spectrum of ${x}_{\rm HeIII}$ (top panel), $\delta T$
(middle panel), and the cross correlation coefficient of ${x}_{\rm
HeIII}$ and $\delta_b$ (bottom panel) in simulation B1.  The thin blue
dotted curve in the middle panel is from a simulation without HeII
reionization.
\label{fig:ps}}
\end{figure}

We have seen that $50$ Mpc fluctuations in the temperature and $x_{\rm
HeIII}$ are present during and after HeII reionization.  These
fluctuations modulate the level of absorption as well as the amount
of small-scale power in the HI Ly$\alpha$ forest. Here, we use power spectrum
statistics to quantify the properties of these spatial fluctuations.

The curves in Figure \ref{fig:ps} are calculated from simulation B1,
which uses the $430$ Mpc box, and represent the power spectrum of
${x}_{\rm HeIII}$ (top panel), $\delta T \equiv T/\bar{T} - 1$ (middle
panel), and the cross correlation coefficient of ${x}_{\rm HeIII}$ and
$\delta_b$ (bottom panel) from simulation D1, which is defined as $r
\equiv P_{\delta_{x_{\rm HeIII}} \delta_b}/ [P_{\delta_{x_{\rm HeIII}}
\delta_{x_{\rm HeIII}}} \, P_{\delta_b \delta_b}]^{-1/2}$, where
$\delta_{x_{\rm HeIII}} \equiv x_{\rm HeIII}/\bar{x}_{\rm HeIII} - 1$.
We have checked that the power spectrum in B1 agrees well with that in
D1, which is identical to B1 except run in the $429$ Mpc box. The
only scales where the two simulations do not agree correspond to $k >
2 \, h \, \Mpc^{-1}$, where there is an upturn owing to shot noise in
B1 that is not present on these scales in D1.
%Also, the $0.97$ power spectra do not agree between the large
%and small boxes (particularly the temperature power spectra); the
%power spectrum from the smaller box does not show as much of a
%decrease at $k \lesssim 0.1$ Mpc$^{-1}$ in power as that from the
%larger box.

The $x_{\rm HeIII}$ fluctuations peak on $\sim 50$ Mpc scales, even
early in the HeII reionization process when $\bar{x}_{\rm HeII}
\approx 0.1$, with the peak in the amplitude of the fluctuations
occurring at $\bar{x}_{\rm HeIII} \approx 0.5$.  This peak scale is
remarkably constant throughout HeII reionization.  This contrasts with
the evolution of the power spectrum in most models of hydrogen reionization, in
which the growth of large HII regions around clusters of
sources leads to the power shifting to larger scales with increasing
$\bar{x}_{\rm HI}$ (e.g., \citealt{mcquinn07}).

The fluctuations in the temperature field are shown in the middle
panel in Figure \ref{fig:ps} at several $\bar{x}_{\rm HeIII}$.  For
comparison, the thin blue dotted curve is from a simulation where HeII
reionization does not occur at $z<6$ and that is initialized to have
$\gamma - 1 \approx 0$ and $T =10^4$ K at $z = 6$.  This curve is at
the same redshift, $z = 4.0$, as the $\bar{x}_{\rm HeIII} = 0.5$
curve.  This illustrates that HeII reionization produces temperature
fluctuations on much larger scales than structure formation.
% At $k \gtrsim 0.3$
%Mpc$^{-1}$, density fluctuations are a significant contribution to the
%$T$ fluctuations at all times during HeII reionization because all
%sources of heating other than photo-ionization of the diffuse HeII
%correlate strongly with density.  HeII reionization affects the
%temperature fluctuations on small scales by modifying the
%$T$-$\Delta_b$ relation.  The steeper this relation, the more small
%scale power.

At $k \lesssim 0.3$ Mpc$^{-1}$, the fluctuations from patchy HeII
reionization are the dominant source of the $T$ fluctuations.  The
scale of the peak in these fluctuations is comparable to the HeII
bubble sizes early on.  However, when $\bar{x}_{\rm HeIII} = 0.3, 0.5$
and $0.97$, large fluctuations are present on all scales captured in our
box.  Interestingly, the amplitude of large-scale temperature fluctuations are
largest at the beginning of HeII reionization, with $(\Delta
T/\bar{T})_{\rm rms} \equiv [k^3 \Delta T/(2\pi^2 T)]^{1/2} = 0.2$. 
% The thick curves in the middle panels in
%Figure \ref{fig:densityfluc} should be contrasted with the thin,
%dotted curve, for which the temperature fluctuations trace the density
%fluctuations.  
Fluctuations in temperature are observable in
the HI Ly$\alpha$ forest \citep{zaldarriaga02, theuns02-fluc}.

The degree of correlation between the ionization and density
fields reveals how important QSO clustering is in shaping the
morphology of HeII reionization.  The bubbles begin to correlate with
the density field on $50$ Mpc scales (bottom panel in
Fig. \ref{fig:ps}). On smaller scales, ${x}_{\rm HeIII}$
is anti-correlated with $\delta_b$. The level of anti-correlation
increases with $\bar{x}_{\rm HeIII}$ as the HeIII regions encompass
more dense neutral systems.
%and we have checked that the same level of
%anti-correlation is present in our higher resolution simulations.

\section{Observational Implications}
\label{observations}
This section discusses how HeII reionization affects observations
of the IGM.  The observables that are addressed are the IGM temperature, the T-$\Delta_b$ relation, and the mean
transmission in the HI and HeII Ly$\alpha$ forests, all of which can
be estimated from high-redshift quasar spectra.
%In due course, we
%will investigate in more detail the impact of HeII reionization on
%Ly$\alpha$ forest measurements as well as on the
%abundance of high-redshift metal absorption lines.

\subsection{Temperature Evolution}
\label{Tevolution}

\begin{figure}
\rotatebox{-90}{\epsfig{file=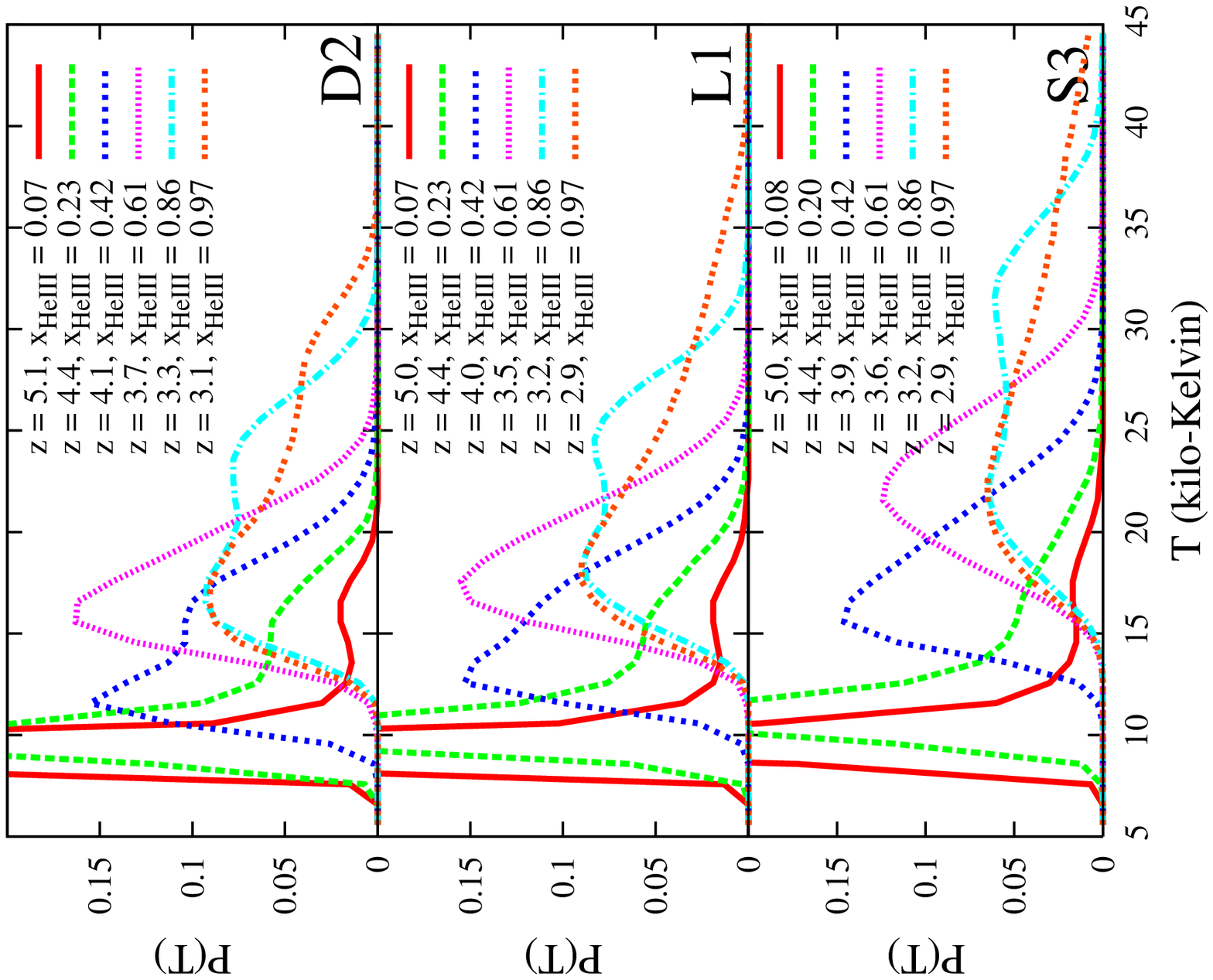, height=8.8cm}}
\caption{PDF of the temperature tabulated from grid cells with
  $-0.1 < \delta_b < 0.1$ in simulations D2, L1, and S3.
\label{fig:Temp_pdf}}
\end{figure}

Figure \ref{fig:Temp_pdf} plots the evolution of the temperature PDF
in simulations D2, L1, and S3, tabulated from grid cells with $-0.1 < \delta_b <
0.1$.  Simulation D2 has no density fluctuations and $\bar{\alpha}_{\rm UV}
= 1.6$; simulation L1 includes density fluctuations, $\bar{\alpha}_{\rm UV}
= 1.6$, and filtering Method A; and simulation S3 is the same as L1
but with $\bar{\alpha}_{\rm UV} = 1.2$.  

The evolution of the PDF in simulation L1 is characteristic of our
simulations with $\bar{\alpha}_{\rm UV} = 1.6$.  At times in L1 for which
$\bar{x}_{\rm HeIII} \ll 1$, most of the IGM has $ T \approx 1\times
10^4$ K, except ionized regions which have $T \approx 1.6 \times 10^4$
K.  The result is a bimodal temperature PDF (see the $\bar{x}_{\rm
HeIII} = 0.1$ curve in the middle panel in Fig. \ref{fig:Temp_pdf}).
As HeII reionization proceeds, the PDF shifts to higher temperatures.
If only the insides of HeIII regions were heated, this PDF would
become broader rather than move towards higher $T$ in the manner seen
in Figure \ref{fig:Temp_pdf}, implying that hard photons are streaming
far from their sources prior to being absorbed.  The bimodality of the
PDF decreases with increasing $\bar{x}_{\rm HeIII}$ until
$\bar{x}_{\rm HeIII} = 0.6$, at which point the PDF is fairly
Gaussian.  For $\bar{x}_{\rm HeIII} > 0.6$, the PDF again becomes
skewed as the harder photon background builds up and starts to
appreciably heat regions far from quasars.  The temperature PDF at
the end of HeII reionization is broad, extending between $12,000 \;
{\rm K}$ and $35,000 \; {\rm K}$ in simulation L1.  The temperature of
a gas parcel at the end of HeII reionization is a result of many
factors, including the redshift(s) it was ionized and the
hardness of the radiation field that ionized the parcel.

 The evolution of the temperature PDF in simulations D2 and S3 are
similar to that in L1.  The PDF in D2 does not extend to as high
temperatures as in L1.  This implies that density inhomogeneities
contribute to the temperature of the hottest gas parcels.  The evolution of
the PDF is also similar in S3 to L1, except the harder
spectrum in S3 results in higher temperatures.

\citet{schaye00} and \citet{ricotti00} estimated the temperature of
the IGM from the widths of the narrowest lines in the HI Ly$\alpha$
forest, calibrating their measurements with numerical simulations that
essentially assumed a power-law $T$-$\Delta_b$ relation at
relevant $\Delta_b$.  These studies claimed to have detected a sudden
increase in the IGM temperature between $z \approx 3.5$ and $3$. A
subsequent analysis using a similar methodology by \citet{mcdonald01b}
did not detect a sudden increase in temperature, but rather a
temperature at mean density of $T_0 \approx 17,000 \pm 2000\; K$ at $z
=2, 3$, and $4$.  \citet{zaldarriaga01} derived similar numbers to
\citet{mcdonald01b} from measurements of the HI Ly$\alpha$ forest
power spectrum.  Although, both \citet{mcdonald01b} and
\citet{zaldarriaga01} measured the temperature in three coarse bins
centered at $z = 2.4, 3.0$ and $3.9$, possibly obscuring a sudden
temperature change.

The bottom panel of Figure \ref{fig:Tdrelation1} plots the
\citet{schaye00}, \citet{ricotti00}, and \citet{mcdonald01b}
measurement values for $T_0$ as well as the evolution of this quantity
in simulations L1, D1, and S4b (thick curves).  In general, our
calculations over-predict the measured values of $T_0$.  However,
since these observations look for the narrowest lines in the forest,
it is probable that they are most sensitive to the coolest
temperatures at a given density as is argued in \citet{furlanetto07a}.
A similar argument applies for measurements using the HI Ly$\alpha$
forest power spectrum.  The thin curves bracketing the thick ones are
$\pm 1$ s.d. of the mean $T_0$.  Note that the agreement with
measurements is better if we compare with the $- 1$ s.d. curves rather
than the mean $T_0$ curve.  A more detailed study is required in order
to determine whether these observations are consistent with our
simulations.

%All of our simulations that use $\bar{\alpha}_{\rm UV} = 1.6$ have
%similar values of $\bar{T}_0$ when comparing at fixed $\bar{x}_{\rm
%HeIII}$, with scatter among them of $\sim 1000$ K.  Among the
%simulations featured in the bottom panel in Figure
%\ref{fig:Tdrelation1}, simulation D1 does not employ an additional
%filtering method whereas simulation L1 uses filtering Method A, and
%the sources have the same efficiency in L1 as in D1. 
%Simulation S4b, which has $\bar{\alpha}_{\rm UV} = 0.6$, results in a
%much larger temperature than the other two and may be excluded by
%these observations.  Although, a small fraction of the volume has $T_0
%< 20,000 \; {\rm K}$, and so it is not obvious that it is excluded.
% This
%trend is clearly seen in simulation D1, but in simulation L1 the
%temperature of the IGM cools more slowly; the HeII-ionizing radiation
%just after HeII reionization is harder in simulation L1 than in D1
%such that the HeII-photo heating is larger because of the increased
%filtering.

Figure \ref{fig:Tdrelation1} also includes two curves for the case in
which HeII reionization happens at $z > 6$ and the simulation is
initialized with temperatures of $T(\Delta_b) = 10^4$ K and
$T(\Delta_b) = 2\times 10^4$ K at $z = 6$ (thin cyan dot-dashed
curves).  By $z \approx 3$ both of these curves have asymptoted to a
temperature of $\approx 7,000$ K as expected \citep{hui97}.  The
curves that represent the simulations with HeII reionization (at least
in models D1 and L1) are more consistent with the measurement points than these
curves.

\subsection{Temperature-Density Relation}
\label{tdrelation}
The temperature of the gas depends primarily on the competition
between photo-heating and adiabatic cooling.  These processes lead to
a power-law relationship between $T$ and $\Delta_b$ for unshocked
gas at times sufficiently after a reionization epoch
\citep{hui97}.  This power-law index asymptotes in time to the value
$\gamma - 1 = 0.6$ \citep{hui97}.\footnote{This value for $\gamma -1$
is just slightly smaller than if only adiabatic cooling is included,
which would yield $\gamma - 1 = 2/3$.}

\begin{figure}
\rotatebox{-90}{\epsfig{file=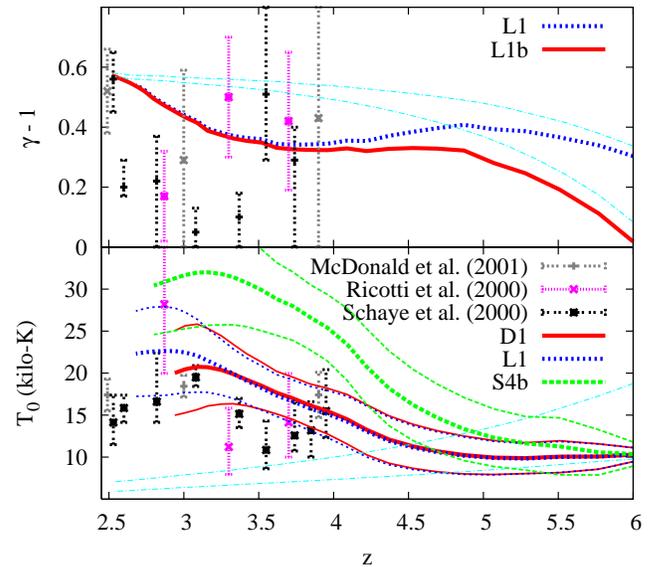, height=8.8cm}}
\caption{Top Panel: Best-fit power-law index of the $T$-$\Delta_b$ relation in
two simulations (thick curves), as well as its evolution in
simulations without a $z\sim 3$ HeII reionization (thin curves).  The
markers with error bars are measurements of this quantity in the
literature, and the references for these measurements are given in the
key in the bottom panel.  Bottom Panel: Evolution of best-fit $T_0$ in
three simulations (thick curves) as well as $\pm 1$ s.d. in this
quantity (the thinner curves with the same line
style).  The thin cyan dot-dashed curves that are increasing with $z$
represent simulations in which HeII reionization occurs at $z >6$.
Also shown are measurements of $T_0$.
\label{fig:Tdrelation1}}
\end{figure}

\begin{figure}
{\epsfig{file=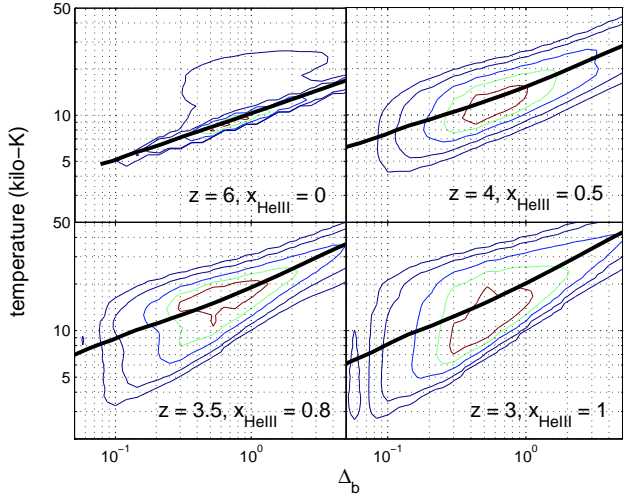, width=9.4cm}}
\caption{$T$-$\Delta_b$ relation in simulation D1. This simulation
is initialized with $\gamma - 1 = 0.3$ at $z =6.1$.  The thick solid
curves are the mean $T(\Delta_b)$, and the contours 
enclose $(33, 67,
90, 99, 99.9)$ percent of the grid cells -- 
encompassing the regions that have the highest densities in $\log T$ - 
$\log \Delta_b$. 
% The evolution of the
%$T$-$\Delta_b$ relation in simulation D1 is characteristic of our
%simulations with $\bar{\alpha}_{\rm UV} = 1.6$. 
\label{fig:Td_relation}}
\end{figure}

However, HeII reionization can lead to a more complicated relation
between $T$ and $\Delta_b$ than a single power-law.  Just after an
\emph{instantaneous} and \emph{homogeneous} HeII reionization process,
the equation of state $\gamma - 1$ would equal zero (an isothermal
IGM), at least if the temperature afterward is much larger
than before.  However, the finite duration of HeII
reionization causes $\gamma - 1$ to deviate from zero, and the
duration as well as the inhomogeneity of the photo-heating introduces
scatter into the $T$-$\Delta_b$ relationship.  Regions that are
ionized earliest have more time to cool, and regions ionized by the
most filtered radiation are the hottest.

Figure \ref{fig:Td_relation}
shows the evolution of the average $T$-$\Delta_b$ relation in simulation D1,
which uses Quasar Model II, at times for which
$\bar{x}_{\rm HeIII} = 0$, $0.5$, $08$, and $1$ (thick solid curves).
The slope of this relation (which was initialized with $\gamma - 1 =
0.3$ at $z = 6$) does not change substantially throughout HeII
reionization.  While the value of $\gamma - 1$ appears fairly
constant, the dispersion in this relationship grows during HeII
reionization.  The contours enclose $(33, 67, 90, 99, 99.9)$ percent
of the cells in the simulation grid -- encompassing the regions that
have the highest densities in $\log T$ - $\log \Delta_b$.  

%If the HeII is ionized first in overdense
%regions in the IGM (causing underdense regions to be heated last and
%have had less time to cool), HeII reionization could result in an
%inverted T-$\Delta_b$ relation (i.e., $\gamma - 1 < 0$).  This clearly
%is not happening in simulation D1 (Fig. \ref{fig:Td_relation}).

The top panel in Figure \ref{fig:Tdrelation1} shows the measured value
of $\gamma - 1$ at various times in simulations L1 and L1b, which use
QSO Model II and filtering Method A.\footnote{We fit a power-law to
the mean value of $T(\Delta_b)$ in bins spanning $0.5 < \Delta_b < 2$
to obtain $\gamma - 1$.  We find that the same procedure applied to
the median of $T(\Delta_b)$ rather than the mean or to a more
restricted range in $\Delta_b$ yields similar results.}  The only
difference between these two simulations is that L1 is initialized
with $\gamma - 1 = 0.3$ at $z =6$ and L1b with $\gamma - 1 = 0$.  In
both of these cases, as well as in all of the simulations listed in
Table \ref{table1}, the $T$-$\Delta_b$ relation does not become
inverted or isothermal (i.e., $\gamma - 1 \leq 0$).  This result owes
to the large QSO bubbles being essentially uncorrelated with
$\Delta_b$; no value of $\Delta_b$ is ionized preferentially at a
given time during HeII reionization.  Unless some $\Delta_b$ is
ionized preferentially, then the inequality $\gamma - 1 > 0$ must
hold.  Our simulations yield $\gamma - 1 \approx 0.3$ during the bulk
of HeII reionization.  After HeII reionization, this index steepens.
Interestingly, the evolution of $\gamma(z)-1$ in simulations S4, which has
$\bar{\alpha}_{\rm UV}= 0.6$, is nearly identical to this in
L1, even though the IGM in S4 is heated to a much higher temperature.

Also shown in Figure \ref{fig:Tdrelation1} are measurements of $\gamma
- 1$ in the literature.  The different estimates are not consistent
with one another, and generally have large error bars.  These
measurements are calibrated using hydrodynamic or hydro-PM simulations
that do not account for inhomogeneous heating owing to HeII
reionization.\footnote{Also, note that the \citet{schaye00} data points --
which are the most discrepant with respect to our predictions -- are
each taken from half of the Ly$\alpha$ forest from a \emph{single}
quasar spectrum (i.e., approximately a $250$ Mpc skewer).  Because of
the large structures during HeII reionization, it is conceivable that
the \citet{schaye00} estimates do not sample enough regions to be
representative.}

Previous studies have found differing results concerning the effect of
HeII reionization on the $T$-$\Delta_b$ relation.  As in our
simulations, the $T$-$\Delta_b$ relation in the studies of
\citet{gleser05} and \citet{paschos07} is not inverted.
In fact, the predictions for $\gamma - 1$ in \citet{gleser05} are quite
steep with $\gamma - 1 \geq 0.38$.  In contrast, \citet{furlanetto07a}
found that the $T$-$\Delta_b$ relation could become inverted by HeII
reionization and that the mean relation could deviate significantly
from a power-law in their excursion set based model.  This result owed
to the underdense regions being ionized later during HeII reionization
and, therefore, being the hottest.  In a toy case considered in
\citet{furlanetto07a}, in which there are no correlations between the
QSO bubble and the density field, the evolution of $\gamma-1$ during
HeII reionization is similar to what we find.  
%This slight difference is probably because
%\citet{furlanetto07a} assume ionized gas has a single temperature at
%the time of ionization -- i.e., that there is no memory of past
%heating.

The 
no-correlation approximation in the \citet{furlanetto07a} toy model is
reasonable during HeII reionization by quasars.  The magnitude of the correlation between HeIII bubbles
and fluctuations on the Jeans scale is relevant for how HeII
reionization affects the $T$-$\Delta_b$ relation.  The linear-theory
correlation coefficient between the overdensity in a region of $10^{12}
\; \Msun$ that hosts a QSO and a Jeans mass region separated from it
by $20$ Mpc (where we take $M_J = 10^{10} \; \Msun$) is $r= 0.02$ ($r
= 0.008$ for a $30$ Mpc separation).  The value of $r$ would be even
smaller if non-linearities are taken into account.  Therefore, the
probability distribution function of $\delta_b$ smoothed on the Jeans
scale within the HeIII region of a QSO is a Gaussian (in linear
theory) centered at $\langle \delta_b \rangle = r \, N_{\sigma}$,
where $N_{\sigma}$ is the quasar overdensity in units of the s.d. in
$\delta_b$.  The variance of this Gaussian is reduced by $1-r^2$ from the
volume-averaged variance.  Therefore, for the small $r$
values quoted above and reasonable values of $N_{\sigma}$, a Jeans mass clump in a
quasar HeIII region has a nearly equal chance to be overdense as
it does to be underdense.  
%Furthermore, since much of the heating takes
%place far from the quasar, the effective $r$ value between the region
%where heat is injected and the quasar could be smaller than in the
%above estimate.

%\citet{furlanetto07b}
%argued that there should be a transition from the density-independent
%phase to the density-dependent phase at some HeIII fraction at which
%multiple sources sit within a bubble.  However, once the bubbles grow
%large enough for this transition to occur, the large-scale modes at
%the bubble scale are already uncorrelated with the modes at the Jeans
%scale.  At times when the growth is collective, collective growth will
%not increase the correlations between the ionized regions and the
%Jeans-scale modes.

\citet{bolton08} and other studies have postulated that radiative
transfer effects can result in an inverted $T$-$\Delta_b$ relation.
The idea is that the radiation which makes it out into the underdense
voids is highly filtered, heating these regions to higher
temperatures.  However, this picture does not apply when quasars are
the sources that ionize the HeII.  Voids on the Jeans scale, the scale
that is most relevant for the Ly$\alpha$ forest transmission, are not
strongly anti-correlated with the giant HeIII bubbles.  Instead,
radiation that is filtered as it traverses a QSO bubble has an
approximately equal chance of ionizing a void or
filament at the bubble edge. Therefore, the filtering leads to scatter
in the relation between $T$ and $\Delta_b$, but not to an inverted
relation.  If the HeII-ionizing sources were more numerous and
much dimmer than QSOs, in this case an inverted relation is possible.  
%The softer
%radiation from a source would ionize its local overdense surroundings,
%and the harder radiation would travel past the overdense surroundings
%and ionize the voids (which would on average be the last regions to be
%ionized).
% To put in
%this numbers, $r = 0.4$ between a $10^{11}\; \Msun$ halo and an $M_J$
%scale region that are separated by $1$ Mpc -- sources with $R_b = 1$ are much more correlated with the 

Interestingly, \citet{bolton08} found that the flux probability
distribution from the $z \sim 3$ HI Ly$\alpha$ forest favors an
inverted T-$\Delta_b$ relation to accommodate the measured number of high
transmission pixels.  We have shown that the observed population of
QSOs cannot be responsible for an inverted relation.  However, we
think that HeII reionization by QSOs could still reconcile the
\citet{bolton08} result.  The \citet{bolton08} analysis did not
consider physically motivated models for the dispersion in this
relation.\footnote{They did consider a model with $\gamma - 1 \approx 0.3$ in
which this scatter was increased artificially by adding a Gaussian
dispersion to the particle photo-heating rates, and they found that
this model resulted in essentially no change in the flux PDF over a
model without scatter.  However, the \citet{bolton08} model with
scatter resulted in similar $90\%$ contours in the $T$-$\Delta_b$
plane compared to their model without scatter (see their Fig. 3 and
compare with our Fig. \ref{fig:Td_relation}).  The width of their
$90\%$ contours in the $T$ direction are factors of several smaller
than what we find.}  The large dispersion in $T(\Delta_b)$ that we find will
act to increase the number of unabsorbed pixels and could potentially
accommodate the \citet{bolton08} result without the need for an inverted
$T$-$\Delta_b$ relation.

\subsection{Mean HeII Ly$\alpha$ Forest Transmission}
\label{forest}

\begin{figure}
\rotatebox{-90}{\epsfig{file=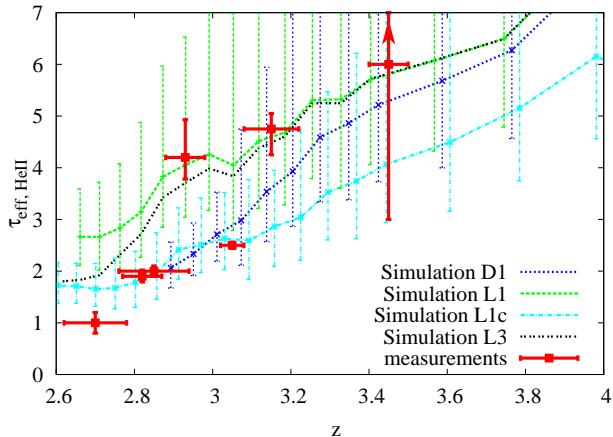, height=8.6cm}}
\caption{Evolution of the effective HeII optical depth.  The error bars on most of the simulation curves bracket
the values in which $\tau_{\rm HeII,eff}$ falls $90\%$ of the time when
measured over a skewer of length $190$ Mpc.  The measurement points
are described in the text. 
\label{fig:taueff}}
\end{figure}

There are currently only a handful of HeII Ly$\alpha$ forest
sight-lines.  The number of such sight-lines is limited because a
bright quasar is required to achieve adequate sensitivity with
existing instruments and because a single HI Lyman-limit system can
absorb much of the flux at wavelengths where HeII Ly$\alpha$
absorption occurs.  The number of HeII Ly$\alpha$ forest sight-lines
is expected to increase once the Cosmic Origins Spectrograph is
installed on Hubble, which is currently scheduled for spring of 2009.

In this section, we calculate $\tau_{\rm eff, HeII}$ along
skewers through our simulation box.  These spectra, as well as those
discussed in Section \ref{forestHI}, are generated by gridding the
N-body particles on a $256^3$ mesh in the $186$ Mpc box and using the
temperature and ionization fields from the simulations. Our
predictions for the HeII-Ly$\alpha$ transmission are reliable to the
extent that our dark matter simulations capture the gas density PDF
(especially for $\delta_b < 0$, where most of the transmission
occurs).  In Appendix \ref{ap:gdist}, we show that the dark matter PDF
measured from our grid reasonably reproduces the expected gas PDF.
Our predictions for the mean HeII Ly$\alpha$ transmission change by
only $\lesssim 20\%$ when we calculate this statistic from the $512^3$
gridded N-body field rather than the $256^3$ one.  This agreement is
reassuring and suggests that our predictions are robust.
%  We ignore
%HeII damping wing absorption in these calculations, which is
%insignificant compared to resonant absorption.

Figure \ref{fig:taueff} plots $\tau_{\rm eff, HeII}$ from our
simulations and from the most constraining observations, where
$\tau_{\rm eff, HeII}$ is defined as minus the log of the HeII
Ly$\alpha$ forest transmission.  The square markers are from
observations of quasars HS1700+6416 \citep{fechner06}, HS1157+3143
\citep{reimers05}, PKS1935-692 \citep{anderson99}, Q0302-003
\citep{heap00}, and SDSSJ2346-0016 \citep{zheng04}. The vertical error
bars on these measurements are just statistical, and the horizontal
error bars signify the redshift range over which each of the values
were estimated.\footnote{Often the span over which a measurement of
$\tau_{\rm eff, HeII}$ is done is chosen such that it encompasses a
very dark (or bright) region in the HeII Ly$\alpha$ forest.
Unfortunately, this practice significantly biases the observational
estimates.}

The curves in Figure \ref{fig:taueff} are $\tau_{\rm eff, HeII}$ from
four simulations, for which each marker value is estimated from
$1000$, $186$ Mpc random skewers through the simulation volume.  The
error bars on most of the simulation curves bracket the values in
which $\tau_{\rm eff, HeII}$ falls $90\%$ of the time when measured
along a single skewer.  Note that the measurement points were taken
from skewers that are comparable in length to the $186$ Mpc box.
There should be an additional cosmic variance error on these
points that is similar in magnitude to the error bars on the
simulation curves.

Our simulated predictions for $\tau_{\rm eff,
HeII}(z)$ provide moderate agreement with the measured values.  The evolution of $\tau_{\rm eff,
HeII}(z)$ in simulations D1 and L1c is more consistent with three of
the four measured points at $z<3$ than the $\tau_{\rm eff,
HeII}(z)$ in simulation L1.  None of the simulations are
consistent with the $z =2.7$ measurement (although, simulation
D1 was terminated at $z = 2.9$ and would have been the most consistent).

The differences in $\tau_{\rm eff, HeII}(z)$ among the simulations
stem in part from reionization ending at $z =3.15$ in D1 and $z =
3.45$ in L1c, whereas it is completed at a later time, at $z =2.95$,
in simulations L1 and L3. (The end of HeII reionization is defined
here as when $\bar{x}_{\rm HeIII}(z_{\rm end}) = 0.95$.)  Another
difference is that, at fixed $\bar{x}_{\rm HeIII}$, the value of
$\bar{\Gamma}_{\rm HeII}$ is smaller in simulations L1, L1c, and L3
than it is in D1 because the former supplement the resolution by using
filtering Method A or B while the latter does not employ additional
filtering.  A larger value for $\bar{\Gamma}_{\rm HeII}$ results in
more transmission.
%\footnote{Note that because recombination radiation
%is not included, the simulations could be underestimating
%$\bar{\Gamma}_{\rm HeII}$ by as much as $\approx 30 \%$ (Appendix
%\ref{ap:correction}), which results in a $\lesssim 30\%$ overestimate
%of $\tau_{\rm eff, HeII}$.}  

Simulations L1 and L3 differ only in the
filtering method that is used, and this also leads to differences in
the predictions for $\tau_{\rm eff, HeII}(z)$ primarily after HeII
reionization.  The uncertainties in the filtering of absorption systems
can be remedied with improvements to the radiative transfer algorithm
and by using high resolution gas simulations.  Even though some
theoretical uncertainty remains, the different trends among the
simulations suggest that $\tau_{\rm eff, HeII}(z)$
could be useful to distinguish between different HeII reionization
models.

%As with the mean value of $\tau_{\rm eff, HeII}$, the dispersion in
%$\tau_{\rm eff, HeII}$ can be used to discern between different HeII
%reionization models.  During HeII reionization, the dispersion in
%$\tau_{\rm eff, HeII}$ when estimated from a $186$ Mpc skewers is large
%because some skewers are completely absorbed by diffuse HeII whereas
%others allow significant transmission (Figure \ref{fig:taueff}). 
%This dispersion is significantly different among our simulations.  Of
%course, $186$ Mpc is a rather arbitrary scale chosen here to roughly match the
%path lengths used to estimate the observational data points.  A more motivated
%choice would perhaps be the dispersion in the transmission on the bubble scale.

\subsection{Mean HI Ly$\alpha$ Forest Transmission}
\label{forestHI}

\begin{figure}
\rotatebox{-90}{\epsfig{file=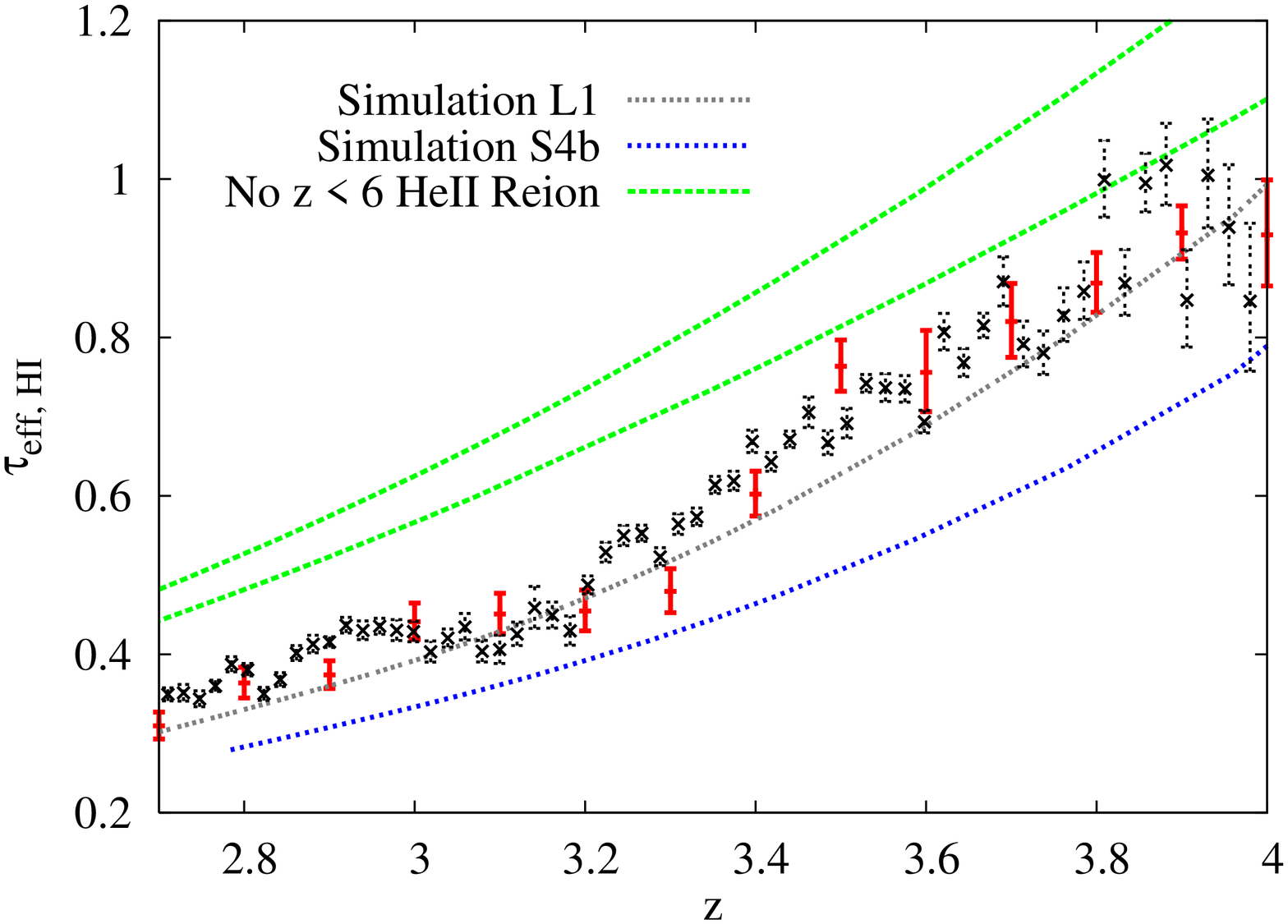, height=8.7cm}}
\caption{Evolution of the effective HI Ly$\alpha$ forest optical
depth.  The red solid points with error bars are the measurement of
$\tau_{\rm eff, HI}$ from \citet{faucher07}, and the black crosses are
the estimates of \citet{bernardi03}.  The features that are present at
$z \approx 3.2$ in the \citet{bernardi03} and \citet{faucher07} data
cannot be explained by HeII reionization in our simulations.  Any
disagreement between the data points and the model curves does not
imply that the model is disfavored, and instead it may indicate that
an incorrect form for $\Gamma_{\rm HI}(z)$ was assumed.
\label{fig:taueff_HI}}
\end{figure}

By increasing the temperature of the IGM, HeII reionization changes
the photo-ionization state of the intergalactic hydrogen (because
$\bar{x}_{\rm HI} \approx \alpha_{\rm A} \, n_e/\Gamma_{\rm HI}$,
where $n_e$ is the electron number density and $\alpha_{\rm A}
\sim T^{-0.7}$), thereby affecting the transmission properties of
the HI Ly$\alpha$ forest.  Interestingly, \citet{bernardi03} found
evidence for a depression in the mean transmission at $z \approx 3.2$
with width $\Delta z \approx 0.3$.  A similar depression was later
discovered by \citet{faucher07}.\footnote{The
\citet{faucher07} estimate uses $84$ high resolution Keck and Magellan
spectra, whereas the \citet{bernardi03} measurement is from $1061$
lower signal to noise and lower resolution Sloan spectra.}  
%Despite
%these differences, the \citet{bernardi03} and \citet{faucher07}
%measurement points are consistent with being drawn from the same
%distribution with $\chi^2/{\rm d.o.f.} = 1.26$ \citep{faucher07}.

The red points with solid error bars in Figure \ref{fig:taueff_HI} are
the \citet{faucher07} measurement of $\tau_{\rm eff, HI}$, which is
defined as minus the log of the HI Ly$\alpha$ forest transmission.
The black crosses with errors are the \citet{bernardi03} measurement.
All errors bracket the $68\%$ confidence level region.  Neighboring points in the \citet{bernardi03} measurement are
correlated.  The dip in absorption in these measurements that has been
the subject of attention is at $z \approx 3.2$ and has width $\Delta z
\lesssim 0.4$.  A dip is present with higher significance in the
\citet{bernardi03} measurement.

The theoretical curves are calculated from our simulations in the
manner described in Section \ref{forest} except for HI Ly$\alpha$
absorption.  These calculations assume a uniform HI-ionizing
background with $\Gamma_{\rm HI}(z) = 10^{-12}$ s$^{-1}$.  The effective optical depth in simulation L1 is representative of most of our runs, and $\tau_{\rm eff, HI}$ in simulation S4b is smaller than is typical
because of the higher IGM temperatures achieved in this simulation.  The dashed
curves give the evolution of the transmission for a simulation that is
initialized with $\gamma - 1 = 0$ and $T_0 = 10^4 \; {\rm K}$ or $2
\times 10^4 \, {\rm K}$ at $z = 6$ and in which HeII reionization does
not happen at $z < 6$.  Any disagreement between the data points and
the model curves does not imply that the model is disfavored.  The
form of $\Gamma_{\rm HI}(z)$ can be adjusted to yield any form for
$\tau_{\rm eff, HI}(z)$.

Our predictions for $\tau_{\rm eff, HI}(z)$ show no feature owing to
HeII reionization.  HeII reionization takes place over a much longer
interval in the simulations than the $\Delta z \lesssim 0.4$ of the
features seen in these measurements.  The duration of HeII
reionization in our simulations is determined primarily by the density
of $L \sim L_*$ quasars, which over the relevant redshift interval is
well constrained by current observations.  In addition,
even once HeII reionization is complete, it takes $\Delta z \sim 1.5$
after HeII reionization for the gas at mean density to cool by a
factor of two.  Therefore, cooling after HeII reionization is unlikely to produce any sharp feature in $\tau_{\rm eff, HI}$.

 \citet{theuns02} found using hydrodynamic simulations that $\tau_{\rm
eff, HI}$ can evolve significantly over a shorter interval than the
time over which the gas cools.  They explained this effect as owing to
velocity gradients that are induced by the heating.  They claimed that
these gradients broaden absorption lines, creating the rise in
$\tau_{\rm eff, HI}$.  Furthermore, they found that that the rise
after HeII reionization in their simulations is similar to the rise
seen at the low redshift end of the \citet{bernardi03} feature.  This
effect is not captured by our method, which misses the backreaction of
the photo-heating on the gas density.  However, \citet{theuns02}
attributed the decrease in $\tau_{\rm eff, HI}$ on the other side of
the \citet{bernardi03} feature to heating from HeII reionization,
which they modeled as a quick, homogeneous process.  In our
simulations, HeII reionization takes place over $\Delta z > 1$
compared to $\Delta z \approx 0.1$ in the \citet{theuns02}
simulations, and, therefore, any trends in $\tau_{\rm eff, HI}$ owing
to heating must be broader in our picture.

\section{Conclusions}

We have run a set of simulations of HeII reionization to understand
the structure of HeII reionization and its effect on several
observables.  We find that for a late reionization of HeII by
quasars:
\begin{itemize}

\item The popular assumptions that the HeIII ionization fronts are
sharp and that there is uniform heating within the front (and no
heating outside of it) do not yield a realistic model for HeII
reionization.  While the ionization fronts are still fairly localized,
hard photons stream far from their sources and are absorbed ahead of
the front.  These photons inject at large distances a significant
fraction of the energy radiated above the HeII Lyman-limit.

\item The average temperature at the mean density is increased by
$\approx 12,000 \; {\rm K}$ over the average temperature of the gas in
the absence of HeII reionization for $\bar{\alpha}_{\rm UV} = 1.6$.
This temperature increase is consistent with estimates that assume
that the IGM absorbs all photons with energies less than a few hundred
eV during HeII reionization. Regions that are ionized last are ionized
by the hardest radiation, reaching $T > 30,000 \; {\rm K}$ in our
fiducial model. 
% There is a broad distribution of temperatures at a
%given density, which has a s.d. of $\Delta T \sim 10,000 \; {\rm K}$
%at the end of HeII reionization.  
If the spectral index of the QSOs is different from our fiducial value
of $\bar{\alpha}_{\rm UV} = 1.6$, the average temperature can be be
significantly altered.

\item Poisson fluctuations in the number of QSOs rather than their
spatial clustering shape the structure of the ionization and
temperature fluctuations on $\lesssim 50$ Mpc scales because rare $L
\sim L_*$ quasars ionize the HeII.  HeII reionization produces large
temperature fluctuations on $50$ Mpc scales, and it results in the
ionization and HeII photo-heating fluctuations being essentially
uncorrelated with the density fluctuations on the Jeans scale, the
scale most relevant for HI and HeII Ly$\alpha$ forest statistics.

\item Measurements of the $z \sim 3$ forest suggest $T_0 \approx
20,000\, {\rm K}$. Without invoking an exotic heating mechanism, a
late HeII reionization epoch is required to produce this temperature.
The amount of additional heat provided by HeII reionization in our
simulations is enough to produce the inferred temperatures.

\item To the extent that there is a $T$-$\Delta_b$ relation, HeII
reionization by quasars leads to a temperature-density equation of
state of $\gamma -1 \approx 0.3$ for $0.1 \lesssim \bar{x}_{\rm HeIII}
\lesssim 0.9$.  HeII reionization by QSOs cannot result in an inverted
relation ($\gamma - 1 < 0$) as has been claimed. 

\item Our simulations of a $z\sim 3$ HeII reionization process produce
  a similar evolution in the HeII Ly$\alpha$ mean transmission to what
  has been estimated.  Better observations of the mean transmission and its
  scatter will be able to rule out models presented here.

\item In our simulations, the heating from HeII reionization by
quasars is unable to produce any semblance of the observed depression
in the $z \approx 3.2$ HI Ly$\alpha$ forest opacity.  The HeII
reionization process is too extended in redshift ($\Delta z > 1$) to
be responsible for this narrow feature ($\Delta z \lesssim 0.4$).

%There have been other reported abrupt changes in the IGM properties.
%\citet{schaye00} and \citet{ricotti00} detected an increase in the
%temperature of the IGM from fitting to line widths in the HI
%Ly$\alpha$ forest, with the increase occurring over $\Delta z \approx 0.5$.
%We have argued that these measurements may not be inconsistent with
%our simulations, but a more detailed comparison is necessary.  In
%addition, \citet{songaila98} measured a decrease in the SIV/CIV
%ratio at $z \approx 3$.  This measurement is consistent with a
%decrease over $\Delta z \approx 1$, comparable to the duration of HeII
%reionization in our simulations.

\end{itemize}

 The morphology of HeII reionization is considerably different than
that of hydrogen reionization (e.g., \citealt{mcquinn07}).  During
hydrogen reionization, current models find that the growth of
ionized bubbles is more collective; hundreds or even thousands of
galaxies within a bubble contribute to its growth, leading to the
bubble structure tracing the distribution of galaxies and to tens of
Mpc HII regions (e.g., \citealt{zahn07}).  During HeII reionization,
the growth is more stochastic.  Regions that happen to host a
``nearby'' quasar (within $\sim 30$ Mpc) are ionized by that quasar.
Because the QSO bubbles are so extended, the ionized structures are
larger than during HI reionization.

Another significant difference stems from the m.f.p. of the ionizing
photons to be absorbed in diffuse gas.  The spectrum from quasars is
harder than that of stars -- our best guess for what ionizes the
hydrogen -- the number density of helium is $70$ times smaller at $z=
3$ than hydrogen at $z= 6$, and the cross section is $4$ times
smaller.  These three factors conspire to make the typical m.f.p. for
a HeII ionizing photon megaparsecs rather than kiloparsecs, as it is
for HI ionizing photons during hydrogen reionization.  We have seen
that some HeII ionizing photons free stream far from their sources,
partially ionizing and heating up these regions.  If stars reionize
the hydrogen, ionization and heating occurs within HII bubbles.

%In future work, we intend to address how HeII reionization affects the
%HI and HeII Ly$\alpha$ forest transmission in more detail, including
%work to understand whether our simulations consistently reproduce the
%large opacity fluctuations observed at lower redshifts in the HeII
%Ly$\alpha$ forest (e.g., \citealt{shull04}).  Furthermore, we plan to
%investigate statistics with which to isolate the effect of HeII
%reionization in the forest.  To capture the effect HeII reionization
%has on pressure-smoothing the gas will require coupling our radiative
%transfer code to a hydrodynamic code.

 A definitive identification of the redshifts of HeII reionization
will place constraints on the sources that produce $> 4$ Ry photons
and will aid studies of the HI Ly$\alpha$ forest.  The data from
previous observations, if analyzed properly, may be able to confirm
whether quasars reionize HeII at $z \approx 3$.  In addition,
future observations of the HI and HeII Ly$\alpha$ forest will soon be
available with Sloan Digital Sky Survey III\footnote{www.sdss3.org}
and the Cosmic Origins Spectrograph on the Hubble Space
Telescope. These telescopes will significantly increase the number of
sight-lines in the HI and HeII Ly$\alpha$ forests.  It is timely to
make predictions for the effect of HeII reionization on these
observations.

\section{acknowledgments}
Many thanks to Mark Dijkstra, Alexandre Tchekhovskoi, and Hy Trac for
interesting discussions, and to James Bolton and Steven Furlanetto for
providing useful comments on the manuscript.  MM is supported by the
NSF graduate student fellowship.  This work was supported in part by
NSF grants ACI 96-19019, AST 00-71019, AST 02-06299, AST 03-07690, and
AST 05-06556, and NASA ATP grants NAG5-12140, NAG5-13292, NAG5-13381,
and NNG-05GJ40G.  Further support was provide by the David and Lucile
Packard, the Alfred P. Sloan, and the John D. and Catherine
T. MacArthur Foundations.
\begin{appendix}

\section{A. HeII Absorbers}
\label{ap:LLS}

In this section, we examine the effect that dense systems have on
filtering the radiation field and whether this effect is adequately
captured in our simulations.  We discuss two methods that we use to
supplement the resolution of the simulations.

\subsection{Analytic Modeling}

The HI column density ($N_{\rm HI}$) distribution is relatively well
measured from the HI Ly$\alpha$ forest, whereas the HeII column
density ($N_{\rm HeII}$) distribution is significantly more
uncertain. Given the well-measured $N_{\rm HI}$ distribution, we would
like to derive the $N_{\rm HeII}$ distribution and then use this to
calculate the effect of filtering.  We follow the method described in
\citet{haardt96}.  This approach uses the value of $N_{\rm HI}$ for an
absorber and the incident values of $\Gamma_{\rm HI}$ and $\Gamma_{\rm
HeII}$ to predict the value of $N_{\rm HeII}$.

If the absorber is optically thin to $4 \, \Ry$ photons, it is trivial to
infer the value of $N_{\rm HeII}$, namely
\begin{equation}
  N_{\rm HeII} = N_{\rm HI}\; \eta_{\rm thin},
\label{eq:NHeII_eq}
\end{equation}
where
\begin{equation}
  \eta_{\rm thin} \approx \frac{5.5 \;Y_{\rm He}}{4 \; (1 - Y_{\rm He})} \; \left( \frac{\Gamma_{\rm HI}}{\Gamma_{\rm HeII}} \right),
\label{eq:eta_thin}
\end{equation}
and where $Y_{\rm He}$ is the mass fraction in helium and the 
factor $5.5$ is the approximate ratio of HeII and HI recombination rates. HI column densities with
\begin{equation}
  N_{\rm HI, thin} < 4 \times \sigma_{\rm HI}^{-1} \;\eta^{-1} = 1.3 \times 10^{16} ~ \left(\frac{50}{\eta} \right) \; \cmeter^{-2},
\label{eq:selfshield}
\end{equation}
yield optical depths of $\tau \lesssim 1$ for  $4 \, \Ry$ photons, where $\eta \equiv N_{\rm
HeII}/N_{\rm HI}$.  

What values of $\eta$ do we expect?  Taking the $\Gamma$-values
inferred from the HI Ly$\alpha$ and HeII Ly$\alpha$ forests at
$z\approx 2.5$ of $\Gamma_{\rm HI} \approx 10^{-12} \; {\rm s}^{-1}$
and $\Gamma_{\rm HeII} \approx 10^{-14} \; {\rm s}^{-1}$ (e.g.,
\citealt{shull04}), equation (\ref{eq:eta_thin}) yields $\eta_{\rm
thin} \approx 50$.  Or, taking the $z \approx 3 $ values of
$\Gamma_{\rm HI} \approx 10^{-12} \; {\rm s}^{-1}$ and $\Gamma_{\rm
HeII} \approx 10^{-15} \; {\rm s}^{-1}$ -- at least in some regions -- 
$\eta_{\rm thin}$ increases to $500$ (e.g., \citealt{reimers97}).

An absorber is optically thick to $4 \; \Ry$ photons while also being
optically thin to $1 \; \Ry$ photons for $N_{\rm HI, thin} \lesssim
N_{\rm HI} \lesssim \sigma_{\rm HI}^{-1} = 1.6 \times 10^{17} \;
\cmeter^{-2}$.  Consequently, $\Gamma_{\rm HeII}$ can be
much lower within an absorber than the incident value, and, therefore,
$N_{\rm HeII}$ can be larger relative to $N_{\rm HI}$ than in the
optically thin case.  Thus, to calculate $N_{\rm HeII}$ we solve
the quadratic formula given in \citet{fardal98} (their eqn. A11),
which determines $N_{\rm HeII}$ assuming a slab of primordial gas.
This equation requires as input $N_{\rm HI}$, $\Gamma_{\rm HI}$,
$\Gamma_{\rm HeII}$, and $n_H$ -- the number density of atomic and
ionized hydrogen in the absorber.  This method is an improvement over the
\citet{haardt96} multi-zoned model for incorporating HeII absorption
\citep{fardal98}.

To proceed with solving the \citet{fardal98} equation for $N_{\rm
HeII}$, a model for $n_H(N_{\rm HI})$ is required.  Simulations of the
HI Ly$\alpha$ forest have had remarkable success at matching the
$N_{\rm HI}$ distribution found in HI Ly$\alpha$ forest observations
(e.g., \citealt{cen94, miralda96, hernquist96, katz96}),
and analytic models have been
constructed to understand these simulations.  In particular,
\citet{schaye01} argued that Ly$\alpha$ forest absorbers have sizes that
are comparable to the Jeans length, $L_j(\delta_b, T)$.
%The reason for this characteristic size is
%simple: Ly$\alpha$ absorbers are gravitationally confined (e.g.,
%\citealt{cen94}), and gravitationally confined absorbers will adjust such
%that the maximum density is smoothed over a scale that is comparable
%to $L_j$ \citep{schaye01}.  
With the assumption that $N_{\rm HI} = n_{\rm HI} \; L_j$ and of
photo-ionization equilibrium, it is trivial to map between $n_H$ and
$N_{\rm HI}$ for $N_{\rm HI} \lesssim \sigma_{\rm HI}^{-1}$, namely
\citep{schaye01}
\begin{equation}
n_H \approx 6 \times 10^{-4} \; {\rm cm}^3 \; \left( \frac{N_{\rm HI}}{10^{16} \; {\rm cm}^2} \right)^{2/3} \; T_4^{0.17} \; \left(\frac{\Gamma_{\rm HI}}{10^{-12} \; {\rm s}^{-1}} \right)^{2/3},
\label{eqn:n_H}
\end{equation}
where $T_4 = T/10^4 \; {\rm K}$.  This formula agrees well with
simulations \citep{schaye01}.  For our purposes it may reasonable to
use equation \ref{eqn:n_H} even when $N_{\rm HI} \gtrsim \sigma_{\rm
HI}^{-1}$, which is beyond its realm of applicability, because such
high column density absorbers play an insignificant role in filtering
the radiation near the HeII Lyman-limit.

\begin{figure}
\begin{center}
\rotatebox{-90}{\epsfig{file=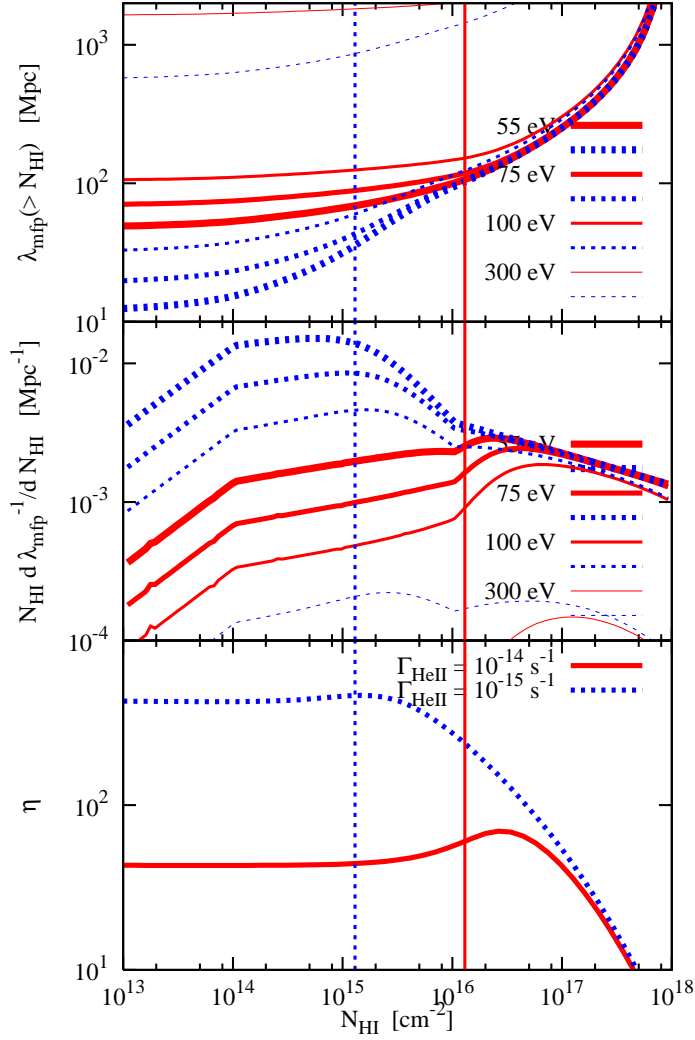, height=9.5cm}}
\end{center}
\caption{Dotted curves in each panel are for $\Gamma_{\rm HeII} =
10^{-15} \; \second^{-1}$ and the solid are for $\Gamma_{\rm HeII} =
10^{-14} \; \second^{-1}$ and all curves use $\Gamma_{\rm HI} = 10^{-12} \;
\second^{-1}$. Top Panel: The m.f.p. for photons at select
energies to be absorbed in a system with column density greater than $N_{\rm
HI}$. Middle Panel: The derivative with respect to $N_{\rm HI}$ of the
inverse of the m.f.p.  This quantity is maximized for the
$N_{\rm HI}$ that contribute most significantly to the absorption.
Bottom Panel: The value of $\eta = N_{\rm HeII}/N_{\rm HI}$ in this
model as a function of $N_{\rm HI}$.  The vertical lines in both
panels correspond to the $N_{\rm HI}$ at which an absorber becomes
optically thick for the $\Gamma_{\rm HeII}$ with the same line style.
\label{fig:abs_prob}}
\end{figure}

Figure \ref{fig:abs_prob} plots the predictions of this model for
$\eta(N_{\rm HI})$ (bottom panel), the m.f.p. to be absorbed in
systems with column density greater than $N_{\rm HI}$ (top panel), and
the contribution of different systems to the m.f.p. ($d \lambda_{\rm
mfp}(> N_{\rm HI})^{-1}/d \log N_{\rm HI}$ (middle panel).  The
function $d \lambda_{\rm mfp}(> N_{\rm HI})^{-1}/d N_{\rm HI}$ is
proportional to the probability a photon is absorbed in a system of
column density $N_{\rm HI}$.  These panels use the A1 fit to the
$N_{\rm HI}$ distribution in
\citet{fardal98}.\footnote{\citet{fardal98} derived best fit
functional forms for the distribution of absorbers as a function of
$N_{\rm HI}$ and $z$ (see Table 1 in \citealt{fardal98}), fitting to a
compilation of Ly$\alpha$ forest observations.  These fits account for
the observed depression in ${\partial^2 N}/{\partial N_{\rm HI}
\partial z}$ centered at $N_{\rm HI} \approx 10^{16} \; \cmeter^{-2}$
\citep{petitjean93}.  Note that the piecewise form for the
curves in the middle panel in Figure \ref{fig:abs_prob} owes to the
piecewise power-law that \citet{fardal98} used to fit the observed
$N_{\rm HI}$ distribution.}
%We employ models A1 and A2 in
%\citet{fardal98} for ${\partial^2 N}/{\partial N_{\rm HI} \partial
%z}$.  Model A1 is the \citet{fardal98} best fit to the data, and Model
%A2 is a slightly less probable fit that is designed to maximize the
%absorption owing to HeII and still agree with the data.
All curves assume $\Gamma_{\rm HI} = 10^{-12} \; \second^{-1}$ and $T
= 10^4$ K.  In the top panel, the solid curves take $\Gamma_{\rm HeII}
= 10^{-15} \; \second^{-1}$, and the dotted curves take $\Gamma_{\rm
HeII} = 10^{-14} \; \second^{-1}$.  The vertical lines in both panels
correspond to the approximate value of $N_{\rm HI}$ at which an
absorber becomes self-shielded to HeII ionizing photons for
$\Gamma_{\rm HeII} = 10^{-14} \; \second^{-1}$ (solid line) and
$\Gamma_{\rm HeII} = 10^{-15}\; \second^{-1}$ (dotted line).  For the
column densities at which an absorber becomes self-shielded, the value of
$\eta$ increases.  However, this increase is compensated by the
monotonically increasing nature of $n_{\rm H}(N_{\rm HI})$, and
ultimately this effect wins out and $\eta(N_{\rm HI})$ decreases.  This decrease is important, allowing $100$ eV photons to have long m.f.p.s to be absorbed in dense systems.
%\footnote{The
%absorbers that limit the m.f.p. for HeII reionization are much better
%understood that limit the m.f.p. for hydrogen reionization.  The
%absorbers that limit the m.f.p. for HeII ionizing photons are well
%observed in the Ly$\alpha$ forest and captured in hydrodynamical
%simulations.  For hydrogen reionization, the relevant absorbers are
%denser (and therefore more difficult to capture) and also this process
%occurs at high redshifts at which column density distributions cannot
%be reliably captured.  Arguments for the abundance of hydrogen
%Lyman-limit systems at $z >6$ rely on extrapolating the observed
%abundance at $z\sim 3-4$ {\bf cites}.  Such extrapolations are dubious
%since they result in much smaller m.f.p.s than are observed at $z\sim 3-4$.}

% At even higher
%$N_{\rm HI}$ (values greater than $\approx 2\times 10^{17} \;
%\cmeter^{-2}$), both helium and hydrogen becomes neutral and $\eta$
%decreases to zero.

 For $\Gamma_{\rm HeII} = 10^{-14} \; \second^{-1}$, absorbers with
$N_{\rm HI} \gtrsim 10^{16} \; {\rm cm}^{-2}$ limit the m.f.p. of $4$
Ry photons to $\approx 40$ Mpc at the HeII Lyman-limit (see thickest,
solid curve in middle panel in Fig. \ref{fig:abs_prob}).  For
$\Gamma_{\rm HeII} = 10^{-15} \; \second^{-1}$, lower column density
absorbers with $ 10^{14} \gtrsim N_{\rm HI} \gtrsim 10^{15} \; {\rm
cm}^{-2}$ limit the m.f.p. of these photons to $\approx 10$ Mpc
(dashed curves in middle and top panels).  Higher energy photons have
longer m.f.p. values. 
% These large m.f.p.s are in
%part attributable to the fall off in $\eta$ seen in the middle panel.
%Further, $\lambda_{\rm mfp} \propto E_\gamma^{2.7}$ between the $100$
%eV and $300$ eV curves; the power-law scaling of $\lambda_{\rm mfp}$
%is weaker at lower $E_\gamma$.  

\subsection{Filtering in our Simulations}

The radiative transfer grid employed in our study may not capture all
of the absorbers with $N_{\rm HI} \gtrsim 10^{15} \; \cmeter^{-2}$
(Appendix D). These absorbers can play an important role in filtering
the radiation field.  Furthermore, even if these absorbers are
resolved on the grid, the code will not properly handle the filtering
when $\Gamma_{\rm HeII} \gtrsim \Delta t^{-1}$.  We describe below two
methods that we use to better capture these effects.

\subsubsection{Filtering Method A}
 The simpler of the two filtering methods uses the density structure
on the radiative transfer grid and a modification of the radiative
transfer code to achieve the appropriate filtering.  Appendix \ref{ap:NHI}
shows that the column density distribution of absorbers measured from
the $256^3$ and $512^3$ N-body simulation grids reproduce to within a
factor of $2$ the column density distribution seen in gas simulations.
Therefore, our gridded N-body simulation will produce the correct
filtering of the radiation field if it passes three tests:
\begin{itemize}
\item  The density structure of the absorbers is
captured to the level that is required.
\item The radiative transfer method accounts for the
filtering of the radiation field as radiation traverses overdense cells.
\item The clustering of the absorbers is taken into account.
\end{itemize}
%For the first test, despite the structure of the absorbers being
%washed out when projected onto the radiative transfer grid, the
%density structure structure might not need to be captured.  
When the absorber is optically thin, $\eta$ is independent of the
absorber density.  As seen in the middle panel in Figure
\ref{fig:abs_prob}, a significant fraction of the filtering owes to
optically thin absorbers when $\Gamma_{\rm HeII} = 10^{-15}$ s$^{-1}$.
Even when an absorber becomes optically thick, the dependence of
$\eta$ on density is weak for the most relevant column densities if
one uses the \citet{schaye01} model for $n_H(N_{\rm HI})$. (Note the
flatness of $\eta$ in the bottom panel in Fig. \ref{fig:abs_prob} for
the $N_{\rm HI}$ that contribute the most to the m.f.p.)  Therefore,
since the $N_{\rm HI}$ column density distribution is captured fairly
well at our grid scale, there is reason to believe that the density is
adequately described for the purpose of filtering the radiation
field.

Passing the second of the three tests requires an addition to the
radiative transfer algorithm beyond what was outlined in the previous
section.  For $\Gamma_{\rm HeII}^{-1} \ll \Delta t$, a cell in the
code outlined in \S \ref{code} would have $x_{\rm HeII} < x_{\rm HeII,
eq}$ once some fraction of the rays that will travel through that cell
in a timestep have done so.  Therefore, the filtering will be
underestimated for the rays that enter after $x_{\rm HeII} < x_{\rm
HeII, eq}$.  To remedy this, once a cell with $\Delta_b > \Delta_F$
has $x_{\rm HeII} < x_{\rm HeII, eq}$, we perform the radiative
transfer as if $x_{\rm HeII} = \tilde{x}_{\rm HeII, eq}$, where
$\tilde{x}_{\rm HeII, eq} = \alpha_{\rm A} \, n_e/\tilde{\Gamma}_{\rm HeII}$
and $\tilde{\Gamma}_{\rm HeII}$ is tabulated from all the previous
rays that have entered the cell within the timestep up to the current
ray.  Note that $x_{\rm HeII}$ always approaches $x_{\rm HeII, eq}$
from below because recombinations are performed prior to ionizations
within a timestep.  We set $\Delta_F = 4$, which for $\Gamma_{\rm HI}
= 10^{-12}$ s$^{-1}$ corresponds to
\begin{equation}
N_{\rm HI} = 2 \times 10^{14}\; \cmeter^{-2} \; \left(\frac{\Delta_F}{4}\right)^2 \; \left(\frac{l_{\rm cell}}{1 \; \Mpc} \right) \; \left(\frac{1 +z}{4}\right)^5. 
\end{equation}

Passing the final of the three tests -- capturing the clustering of
the absorbers -- is the most difficult and is not achieved with this
method.  Even though the column density distribution is maintained,
the bias of the absorbers is changed by the coarse gridding (Appendix
D).  It is conceivable that capturing the bias is not of crucial
importance because the m.f.p. for ionizing photons is typically larger
than the correlation length of the absorbing systems.  This method
does not filter the background radiation ($\gtrsim 500 \; {\rm eV}$).
Figure \ref{fig:abs_prob} suggests that such filtering is not
important because the m.f.p. for these photons to be absorbed in a
dense system is gigaparsecs.

% The optical depth of an absorber with $N_{\rm HI}$ is
%\begin{equation}
%\tau(E_{\gamma}) = \sigma_{\rm HI}(E_{\gamma}) \;N_{\rm HI} + \sigma_{\rm HeII}(E_{\gamma})  ~N_{\rm HeII}.
%\end{equation}
\subsubsection{Filtering Method B}
The second method for filtering the radiation is based on the model
proposed in \citet{haardt96} and improved in \citet{fardal98} to infer
$N_{\rm HeII}$ from $N_{\rm HI}$, which is described in the beginning of
this section.  This inference also requires as input $\Gamma_{\rm
HeII}$.  We take it to be the maximum of the current tabulated value of
$\Gamma_{\rm HeII}$ in the cell and the value of $\Gamma_{\rm HeII}$ from
the previous timestep.  

In addition, this method assumes that Ly$\alpha$ absorption
arises from gas at the outskirts of dark matter halos. We assume that
the cross section for a photon to intersect an absorber of column
density $N_{\rm HeII}$ associated with a halo of mass $m$ has the
functional form
\begin{equation}
\sigma(m,N_{\rm HeII})
= \sigma_{m_*} \, P(N_{\rm HeII}) \, (m/m_*)^{2/3},
\label{eqn:cross_section}
\end{equation}
 where $P(N_{\rm HeII})$ is taken to be the probability distribution
of $N_{\rm HeII}$ that corresponds to lines with $\log_{10}(N_{\rm
HI}) > 14.5$ and is zero for smaller column densities, and
$\sigma_{m_*}$ sets the normalization.  In this method, rays that have
photons of energy $E_{\gamma}$ and that travel through a cell with a
halo of mass $m$ are attenuated by the factor $\int dN_{\rm HeII}
\sigma(m,N_{\rm HeII}) \exp[-\tau(E_{\gamma}, N_{\rm HeII}, N_{\rm
HeII})]/l_{\rm cell}^2$, where $\tau = \sigma_{\rm HI}(E_{\gamma}) \,
N_{\rm HI} + \sigma_{\rm HeII}(E_{\gamma}) \, N_{\rm HeII}$.

 The $m^{2/3}$ scaling in equation (\ref{eqn:cross_section}) is chosen
such that the cross section is proportional to the square of the halo
virial radius.  The values of $\sigma_{m_*}$ and $m_*$ are assigned so
that the mass integral over $\sigma(m,N_{\rm HeII}) \, n_h(m)$, where
$n_h(m)$ is the halo mass function, yields $d{\it N}/d{\it l} d N_{\rm
HeII}$ -- the number of absorbers per proper length ${\it l}$ per
$N_{\rm HeII}$.  To derive $d{\it N}/d{\it l} d N_{\rm HeII}$, we use
the A1 fit to $d{\it N}/d{\it l} d N_{\rm HI}(z)$ provided in
\citet{fardal98} and that $d{\it N}/d{\it l} d N_{\rm HeII}(z) \equiv
\eta(N_{\rm HI})^{-1} \, d{\it N}/d{\it l} d N_{\rm HI}(z)$. While we
use the observations to construct $P(N_{\rm HeII})$, physically it has
to do with the density profile around halos.  The motivation for this
filtering method is that the high column density absorbers are shown
to be associated with halos \citep{cen97, miralda98}.  This method filters both
the rays and the background radiation.

\section{B. Tests of Code}
\label{ap:tests}

\subsection{Number of Photon Bins}
\label{ap:photon_bins}

\begin{figure}
\begin{center}
\rotatebox{-90}{\epsfig{file=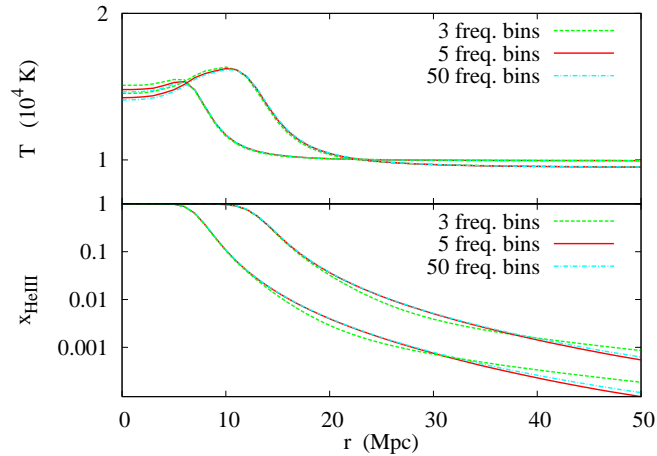, height=9cm}}
\end{center}
\caption{Test of convergence with the number of frequency channels for
the case of a single QSO in a homogeneous IGM with $x_{\rm HeII} = 1$.  The top
panel shows the temperature, and the bottom shows the HeIII
fraction.  The left-most curves are for $t = 50 \; \Myr$ after a
source with $\alpha_{\rm UV} = 1.5$ and $\dot{N} = 10^{54}$ ionizing
photons s$^{-1}$ turns on and the right-most curves are $t = 250 \;
\Myr$ after.
\label{fig:num_freq}}
\end{figure}

Since the computation time scales linearly with the number of
frequency bins $n_{\nu}$ and since memory requirements also increase
linearly with $n_{\nu}$, it is important to find the minimum value of
$n_{\nu}$ that achieves acceptable convergence.  To test how our code
converges with the number of frequency bins, we place a quasar in a
homogeneous IGM initialized with $x_{\rm HeII} = 1$ and solve for
$x_{\rm HeII}$ and $T$ while varying $n_{\nu}$. Figure
\ref{fig:num_freq} displays the temperature (top panel) and $x_{\rm
HeIII}$ (bottom panel) for $n_{\nu} = 3, 5$, and $50$.  For
simplicity, we have set the speed of light to infinity.  The left-most
curves are for a source with $\alpha_{\rm UV} = 1.5$ and $\dot{N}
=10^{54}$ ionizing photons s$^{-1}$ after $t = 50~ \Myr$ and the
right-most curves are the same after $t = 250~ \Myr$, where $t=0$
corresponds to $z=4$.  The curves at fixed $t$ have only minor
differences on scales that have $x_{\rm HeIII} \gtrsim 0.01$.
However, for ${x}_{\rm HeIII} < 0.01$, $3$ frequency bins provides
fairly poor convergence to the true solution (which essentially is
represented by the $n_{\nu} = 50$ case), whereas $5$ frequency bins
shows much better convergence.  The cosmological simulations presented
in this paper use $n_{\nu} = 5$ unless otherwise specified.

\begin{figure}
\begin{center}
{\epsfig{file=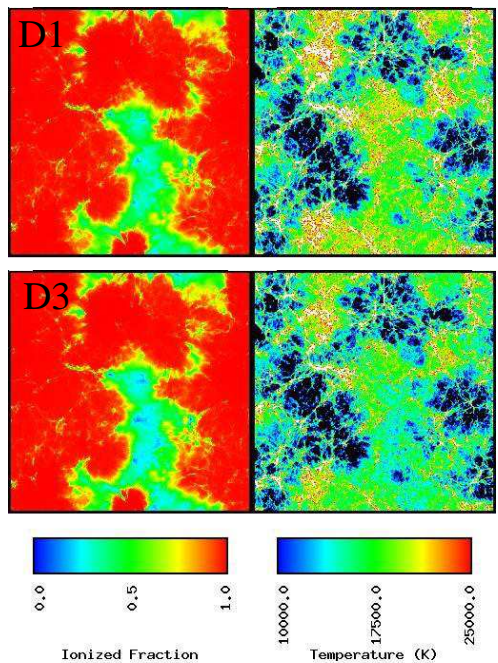, height=9cm}}
\end{center}
\caption{Comparison of the fiducial $5$ frequency bin simulation --
simulation D1 -- to a simulation with $10$ frequency bins -- D3.  The
panels are taken from the $z = 3.3$ snapshot, for which $\bar{x}_{\rm
HeIII} = 0.85$.
\label{fig:num_freq_3D}}
\end{figure}

Even though for a single source the solution is well converged to
fairly large radii for $n_\nu = 5$, in the simulations the small
inaccuracies at large radii add up and can result in a significant
error.  Figure \ref{fig:num_freq_3D} compares at $\bar{x}_{\rm HeIII}
= 0.85$ of the fiducial $5$ frequency bin simulation -- simulation D1
-- to one with $10$ frequency bins -- D3 -- with the bins centered at
$(57, 66, 76, 90, 107, 129, 162, 209, 283, 409)$ eV such that each bin carries an equal energy.  We have checked that the latter simulation
is well converged to the full solution by running simulations between
$5$ and $10$ bins and with different spacings in energy. The
ionization field is similar between simulations D1 and D3, with the
voids being slightly more ionized in D1.  The value of $\bar{T}$ is
$300$ K greater in D1, and it can be seen that some of the hotter
regions are a bit hotter in D1 than in D3.  However, these differences
are small compared to the differences that arise owing to, for
example, the uncertainty in $\bar{\alpha}_{\rm UV}$.  We have also
compared the temperature PDF between D1 and D3, and the PDFs are
almost identical.  The temperature PDF still extends to $35,000\;{\rm
K}$ in D3 as in D1 at the end of HeII reionization.

%While ${x}_{\rm HeIII} \lesssim 0.01$ may seem unimportant to capture
%accurately, a significant fraction of energy during HeII reionization
%is deposited at these ionized fractions, where the ionizations owe to the
%hardest photons.  Small contributions to the ionization fraction from
%hard photons from a single source can sum when many sources are
%present within the m.f.p. of the photons to yield substantial
%photo-ionization and photo-heating rates.

\subsection{Code Comparison}
\label{ap:code_comp}

Analytic solutions to multi-frequency radiative transfer problems
generally do not exist.  To test our code, we compare it with the 1-D
radiative transfer code presented in \citet{lidz07}.  To simplify the
comparison, we use a 1-D version of our code that is identical to our
3-D code except that it sends a single ray during a timestep to
calculate the spherically symmetric solution.  The \citet{lidz07} code
was written independently by Adam Lidz who was not involved in the
development of our new code.  The methodology used in our code is
significantly different from that used in the
\citet{lidz07} code.  The latter code assumes a power-law spectrum, and
uses this power-law and the intervening column densities of HI, HeI,
and HeII to infer the incident spectrum at some radius from the
source.  Furthermore, it uses much smaller timesteps than our code and
iterates to achieve convergence.

We analyze the solutions provided by these codes for the simple case of
a quasar that produces $10^{54}$ s$^{-1}$ HeII ionizing photons with a
spectral index of $1.5$ in a homogeneous IGM with $x_{\rm HeII} = 1$.  We solve for the ionization front of this
quasar assuming a homogeneous, static (non-expanding) universe, with
density equal to the mean density at $z =4$.  Initially, the hydrogen
is assumed to be ionized and the helium to be singly ionized.  For
this calculation, the speed of light is taken to be infinite, the code
used in this paper is set to have $1$ Myr timesteps, and the same ray-casting
parameters as used in the cosmological runs presented in this paper.

\begin{figure}
\begin{center}
\rotatebox{-90}{\epsfig{file=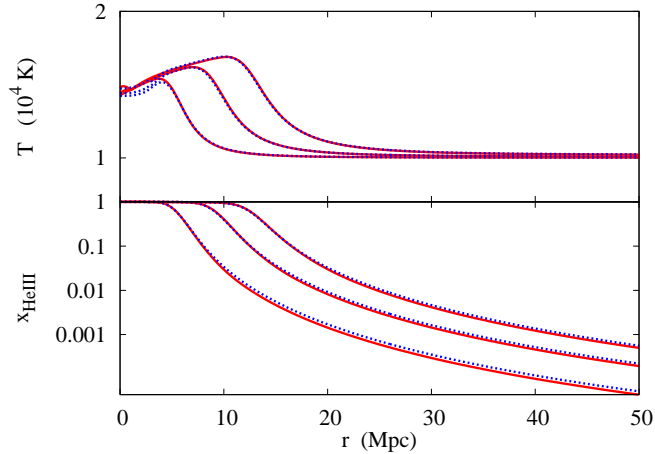, height=9cm}}
\end{center}
\caption{Comparison of the temperature (top panel) and HeII ionized
fraction (bottom panel) between our code (dotted
curves) and the code in \citet{lidz07} (solid curves).  The QSO is
shining in a homogeneous medium at $z = 4$ with $\dot{N} = 10^{54}$
HeII ionizing photons s$^{-1}$.  The left-most curves are $T(r)$ and
$\bar{x}_{\rm HeIII}(r)$ after $2$ Myr, the middle curves are these
functions after $8$ Myr, and the right-most curves are after $20$ Myr.
\label{fig:code_comp}}
\end{figure}

Figure \ref{fig:code_comp} compares the evolution of the temperature
and $\bar{x}_{\rm HeIII}$ front as a function of time between a $1$-D
version of our code (dotted curves) and the code in \citet{lidz07}
(solid curves).  The left-most curves are the temperature, $T$, and
$\bar{x}_{\rm HeIII}$ values as a function of radius after $2$ Myr
($2$ timesteps for our code), the middle curves are those after $8$
Myr ($8$ timesteps), and the right-most curves are those after $20$
Myr ($20$ timesteps).  At $r <5$ Mpc, the agreement between the $T(r)$
produced from the two codes is not perfect (top panel).  The slight
disagreement owes to a simplification our algorithm makes: it groups
all of the photons released in a timestep into a single photon bundle.
In reality, harder photons will travel ahead of softer photons because
softer photons are absorbed first.  We have verified that if we
decrease our timestep the solution from our code converges to that of
the \citet{lidz07} code.  However, our 3-D code sends multiple rays
for each cell at the radii where the discrepancy exists, capturing
better the region where the 1-D code has not converged.  We have
verified that this minor discrepancy essentially disappears in the 3-D
code.  Furthermore, even though timesteps are larger in the
cosmological runs than in this test, the convergence of this code is
most dependent on the number of timesteps rather than the timestep
duration or source luminosity.

The bottom panel in Figure \ref{fig:code_comp} compares the HeIII
ionization fronts in the two codes.  We plot the y-axis in log because
the solutions for the fronts are essentially indistinguishable on
linear scales.  At large radii, there is a minor systematic difference
in the ionization level.  This minor difference is much smaller than
other uncertainties inherent in HeII reionization calculations, in
particular the value of $\bar{\alpha_{\rm UV}}$.  This systematic does
not depend on the number of frequency bins (for $n_{\nu} \geq 5$) or
decreasing the timestep in our code.  Also, note that the output of
the \citet{lidz07} curve is at $1.9$ Myr rather than $2$ Myr for its
left-most curve, which accounts for some of the discrepancy.  The
maximum differences between the $8$ Myr curves is $3\%$.  The small
differences at large $r$ could arise because the \citet{lidz07} code
follow the full three level system, whereas our code ignores HeI.

\subsection{Temperature-Density Relation}
\label{ap:tdrelation}

Here, we investigate how well our scheme for computing the temperature evolution in
N-body simulations works.  The most
difficult aspect of such a scheme is to evaluate the convective
derivatives that account for the flow of matter through cells.  Our
scheme for capturing this is discussed in Section \ref{temperature}.  To test our code, we compare the resulting $T-\Delta_b$ relation in a simulation without HeII reionization to that of the
\citet{hui97} analytic model. This analytic model has been shown to
reasonably match the relation found in semi-analytic calculations
that use the Zel'dovich approximation and the evolution of the
temperature density relation found in hydrodynamic simulations (although,
the \citet{hui97} comparison is rather limited for the latter case).

\begin{figure}
\begin{center}
\rotatebox{-90}{\epsfig{file=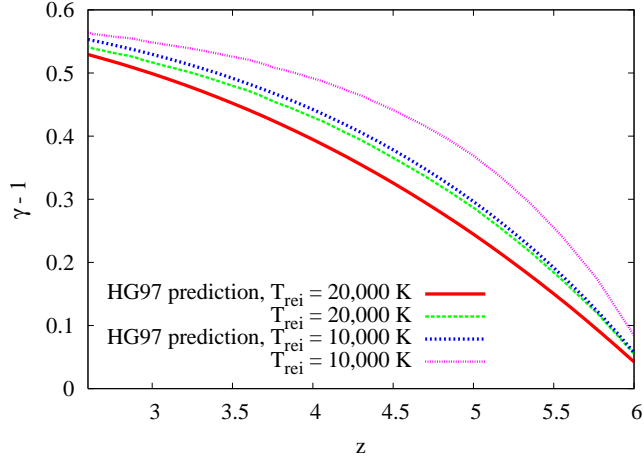, height=9cm}}
\end{center}
\caption{Evolution of $\gamma - 1$ in our code compared to that
predicted by \citet{hui97} (HG97) for $T_{\rm rei} = 20,000$
K and $T_{\rm rei} = 10,000$ K. 
\label{fig:hui_comp}}
\end{figure}

Figure
\ref{fig:hui_comp} compares the evolution of the $T-\Delta_b$ equation
of state $\gamma -1$ in our code using the $256^3$ grid in the $186$
Mpc box to the $\gamma -1$ predicted by the analytic formula in
\citet{hui97} for models where $z_{\rm rei} = 6.2$.  Our $\gamma -1$
are calculated by fitting the $T-\Delta_b$ relation to a power-law,
weighting the fit by the number of grid cells at each $\Delta_b$.  We
have checked that our results do not change if we instead fit only to
$\Delta_b$ values near unity, which is closer to what \citet{hui97} calculate.

The agreement is acceptable between the two predictions for $\gamma
-1$.  We would not expect exact agreement.  Our code calculates the
temperature evolution on a non-linear scale (the cell scale) whereas
the \citet{hui97} formula uses linear theory.  Therefore, our value
for $\gamma -1$ should evolve more quickly than the \citet{hui97}
estimate since $\Delta_b$ on average grows faster on non-linear
scales.
%\footnote{The analytic formula in \citet{hui97} ignores
%  radiative cooling, but we have checked that the evolution for
%  $\gamma -1$ from our code does not change appreciably if we also
%  ignore these processes.}

Our scheme misses the effects of shocks, which add dispersion to the
$T-\Delta_b$ relation.  The s.d. in the $T-\Delta_b$ relation from
shocks is $\lesssim 500$ K at
$\Delta_b = 0.5$ and $ \lesssim 1000$ K at $\Delta_b =1$, and it
increases steadily as $\Delta_b$ increases (see Figure 1 in
\citealt{hui97}).  The dispersion in temperature at low densities in
simulations that include shocks (and a uniform ionizing background) is
much smaller than the dispersion that results from HeII reionization.

\section{C. Ionization Correction}
\label{ap:correction}
The fraction of helium in HeII evolves according to the equation
(assuming HeI is in ionization equilibrium):
\begin{equation}
\frac{dx_{\rm HeII}}{dt} = -\Gamma_{\rm HeII}\; x_{\rm HeII} + \alpha_{\rm A} \; n_e \; x_{\rm HeIII}.
\label{eqn:xHeIIdt}
\end{equation}
If we take $n_e$ to be constant (a relatively good approximation during
$z \sim 3$ HeII reionization because hydrogen is
ionized), equation (\ref{eqn:xHeIIdt}) yields
\begin{equation}
  x_{\rm HeII}(t) = x_{\rm HeII, eq} + \left(x_{\rm HeII}(0)-  x_{\rm HeII, eq} \right) \; \exp\left[-t /t_{\rm eq}\right],
\label{eqn:xHeII}
\end{equation}
where $t_{\rm eq} = (\Gamma_{\rm HeII} + 1/t_{\rm rec})^{-1}$, $x_{\rm HeII}(0)$ is the initial HeII fraction, and
\begin{equation}
x_{\rm HeII, eq} = \frac{\alpha_{\rm A}(T) \; n_e}{\Gamma_{\rm HeII} +
\alpha_{\rm A}(T) \; n_e}.
\label{eqn:equil}
\end{equation}

Compare the exact solution for $x_{\rm HeII}(t)$ given by equation
(\ref{eqn:xHeII}) to that without recombinations (and where an
estimate for the number of recombinations $\alpha_{\rm A} \, n_e \, x_{\rm
HeII}(0) \, \Delta t$ is added to $x_{\rm HeII}$ at the
beginning of the timestep):
\begin{equation}
  \tilde{x}_{\rm HeII}(t) = \left[x_{\rm HeII}(0) + \alpha_{\rm A} \,  n_e \, x_{\rm HeIII}(0)\right] \; \exp\left[-\Gamma_{\rm HeII} \; t\right].
\label{eqn:xHeII_approx}
\end{equation}
The value $\tilde{x}_{\rm HeII}(t)$ is what is computed by
our HeII reionization code over a timestep $\Delta t$.  The
computation of $\tilde{x}_{\rm HeII}(t)$ allows us to determine the
effect of all sources on a single cell independently -- $\Gamma_{\rm HeII}$ in equation (\ref{eqn:xHeII_approx}) is the sum of the  $\Gamma_{\rm HeII}$ from all the incident rays -- aside from the issue of the order rays from different sources reach a cell which affects the heating rate in the cell and how the cell shadows other cells.
%\footnote{The ordering is achieved in our code by randomizing the order of rays, the number of timesteps, and by having multiple rays from a source hit cells that the source can substantially ionized.}

\begin{figure}
\begin{center}
\epsfig{file=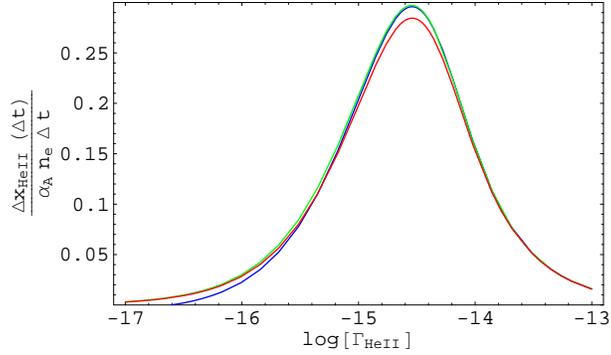, width=8cm}
\end{center}
\caption{Error in our code's calculation of $x_{\rm HeII}(\Delta
t)$ in units of the number of recombinations plotted as a function of
$\Gamma_{\rm HeII}$.  This calculation assumes $\Delta t = 20 \;
\Myr$.  The blue line represents the case $x_{\rm HeII}(0) = 0.01$ and
$\delta_b =0$, the green line is for $x_{\rm HeII}(0) = 1$ and
$\delta_b =0$, and the red line that for $x_{\rm HeII}(0) = 1$ and
$\delta_b =10$.
\label{fig:code_error}}
\end{figure}

Figure \ref{fig:code_error} shows that the error generated by using
equation (\ref{eqn:xHeII_approx}) rather than the true solution,
equation (\ref{eqn:equil}), is minor.  This figure plots the ratio of
$\Delta x_{\rm HeII}$ to an estimate for the number of recombinations
in a timestep $\alpha_{\rm A} \, n_e \, \Delta t$ as a function of
$\Gamma_{\rm HeII}$ for $\Delta t = 20 \; \Myr$.  The blue line is for
$x_{\rm HeII}(0) = 0.01$ and $\delta_b =0$, the green line for $x_{\rm
HeII}(0) = 1$ and $\delta_b =0$, and the red line for $x_{\rm HeII}(0)
= 1$ and $\delta_b =10$.  The error in $\Delta x_{\rm HeII}$ in all
cases is small, less than $0.3$ in units of the number of
recombinations (and $\Delta x_{\rm HeII} \lesssim0.005$).  For other
values of $\Delta t$, we find that $\Delta x_{\rm HeII}/(\alpha_{\rm A} \,
n_e \, \Delta t)$ peaks at roughly the same amplitude of $0.3$, but
with the peak at $\Gamma_{\rm HeII} \approx 1.8 \; {\Delta t}^{-1}$.

This method also reproduces the correct heating rate with minimal
error.  It is easy to check this in the case of ionization
equilibrium, the regime where one might suspect that this code is
unreliable.  In ionization equilibrium, the amount of heating is
\begin{equation}
\Delta Q = \Gamma_{\rm HeII} \, \langle \Delta E \rangle \, x_{\rm HeII, eq} \Delta t = \alpha_{\rm A} \; n_e \; \langle \Delta E \rangle \; \Delta t,
\label{eqn:deltaQ}
\end{equation}
where $\langle \Delta E \rangle$ is the average excess energy above $4$ Ry of a photon that enters a cell and ionized a HeII ion, and we have used equation \ref{eqn:equil} for $x_{\rm HeII, eq}$ in the limit $\Gamma_{\rm HeII}
\gg \alpha_{\rm A} \, n_E$.  Comparing $\Delta Q$ in equation \ref{eqn:deltaQ}
to the value our algorithm provides
\begin{equation}
\Delta Q = \Gamma_{\rm HeII} \, \langle \Delta E \rangle \, \Delta x_{\rm HeII, rec} = \alpha_{\rm A} \; n_e \; \langle \Delta E \rangle \, \Delta t,
\label{eqn:deltaQour}
\end{equation}
where we have used that $\Delta x_{\rm HeII, rec} \approx \alpha_{\rm
A, HeII} \; n_e \; \Delta t$ (since $x_{\rm HeII} \approx 1$) and that
all the HeIII ions that recombine are reionized (i.e., $\Gamma_{\rm
HeII} \gg 1/t_{\rm rec}$).
%Of course, the value of $\Delta E$ depends
%on the spectrum of the incident rays, which in our case are typically
%the first rays that enter during a timestep. 
%While the first rays that enter can come from any source owing to our
%randomization procedure for shooting rays, it is unclear whether the
%$\Delta E$ is accurately captured with this scheme compared to what
%happens in nature -- all incident rays contribute more equally to
%keeping a cell ionized.  {\bf look into this more}

Since an accurate estimate for $x_{\rm HeII}$ is also necessary for HeII
Ly$\alpha$ forest calculations, we correct the mistake our approximate method
incurs by substituting the true solution, i.e. equation (\ref{eqn:xHeIIdt}),
for cells that have $\tau_{\rm HeII,Ly\alpha}(\tilde{x}_{\rm HeII}, z,
\delta_b) > 10$ (which correspond to larger values of $\tau_{\rm
HeII,Ly\alpha}({x}_{\rm HeII}, z, \delta_b)$), using the formula
\begin{equation}
 \tau_{\rm HeII,Ly\alpha}(\tilde{x}_{\rm HeII}, z, \delta_b) = 3.4 \; \left(\frac{x_{\rm HeII}}{10^{-3}}\right)\left( \frac{1+z}{4} \right)^{3/2} \Delta_b.
 \label{eqn:correction}
\end{equation}
This correction might be worrisome because the algorithm no longer
consistently calculates the balance between recombinations and
ionizations. However, a minor correction of $\Delta x_{\rm HeII} =
|x_{\rm HeII}(\Delta t) - \tilde{x}_{\rm HeII}(\Delta t)|$ in a cell
is acceptable because it is smaller than other uncertainties, such as
uncertainties in the recombination coefficient (Section
\ref{ss:rec_rates}).
%For
%example, the number and spectrum of HeII ionizing photons from QSOs is
%not well-constrained.  Furthermore, some recombinations to the ground
%state will redshift out of resonance and not be reabsorbed and most of
%the recombination photons are absorbed in dense systems (Section
%\ref{ss:rec_rates}).  These effects are not modeled by our code.

While we have used Case A recombination coefficients in the tests
above, the equivalent discussion applies if we replace Case A with
Case B, as is normally done by our code (except $\Delta x_{\rm
HeII}$ will be smaller for Case B).  We always use the Case A
coefficient to correct our code's calculation because ground state
recombination photons are not absorbed locally in regions with
significant transmission.  Rather, these photons are absorbed in dense
(probably neutral) systems.  As a consequence, our calculation of
$\tau_{\rm HeII,Ly\alpha}$ misses the contribution to $\Gamma$ from
ground state recombinations and, therefore, underestimates the amount
of transmission.  We can put an upper limit on the contribution to
$\Gamma_{\rm HeII}$ from such recombinations.  If we assume that the
volume-averaged photo-ionization rate is just large enough to balance
the number of recombinations aside from those to the ground state and
that $55$ eV recombination photons have the same m.f.p. as the typical
ionizing photons from quasars, then $\bar{\Gamma}_{\rm rec, HeII} =
(\alpha_{\rm A} - \alpha_{\rm B})/\alpha_{\rm A} \; \bar{\Gamma}_{\rm
HeII} \approx 0.3 \, \bar{\Gamma}_{\rm HeII}$.  In our simulations,
the photo-ionization rate is larger than what is required to balance
the number of recombinations (except in select regions) and the
m.f.p. for HeII-ionizing recombination photons is shorter than for
other photons, such that $\bar{\Gamma}_{\rm rec, HeII} < 0.3 \,
\bar{\Gamma}_{\rm HeII}$.

\section{D. Dark Matter versus Gas Distribution and the $N_{\rm HI}$ distribution }
\label{ap:gdist}
\label{ap:NHI}

\begin{figure}
\begin{center}
\rotatebox{-90}{\epsfig{file=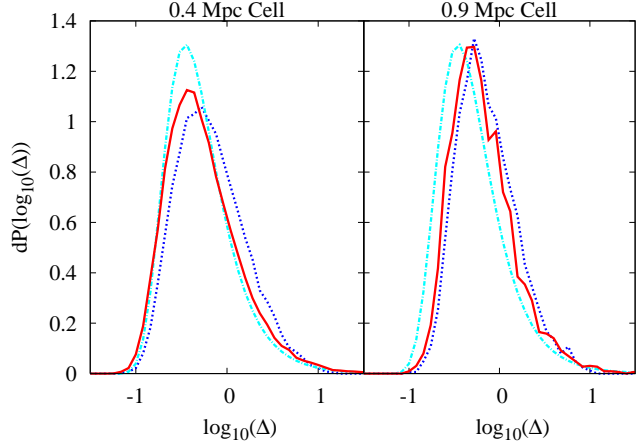, height=8.8cm}}
\end{center}
\caption{Comparison of the dark matter and gas volume-averaged PDFs for $z = 3.5$.
The solid curves are the dark
matter PDFs and the dashed curves are the gas PDFs measured in the
specified cell sizes.  The dot-dashed curves are the MHR gas
PDF.  
\label{fig:gasdist}}
\end{figure}

The radiative transfer computations in this study are performed as post
processing on top of gridded N-body fields.  This assumes that the gas fluctuations are pressure smoothed at
roughly the simulation grid scale and trace the dark matter on larger
scales.  Cold dark matter clumps on extremely small scales, and,
therefore, it is important to properly choose the scale at which to
grid the N-body field to approximate the gas pressure smoothing.
Here, we justify our grid sizes.

We use the $Q6$ simulation of
\citet{springel03}.  This simulation consists of a $14.3$ Mpc box with
$486^3$ dark matter particles and initially $486^3$ SPH particles.
The Q6 simulation includes gas cooling, star formation, the
\citet{haardt96} model for the ionizing background
\citep{katz96a, dave99},
as well as a
prescription for galactic winds \citep{springel02}.
It resolves the Jeans scale at mean
density with $\approx 10^4$ SPH particles, and it has a force
resolution of $1.2$ proper kpc. The following analysis uses the $z =
3.5$ snapshot from this simulation.

We first calculate the dark matter and gas overdensity PDFs to
quantify how well the gridded dark matter traces the gas.  This is
shown in Figure \ref{fig:gasdist} for $0.45$ and $0.89$ comoving Mpc
grid cells.  These cell sizes are comparable to the respective cell
sizes for the $512^3$ and $256^3$ grids in the $186$ Mpc simulation
box.  The solid curves are the dark matter PDFs on the grid and the
dotted are the same for the gas.  The dot-dashed curve is the
\citet{miralda00} (henceforth MHR) gas PDF.  This PDF represents that
of the pressure smoothed gas (grid artifacts should be unimportant).
The extent to which the gridded dark matter PDF agrees with the MHR
gas PDF reflects how well the clumpiness of the gas is captured in the
simulations.\footnote{Note that the MHR PDF is generated using
Eulerian simulations that employ a different cosmology and thermal
history from simulation Q6.  While the gas PDF is fairly robust to the
cosmology, the corresponding PDF of the pressure smoothed gas in
simulation G4 will be slightly different than the MHR PDF.  We choose
to use the MHR PDF because we find that particle noise in SPH
simulations makes it difficult to capture the low density PDF.}  The
low density gas PDF is important for accurately calculating from the
simulations the absorption in the HeII Ly$\alpha$ forest, the number
of recombinations per HeII-ionizing photon, as well as the filtering
of the radiation field.  Figure \ref{fig:gasdist} demonstrates that
the gas clumpiness is described reasonably well with the chosen grid
scales.

%Clumping factors are often used in studies of reionization.  However,
%the factor that is used is often the average clumping factor -- which
%amounts to the clumping factor needed to keep the densest blobs
%ionized (e.g., \citealt{springel03} and \citealt{Iliev06b}).  The
%average clumping factor at $z \sim 8$ is typically estimated to by
%$\approx 30$.  In reality, these dense blobs are not ionized during
%reionization, and this method vastly over-estimates the number of
%photons required to ionize the Universe.

 The high-density density regions, in particular densities responsible
for absorbers with $N_{\rm HI} \gtrsim 10^{15} \cmeter^{-2}$, are
important for filtering the HeII radiation field (as shown in Appendix
A).  Figure \ref{fig:NHI_dist} compares the $N_{\rm HI}$ column
density distribution of the coarse cell-smoothed dark matter (thin
curves) to the column density distribution of the pressure-smoothed
gas (thick curves), with both column densities measured along skewers
that have the same length as a coarse cell. Note that $N_{\rm HI} =
x_{\rm HI, eq} \; l_{\rm cell} \; \bar{n}_H(z) \; \Delta_b$, where we
use $\Gamma_{\rm HI} = 10^{-12}$ s$^{-1}$.  The blue dotted curves
represent the case in which $N_{\rm HI}$ is measured with $0.9$ Mpc
coarse cells.  The green dashed and red solid curves are the same but
measured in $0.45$ and $0.22$ Mpc coarse cells, respectively.  The
dark matter field, which for coarser grids underpredicts the number of
$N_{\rm HI} > 10^{15} \; \cmeter^{-2}$ absorbers, overproduces the
abundance of absorbers for $0.22$ Mpc cells.  Interestingly, the
$N_{\rm HI}$ distribution of the $0.45$ Mpc gridded field (which
corresponds roughly to the $512^3$ grid simulation in the $186$ Mpc
box) agrees well with the gas distribution.
%This may not be too surprising since these cells have
%mass equal to the Jeans mass for $T \approx 10^4$ K gas and
%overdensities of several (eqn. \ref{eqn:jeans}), overdensities
%comparable to where these $N_{\rm HI}$ column densities are found in
%the forest \citep{schaye00}.

%The cyan dot-dashed curves in Figure \ref{fig:NHI_dist} are the two
%best fit models for the $N_{\rm HI}$ distribution at $z = 3.5$ presented
%in \citet{fardal98} (their fits A1 and A2 using compilations of Voigt
%profile fits to the HI Ly$\alpha$ forest spectra).  These curves are
%in rough agreement with the curves computed from the Q6 simulation.  Note
%that the simulation curves should not agree exactly with these fits
%owing to the different method for measuring the column density
%distribution.  SPH simulations are able to reproduce the observed
%$N_{\rm HI}$ distribution over the relevant range when
%this distribution is measured by fitting Voigt profiles to mock spectra \citep{hernquist96}.

\begin{figure}
\begin{center}
\rotatebox{-90}{\epsfig{file=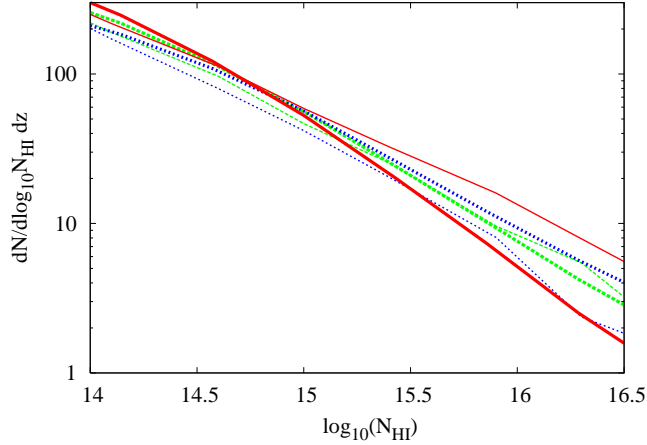, height=8.8cm}}
\end{center}
\caption{Number of absorbing columns per $z$ per
$\log_{10} N_{\rm HI}$.  The blue dotted curves represent the case in
which this quantity is measured in $0.89$ Mpc coarse cells.  The
thicker curve uses the average dark matter density in the coarse cell
to measure $N_{\rm HI}$, and the thin curve uses the gas density along a skewer the length of the coarse cell of the pressure-smoothed gas.  The green dashed and red solid curves are the same as
the blue dotted but measured in $0.45$ and $0.22$ Mpc coarse cells,
respectively. 
\label{fig:NHI_dist}}
\end{figure}

\section{E. QSO Modelling}
\label{ap:qso}
\subsection{QSO Spectral Index}
Both methods for populating the box with QSOs discussed in Section
\ref{sources} return the specific luminosity at the HI Lyman-limit.
We extrapolate from this specific luminosity to higher energies using
a power-law with index $\alpha_{\rm UV}$.  The parameter $\alpha_{\rm
UV}$ has been measured by \citet{telfer02} to be $\approx 1.6$, but
with an approximately Gaussian distribution with a large standard
deviation (s.d.)  of $0.86$ \citep{telfer02}.  The large s.d. in
$\alpha_{\rm UV}$ among QSOs may help explain transmission
fluctuations in the HeII Ly$\alpha$ forest at $z \approx 2.5$
\citep{shull04}.
%In addition, \citet{telfer02} finds no statistically
%significant correlation between $\alpha_{\rm UV}$ and the QSO
%luminosity, between $\alpha_{\rm UV}$ and the power-law index redward
%of the HI Lyman-limit, nor between $\alpha_{\rm UV}$ and the QSO's
%redshift.  
More recently, \citet{scott04} measured a much smaller mean
$\alpha_{\rm UV}$ from a sample of both FUSE and HST quasars.  They
derived an average value of $\alpha_{\rm UV} \approx 0.6$ and with a variance comparable to that of \citet{telfer02}. \citet{scott04} also find a strong correlation between $\alpha_{\rm UV}$
and QSO luminosity.  \citet{scott04} argued that the discrepancy
between their mean $\alpha_{\rm UV}$ and that of \citet{telfer02} likely owed to their sample containing lower luminosity quasars.

The \citet{telfer02} and \citet{scott04} values of $\alpha_{\rm UV}$
are derived from fitting $1  \; {\rm Ry} \lesssim E_{\gamma} \lesssim 4 \; {\rm Ry}$.  For HeII
reionization, the spectrum between $4$ Ry and $1$ keV is relevant.  If
we naively extrapolate from $1$ Ry to the soft X-ray using the
\citet{telfer02}, or particularly the \citet{scott04}, $\alpha_{\rm
UV}$ distribution this would result in an over-estimate of the soft
X-ray luminosity function for quasars because of the large variance
these studies measure in $\alpha_{\rm UV}$ (and in the case of
\citet{scott04}, the low value of $\bar{\alpha}_{\rm UV}$).  The
average spectral index $\alpha_{0,x}$ needed to join the luminosity at
$2500\; \AA$ and $2$ keV has been measured from a sample of optically selected quasars to
be $\sim 1.6$ \citep{steffen06}.\footnote{\citet{steffen06} found that
this power-law was correlated with luminosity, ranging from
$\alpha_{0,x} = 1.62$ for $L_{\rm bol} = 10^{43}$ erg s$^{-1}$ and
$\alpha_{0,x} = 1.25$ for $L_{\rm bol} = 10^{46}$ erg s$^{-1}$.  This
trend with $L_{\rm bol}$ is the opposite of what was found for
$\alpha_{\rm UV}$ in \citet{scott04}.}  Furthermore, \citet{steffen06}
measured a dispersion in this index $0.08-0.14$, significantly
smaller than the dispersion in $\alpha_{\rm UV}$ that \citet{telfer02}
and \citet{scott04} infer.

\subsection{Escape Fraction}
Not all of the ionizing radiation escapes from a quasar.  The fraction
of HeII-ionizing radiation that is obscured is denoted by $f_{\rm
cov}$.  Measurements in the X-ray suggest $f_{\rm cov} \approx 0.5$,
with evidence that $f_{\rm cov}$ increases with luminosity
\citep{gilli07}.  We use the \citet{gilli07} fitting function for
$f_{\rm cov}$ in calculations that require it.\footnote{However, $f_{\rm cov}$ is defined in \citet{gilli07}
as the fraction of QSOs with $\log_{10}(N_{\rm HI}) > 21$ in
c.g.s. units, which is not exactly the fraction of radiation that
escapes, which for is sensitive to the number of sight-lines with
$\log_{10}(N_{\rm HeII}) > 17.9$.}
%\citet{gilli07} parametrize $f_{\rm cov}
%= [1 + R(L)^{-1}]^{-1}$ where
%\begin{equation}
%R(L_X) = R_S \; \exp \left(-L_X/L_c \right) + R_Q \;\left(1-  \exp \left(-L_X/L_c \right) \right),
%\label{eqn:gilli}
%\end{equation}
%with $\log_{10} L_c = 43.5$ and $L_X$ is the luminosity in the $0.5-2$
% keV band.  The best fit model in \citet{gilli07} is $R_S = 3.7$ and
 %$R_Q = 1$. 

The fraction of ionizing radiation that escapes from QSOs is almost certainly
directionally dependent -- it will be easiest for ionizing photons to
escape on trajectories that do not intersect infalling material or
galactic disks.  In select simulations, we take a beam for the
ionizing radiation from the QSO with solid angle $\Omega_b = 2 \pi \; (1 -
f_{\rm cov})$ along both axes.  We assume the symmetry axis of the beam is
randomly oriented.  For the other simulations, we uniformly suppress the
intensity along all sight-lines from the QSO by the factor $f_{\rm
cov}$.  This approximates the scenario in which the obscuration owes to
infalling material and (when averaged over the QSO lifetime) is
isotropic.   The value of the covering fraction is only required for Method I if the quasars are beamed.  It affects the quasars in Method II by suppressing their ionizing luminosity.

%for optically selected quasars, \citet{scott04} finds no
 %evidence for a break at the HI Lyman-limit in their composite
 %spectrum, suggesting that $f_{\rm esc} \approx 1$ for this population.

\end{appendix}

\bibliographystyle{aastex}
%\bibliography{References}

\label{lastpage}
\end{document}